\PassOptionsToPackage{prologue,dvipsnames}{xcolor}
\documentclass[acmsmall,screen]{acmart}

\setcopyright{rightsretained}
\acmDOI{10.1145/3704895}
\acmYear{2025}
\acmJournal{PACMPL}
\acmVolume{9}
\acmNumber{POPL}
\acmArticle{59}
\acmMonth{1}
\received{2024-07-11}
\received[accepted]{2024-11-07}

\usepackage{ifthen}
\usepackage{xparse}

\ifx\beamer\undefined
\else
\fi

\usepackage[utf8]{inputenc}
\usepackage{amsmath}
\usepackage{stmaryrd}
\usepackage{amsthm}
\usepackage{mathtools}
\usepackage{proof}
\usepackage{comment}
\usepackage{textcomp}
\usepackage[us]{optional}
\usepackage{color}
\usepackage{url}
\usepackage{verbatim}
\usepackage{mathpartir}
\usepackage{booktabs}
\usepackage{multirow}
\usepackage{tikz}
\usetikzlibrary{automata, positioning, arrows}

\usepackage{stackengine}
\usepackage{scalerel}

\usepackage{hyperref}

\usepackage{array}

\usepackage[clock]{ifsym}

\renewrobustcmd{\path}[1]{\colorbox{green!60!black!12}{\texttt{\detokenize{#1}}}}

\usepackage{listings}

\makeatletter
\lst@Key{matchrangestart}{f}{\lstKV@SetIf{#1}\lst@ifmatchrangestart}
\def\lst@SkipToFirst{%
    \lst@ifmatchrangestart\c@lstnumber=\numexpr-1+\lst@firstline\fi
    \ifnum \lst@lineno<\lst@firstline
        \def\lst@next{\lst@BeginDropInput\lst@Pmode
        \lst@Let{13}\lst@MSkipToFirst
        \lst@Let{10}\lst@MSkipToFirst}%
        \expandafter\lst@next
    \else
        \expandafter\lst@BOLGobble
    \fi}
\makeatother

\definecolor{background_color}{RGB}{240, 240, 240}
\definecolor{string_color}    {RGB}{180, 156,   0}
\definecolor{keyword_color}   {RGB}{ 64, 100, 255}
\definecolor{comment_color}   {RGB}{  0, 117, 110}
\definecolor{number_color}    {RGB}{ 84,  84,  84}
\lstdefinelanguage{sml}{
    language=ML,
    morestring=[b]",
    morecomment=[s]{(*}{*)},
    morekeywords={
        bool, char, exn, int, real, string, unit, list, option,
        EQUAL, GREATER, LESS, NONE, SOME, nil,
        andalso, orelse, true, false, not,
        if, then, else, case, of, as,
        let, in, end, local, val, rec,
        datatype, type, exception, handle,
        fun, fn, op, raise, ref,
        structure, struct, signature, sig, functor, where,
        include, open, use, infix, infixr, o, print
    }
}

\newrobustcmd{\code}[2][]{{\sloppy
\ifmmode
    \text{\colorbox{background_color}{\lstinline[language=sml,#1]`#2`}}
\else
    {\colorbox{background_color}{\lstinline[language=sml,#1]`#2`}}%
\fi}}
\lstnewenvironment{codeblock}[1][]{\lstset{language=sml,numbers=none,#1}}{}
\newrobustcmd{\codefile}[2][]{%
  \lstinputlisting%
    [language=sml,mathescape=false,frame=single,xleftmargin=3pt,xrightmargin=3pt,title={\path{#2}},numbers=left,#1]%
    {../#2}
}

\usepackage{newunicodechar}
\newunicodechar{λ}{$\lambda$}

\NewDocumentCommand{\Infer}{s o m m}{%
  \IfBooleanTF{#1}%
    {\IfNoValueTF{#2}%
      {\inferrule{#3}{#4}}%
      {\inferrule*[right={#2}]{#3}{#4}}}%
    {\IfNoValueTF{#2}%
      {\inferrule{#3}{#4}}%
      {\inferrule*[Right={#2}]{#3}{#4}}}%
}

{\begin{equation*}
      \label{eqn:#1}
      \renewcommand{\bnfalt}{\mathrel{\phantom{\mid}}}
      \begin{array}{llc@{\quad\extracolsep{\fill}}lll}
}
{\end{array}\end{equation*}\ignorespacesafterend}

\newenvironment{synchart*}[1]%
{\begin{equation*}
      \label{syn:#1}
      \renewcommand{\bnfalt}{\mathrel{\phantom{\mid}}}
      \begin{array}{llc@{\quad\extracolsep{\fill}}lll}
        \textit{Sort} & & & \textit{Abstract} & \textit{Concrete} & \textit{Description} \\
}
{\end{array}\end{equation*}\ignorespacesafterend}

\newenvironment{synchartmini}[1]%
{\begin{equation*}
      \label{eqn:#1}
      \renewcommand{\bnfalt}{\mathrel{\phantom{\mid}}}
      \begin{array}{llc@{\quad\extracolsep{\fill}}ll}
}
{\end{array}\end{equation*}\ignorespacesafterend}

\newtheorem*{notation}{Notation}

\DeclareDocumentCommand{\coloncolon}{ }{::}
\DeclareDocumentCommand{\coloncolonequals}{ }{::=}

\makeatletter
\newcommand{\xMapsto}[2][]{\ext@arrow 0599{\Mapstofill@}{#1}{#2}}
\def\Mapstofill@{\arrowfill@{\Mapstochar\Relbar}\Relbar\Rightarrow}
\makeatother

\lstdefinelanguage{DSL}{
    morestring=[b]",
    morecomment=[s]{/*}{*/},
    morekeywords={
        bool, u128, string, unit,
        pub, fn, type, where
    }
}

\newrobustcmd{\dsl}[2][]{{\sloppy
\ifmmode
    \text{\colorbox{background_color}{\lstinline[language=DSL,#1]`#2`}}
\else
    {\colorbox{background_color}{\lstinline[language=DSL,#1]`#2`}}%
\fi}}
\lstnewenvironment{dslblock}[1][]{\lstset{language=dsl,numbers=none,#1}}{}

\newrobustcmd{\rust}[2][]{{\sloppy
\ifmmode
    \text{\colorbox{background_color}{\texttt{#2}}}
\else
    {\colorbox{background_color}{\texttt{#2}}}%
\fi}}

\newcommand{\xref}{\nameref}
\makeatletter
\newcommand\xlabel[2][]{\phantomsection\def\@currentlabelname{#1}\label{#2}}
\makeatother

\NewDocumentCommand{\defrule}{o o m m}{
  \inferrule*[rightskip=2em,lab=\IfNoValueTF{#1}{}{[#1]}]
  { #3 }
  { #4 }
  \IfNoValueTF{#2}{}{\xlabel[#1]{\ifdefined\InApx{apx:}\else\fi#2}}
}

\NewDocumentCommand{\defsteprule}{o o m m m m}
{\IfNoValueTF{#1}{}{{\hbox to 2cm {\textsc{[#1]}\hfil}}}
\ensuremath{{#3}\LLStepsTo{#4}{#5} {#6}}
\IfNoValueTF{#2}{}{\xlabel[#1]{#2}}}

\NewDocumentCommand{\defsensorrule}{o o m m}
{\IfNoValueTF{#1}{}{\hbox to 1.5cm {\textsc{[#1]}\hfil}}
\ensuremath{{\hbox to 3cm {\ensuremath{#3}\hfil}} \quad\slash\quad {#4}}
\IfNoValueTF{#2}{}{\xlabel[#1]{#2}}}

\NewDocumentCommand{\defruler}{o o m m}{
  \inferrule*[right={\IfNoValueTF{#1}{}{[#1]}}]
  { #3 }
  { #4 }
  \IfNoValueTF{#2}{}{\xlabel[#1]{#2}}
}

\NewDocumentCommand{\ruleref}{m}{{[\textsc{\xref{#1}}]}}

\NewDocumentCommand{\itemheader}{m}{\vspace{1em}\textsc{\textbf{#1}}\vspace{1em}}

\newif\ifrevisionmarker

\NewDocumentEnvironment{revision}{s O{blue}}
  {\ifrevisionmarker\bgroup \color{#2}\fi}
  {\ifrevisionmarker\egroup\fi}
\NewDocumentCommand{\irevision}{s O{blue} m}
  {\ifrevisionmarker\bgroup\color{#2}{#3}\egroup\else{#3}\fi}

\newif\ifappendix

\NewDocumentCommand{\ifextended}{m}{\ifappendix{#1}\fi}

\NewDocumentCommand{\apxname}{s}{%
  \IfBooleanTF{#1}%
    {the extended version of this paper~\citep{fullpaper}}%
    {the full paper~\citep{fullpaper}}%
}

\NewDocumentCommand{\refapx}{s o}{%
  \ifappendix%
    \IfBooleanTF{#1}{the appendix}{\Cref{#2}}%
  \else%
    \IfBooleanTF{#1}{\apxname*}{\apxname}\fi%
  \xspace%
}

\usepackage{import}

\newboolean{live}             
\newboolean{solutions}          
\newboolean{solutionsonly} 
\newboolean{judgmentmodes}      
\newboolean{restrictedpoly}     
\newboolean{polarization}       
\newboolean{deptypes}           
\newboolean{canonization}       
\newboolean{patterns}      
\newboolean{situated}  
\newboolean{past}  
\newboolean{doublespace} 
\newboolean{modalexcs} 
\newboolean{bisim} 
\newboolean{abbreviated} 
\newboolean{symops} 
\newboolean{infixprodsum}  
\newboolean{3rded} \setboolean{3rded}{false}

\AtBeginDocument{\mathcode`\;=\numexpr\mathcode`\;-\string"4000\relax}

\usepackage{stackengine}
\usepackage{scalerel}

\stackMath

\newcommand{\ifnull}[3]{\ifthenelse{\equal{#1}{}}{#2}{#3}}

\newcommand{\bnfdef}{\coloncolonequals}

\newcommand{\bnfalt}{\mathrel{\mid}}

\newcommand{\entails}[1][]{\vdash_{#1}}

\DeclarePairedDelimiter{\parens}{(}{)}
\DeclarePairedDelimiter{\braces}{\{}{\}}
\DeclarePairedDelimiter{\sqbracks}{[}{]}

\newcommand{\kw}[1]{\ensuremath{\mathtt{#1}}}
\newcommand{\kwop}[1]{\ensuremath{\mathop{\mathtt{#1}}\nolimits}}
\newcommand{\kwbin}[1]{\ensuremath{\mathbin{\mathtt{#1}}}}

\newcommand{\cdperiod}{\kw{.}}
\newcommand{\cddot}{\cdperiod}

\newcommand{\cdbar}{\kwbin{|}} 

\newcommand{\cdparens}[1]{\kw{(}\,#1\,\kw{)}}

\newcommand{\cdbraces}[1]{\kw{\{}{#1}\kw{\}}}

\newcommand{\calA}{\mathcal{A}}

\newcommand{\calG}{\mathcal{G}}

\newcommand{\calR}{\mathcal{R}}

\ifthenelse{\boolean{restrictedpoly}}{

}{}

\ifthenelse{\boolean{modalexcs}}{}{}

\newcommand{\Sort}[1]{\textsf{#1}}

\ifthenelse{\boolean{3rded}}{}{}

\newcommand{\OpABT}[2]{{#1}\cdparens{#2}}

\newcommand{\ABS}[2]{{#1}\mathbin{\cddot}{#2}}
\newcommand{\AbsABT}[2]{\ABS{#1}{#2}}

\newcommand{\TypeSort}{\Sort{Typ}}

\newcommand{\Subst}[3]{\braces{{#1}\mathord{/}{#2}}{#3}}

\newcommand{\JInfix}[3]{{#1}\mathrel{#2}{#3}}

\newcommand{\IsOf}[2]{\JInfix{#1}{:}{#2}}

\newcommand{\IsOfKd}[2]{\JInfix{#1}{\coloncolon}{#2}}

\newcommand{\StepsTo}[1][]{\xmapsto[\,#1\,]{\hspace*{1em}}}

\newcommand{\LLStepsTo}[2]{\xmapsto[\,#2\,]{\,#1\,}}

\newcommand{\NFold}[2]{#1^{#2}}

\newcommand{\MultiStepsTo}[1][\ast]{\NFold{\StepsTo}{#1}}

\ifthenelse{\boolean{symops}}{
  
}{
  
}

\newcommand{\landcst}[2]{#1\wedge #2}

\newcommand{\lorcst}[2]{#1\vee #2}

\newcommand{\limpcst}[2]{#1\supset #2}

\newcommand{\lnotcst}[1]{\neg #1}

\newcommand{\newchcmdcst}[1]{\kw{newch}}

\newcommand{\nullevtexcst}[1]{\kw{never}}

\newcommand{\exchandlecst}[7]%
{\kwop{tryexc}{#1}\kwop{ow}{\varinexcst{{#3}_{1}}{{#4}_{1}}\casebrcst{{#5}_{1}}\casesepcst\dots\casesepcst\varinexcst{{#3}_{#2}}{{#4}_{#2}}\casebrcst{#5}_{#2}\casesepcst{#6}\casebrcst{#7}}}

\newcommand{\dzerocst}{\dzeroabt}

\newcommand{\evalstosymbol}{\ensuremath{\Downarrow}}
\newcommand{\evalsto}{\mathrel{\evalstosymbol}}

\newcommand{\numcst}[1]{#1}

\ifthenelse{\boolean{symops}}{
  
}{
  
}

\ifthenelse{\boolean{symops}}{
  
}{
  
}

\ifthenelse{\boolean{symops}}{
  
}{
  
}

\ifthenelse{\boolean{canonization}}{

}{}

\ifthenelse{\boolean{symops}}{
  
}{
  
}

\ifthenelse{\boolean{symops}}{
  
}{
  
}

\ifthenelse{\boolean{restrictedpoly}}{

}{}

\ifthenelse{\boolean{symops}}{

}{

}

\newcommand{\pairexcst}[2]{\langle #1, #2\rangle}

\ifthenelse{\boolean{infixprodsum}}{
  
}{

}

\ifthenelse{\boolean{infixprodsum}}{

}{

}

\ifthenelse{\boolean{symops}}{

}{

}

\newcommand{\inttyabt}{\kw{int}}
\newcommand{\inttycst}{\inttyabt}

\newcommand{\casesepcst}{\ensuremath{\mathbin{\cdbar}}}
\newcommand{\casebrcst}{\ensuremath{\mathbin{\hookrightarrow}}}

\ifthenelse{\boolean{symops}}{

}{

}

\newcommand{\inexcst}[3]{{#2}\mathbin{\cdot}{#3}}
\newcommand{\varinexcst}[2]{\inexcst{}{#1}{#2}}

\ifthenelse{\boolean{infixprodsum}}{
  
}{

}

\newcommand{\booltycst}{\kw{bool}}
\newcommand{\trexabt}{\kw{true}}
\newcommand{\trexcst}{\trexabt}
\newcommand{\faexabt}{\kw{false}}
\newcommand{\faexcst}{\faexabt}

\ifthenelse{\boolean{infixprodsum}}{

}{

}

\newcommand{\toprefabt}[1]{\top}
\newcommand{\toprefcst}[1]{\top}

\DeclareDocumentCommand{\calA}{}{\mathcal{A}}

\DeclareDocumentCommand{\calF}{}{\mathcal{F}}
\DeclareDocumentCommand{\calG}{}{\mathcal{G}}

\DeclareDocumentCommand{\calR}{}{\mathcal{R}}

\DeclareDocumentCommand{\hatA}{}{\hat{A}}
\DeclareDocumentCommand{\hatB}{}{\hat{B}}
\DeclareDocumentCommand{\hatC}{}{\hat{C}}

\DeclareDocumentCommand{\hatP}{}{\hat{P}}
\DeclareDocumentCommand{\hatQ}{}{\hat{Q}}

\DeclareDocumentCommand{\hatT}{}{\hat{T}}

\DeclareDocumentCommand{\hatDelta}{}{\hat{\Delta}}

\DeclareDocumentCommand{\hatp}{}{\hat{p}}
\DeclareDocumentCommand{\hatq}{}{\hat{q}}

\DeclareDocumentCommand{\bfOmega}{}{\mathbf{\Omega}}

\DeclareDocumentCommand{\bfw}{}{\mathbf{w}}

\newcommand{\DStepsTo}[1][]{\xMapsto[\,#1\,]{\hspace*{1em}}}

\DeclareDocumentCommand{\SetCmp}{m m}{\{{#1}\mid{#2}\}}

\NewDocumentCommand{\Rel}{m m}{\mathcal{R}({#1};{#2})}

\NewDocumentCommand{\Dom}{m}{\mathsf{Dom}(#1)}

\DeclareDocumentCommand{\unitstycst}{}{\ensuremath{\mathbf{1}}}

\DeclareDocumentCommand{\larrstycst}{ m m }{{#1}\mathrel{\multimap}{#2}}

\DeclareDocumentCommand{\tensorstycst}{ m m }{{#1}\mathrel{\otimes}{#2}}

\DeclareDocumentCommand{\bsumstycst}{ m m }{\sumtyabt{#1}{#2}}
\DeclareDocumentCommand{\bsumstycst}{ m m }{{#1}\mathrel{\oplus}{#2}}

\DeclareDocumentCommand{\bwithstycst}{ m m }{{#1}\mathrel{\&}{#2}}

\DeclareDocumentCommand{\ProcSort}{}{\Sort{Proc}}

\DeclareDocumentCommand{\fwdcst}{ m }{\OpABT{\kw{fwd}}{#1}}

\DeclareDocumentCommand{\bangstycst}{ m m }{{\,!\,}_{#1}\cdot\,{#2}}
\DeclareDocumentCommand{\querystycst}{ m m }{{\,?\,}_{#1}\cdot\,{#2}}

\DeclareDocumentCommand{\ConfSort}{}{\Sort{Conf}}

\DeclareDocumentCommand{\IsOfProc}{ m m }{\IsOfKd{#1}{#2}}

\DeclareDocumentCommand{\fshiftcst}{ m m }{{#1} + {#2}}

\DeclareDocumentCommand{\finitcst}{ }{\kw{init}}

\DeclareDocumentCommand{\dzerocst}{ }{\bar{0}}

\DeclareDocumentCommand{\dunitcst}{ }{\bar{1}}

\DeclareDocumentCommand{\dsumcst}{ m m }{{#1} + {#2}}

\DeclareDocumentCommand{\ddiffcst}{ m m }{{#1} - {#2}}

\DeclareDocumentCommand{\tpunitstycst}{ m m }{\ensuremath{\mathbf{1}^{\AbsABT{#1}{#2}}}}

\DeclareDocumentCommand{\tplarrstycst}{ m m m m }{{#3}\mathrel{\multimap^{\AbsABT{#1}{#2}}}{#4}}

\DeclareDocumentCommand{\tptensorstycst}{ m m m m }{{#3}\mathrel{\otimes^{\AbsABT{#1}{#2}}}{#4}}

\DeclareDocumentCommand{\tpsumstycst}{ m m m m }{{#3}\mathrel{\oplus^{\AbsABT{#1}{#2}}}{#4}}

\DeclareDocumentCommand{\tpwithstycst}{ m m m m }{{#3}\mathrel{{\&}^{\AbsABT{#1}{#2}}}{#4}}

\DeclareDocumentCommand{\tpbangstycst}{ m m m m }{{\,!\,}_{#3}^{\AbsABT{#1}{#2}}.\,{#4}}
\DeclareDocumentCommand{\tpquerystycst}{ m m m m }{{\,?\,}_{#3}^{\AbsABT{#1}{#2}}.\,{#4}}

\DeclareDocumentCommand{\ProcSort}{}{\Sort{Proc}}

\DeclareDocumentCommand{\lblleft}{}{\kw{left}}
\DeclareDocumentCommand{\lblright}{}{\kw{right}}

\DeclareDocumentCommand{\tpfwdcst}{ m m }{\kw{fwd}^{#1}{#2}}

\DeclareDocumentCommand{\tpclosecst}{ m m }{\kw{close}^{\AbsABT{#1}{#2}}}

\DeclareDocumentCommand{\tpwaitcst}{ m m m }{\kw{wait}^{#1}{#2};{#3}}

\DeclareDocumentCommand{\tplarrrcvcst}{ m m m m m }{\lambda^{\AbsABT{#1}{#2}}{\parens{\IsOf{#4}{#3}}{#5}}}

\DeclareDocumentCommand{\tplarrsndcst}{ m m m m }{\kw{app}^{#2}{#1}\,\parens{#3};{#4}}

\DeclareDocumentCommand{\tptensorsndcst}{ m m m m }{{#3}\mathrel{\otimes^{\AbsABT{#1}{#2}}}{#4}}

\DeclareDocumentCommand{\tptensorrcvcst}{ m m m m }{\kw{split}^{#2}\,{#1};\AbsABT{#3}{#4}}

\DeclareDocumentCommand{\tpinlcst}{ m m m m m }{\kw{switchL}^{\AbsABT{#1}{#2}};{#5}}
\DeclareDocumentCommand{\tpinrcst}{ m m m m m }{\kw{switchR}^{\AbsABT{#1}{#2}};{#5}}

\DeclareDocumentCommand{\tpcasecst}{ m m m m }{\kw{case}^{#1}\,{#2}\,\cdbraces{{#3} \mid {#4}}} 

\DeclareDocumentCommand{\tpoffercst}{ m m m m }{\kw{offer}^{\AbsABT{#1}{#2}}\,\cdbraces{{#3} \mid {#4}}}

\DeclareDocumentCommand{\tpselectlcst}{ m m m }{\kw{selL}^{#2}{#1};{#3}}

\DeclareDocumentCommand{\tpselectrcst}{ m m m }{\kw{selR}^{#2}{#1};{#3}}

\DeclareDocumentCommand{\tpspawnproccst}{ m m m m }{\kw{spawn}^{#1}{#2};{\AbsABT{#3}{#4}}}
\DeclareDocumentCommand{\tpproduceproccst}{ m m m m }{\kw{produce}^{\AbsABT{#1}{#2}}\,{#3};{#4}}
\DeclareDocumentCommand{\tpconsumeproccst}{ m m m m}{\kw{consume}^{#1}\,{#2};\AbsABT{#3}{#4}}
\DeclareDocumentCommand{\tpqueryproccst}{ m m m m }{\kw{query}^{\AbsABT{#1}{#2}};\AbsABT{#3}{#4}}
\DeclareDocumentCommand{\tpsupplyproccst}{ m m m m }{\kw{supply}^{#1}\,{#2}({#3});{#4}}

\DeclareDocumentCommand{\ConfSort}{}{\Sort{Conf}}

\DeclareDocumentCommand{\ClockProc}{m}{\ensuremath{{\,\,\VarClock\,\,}_{#1}}}
\DeclareDocumentCommand{\RunProcAbt}{ m m }{\OpABT{\kw{proc}\sqbracks{#1}}{#2}}
\DeclareDocumentCommand{\RunProc}{ m m }{\RunProcAbt{#1}{#2}}
\DeclareDocumentCommand{\FwdProc}{ m m }{\OpABT{\kw{fwd}\sqbracks{#1}}{#2}}
\DeclareDocumentCommand{\StopConf}{}{\mathop{\mathbf{1}}}

\DeclareDocumentCommand{\actsnd}{m m}{{#1}\mathbin{!}{#2}}
\DeclareDocumentCommand{\actrcv}{m m}{{#1}\mathbin{?}{#2}}
\DeclareDocumentCommand{\actsil}{}{\varepsilon}

\DeclareDocumentCommand{\actsndchancst}{m m}{\actsnd{#1}{#2}}
\DeclareDocumentCommand{\actrcvchancst}{m m}{\actrcv{#1}{#2}}
\DeclareDocumentCommand{\actsndlblcst}{m m}{\actsnd{#1}{#2}}
\DeclareDocumentCommand{\actrcvlblcst}{m m}{\actrcv{#1}{#2}}
\DeclareDocumentCommand{\actsndclosecst}{m}{\actsnd{#1}{\kw{cls}}}
\DeclareDocumentCommand{\actrcvclosecst}{m}{\actrcv{#1}{\kw{cls}}}
\DeclareDocumentCommand{\actsndvalcst}{m m}{\actsnd{#1}{\kw{val}(#2)}}
\DeclareDocumentCommand{\actrcvvalcst}{m m}{\actrcv{#1}{\kw{val}(#2)}}

\DeclareDocumentCommand{\IsOfProc}{ m m m }{\IsOfKd{{#1}\mathrel{@}{#2}}{#3}}

\DeclareDocumentCommand{\Interp}{m}{\left\lceil{#1}\right\rceil}

\DeclareDocumentCommand{\FwdComp}{m m m}{{#1}\rtimes{#2}\mathrel{@}{#3}}
\DeclareDocumentCommand{\CutComp}{m m m}{{#1}\ltimes{#2}\mathrel{@}{#3}}

\DeclareDocumentCommand{\SchedStepsTo}{ m }{\DStepsTo[]{}}

\DeclareDocumentCommand{\topcst}{ }{\top}

\DeclareDocumentCommand{\botcst}{ }{\bot}

\DeclareDocumentCommand{\eqcst}{ m m }{{#1}={#2}}

\DeclareDocumentCommand{\leqcst}{ m m }{{#1}\leq{#2}}

\DeclareDocumentCommand{\fbetween}{m m m}{{#1}\leq{#2}\leq{#3}}

\DeclareDocumentCommand{\Interp}{m}{\llbracket{#1}\rrbracket}

\DeclareDocumentCommand{\expose}{m m}{{{#1}{\,|}_{#2}}}
\DeclareDocumentCommand{\wp}{m m m}{{#1}^{\AbsABT{#2}{#3}}}

\DeclareDocumentCommand{\LInt}{m m}{\mathcal{L}\llbracket{{#1}}\rrbracket\mathrel{@}{#2}}
\DeclareDocumentCommand{\VInt}{m m}{\mathcal{V}\llbracket{{#1}}\rrbracket\mathrel{@}{#2}}
\DeclareDocumentCommand{\LSInt}{m m}{\mathcal{L}^\star\llbracket{{#1}}\rrbracket\mathrel{@}{#2}}
\DeclareDocumentCommand{\VSInt}{m m}{\mathcal{V}^\star\llbracket{{#1}}\rrbracket\mathrel{@}{#2}}

\DeclareDocumentCommand{\LAnyInt}{m m}{\mathcal{L}^{(\star)}\llbracket{{#1}\rrbracket}{\mathrel{@}{#2}}}
\DeclareDocumentCommand{\VAnyInt}{m m}{\mathcal{V}^{(\star)}\llbracket{{#1}\rrbracket}{\mathrel{@}{#2}}}

\DeclareDocumentCommand{\sementails}{m}{\gg{#1}}

\DeclareDocumentCommand{\gl}{m m}{\langle{#1}, {#2}\rangle}

\DeclareDocumentCommand{\StepRefl}{m m}{\kw{refl}_{#1}^{#2}}
\DeclareDocumentCommand{\StepT}{m m m m}{\OpABT{\kw{stepT}_{{#1};{#2}}^{#3}}{#4}}
\DeclareDocumentCommand{\StepC}{m m m m}{\OpABT{\kw{stepC}_{{#1}}^{#2;#3}}{#4}}

\DeclareDocumentCommand{\TInt}{m m}{\left[\,{#1},{#2}\,\right)}
\DeclareDocumentCommand{\TIntInf}{m}{\left[\,{#1}, \infty\,\right)}

\DeclareDocumentCommand{\Trj}{m}{\kw{C}_{#1}}

\DeclareDocumentCommand{\trjTriv}{m}{\kw{i}_{#1}}
\DeclareDocumentCommand{\trjConst}{m m}{\kw{c}_{#1}^{#2}}

\DeclareDocumentCommand{\trjExt}{m m m}{\kw{Ext}_{#1}^{#2}\parens{#3}}
\DeclareDocumentCommand{\trjCat}{m m}{{#1}\mathbin{@}{#2}}
\DeclareDocumentCommand{\trjConc}{m m}{{#1}\otimes{#2}}

\DeclareDocumentCommand{\cwl}{m m}{S_{\TIntInf{#1}}({#2})}
\DeclareDocumentCommand{\cwls}{m m m m}{S_{\TInt{#1}{#2}}({#3; #4})}
\DeclareDocumentCommand{\ctrjI}{m m m}{S_{#1}({#2; #3})}
\DeclareDocumentCommand{\ctrj}{m m m m}{\ctrjI{\TInt{#1}{#2}}{#3}{#4}}

\DeclareDocumentCommand{\constcwl}{m m}{\OpABT{\kw{k}_{#1}}{#2}}

\DeclareDocumentCommand{\concat}{m m}{{#1}\mathbin{@}{#2}}

\DeclareDocumentCommand{\opconc}{ }{\otimes}
\DeclareDocumentCommand{\conc}{m m}{{#1}\otimes{#2}}
\DeclareDocumentCommand{\concfam}{m m}{\bigotimes_{#1}{#2}}

\DeclareDocumentCommand{\ilfam}{m m}{\bigotimes_{#1}{#2}}
\DeclareDocumentCommand{\il}{m m}{{#1}\otimes{#2}}

\DeclareDocumentCommand{\lpar}{m m}{{#1}\downarrow^{#2}}
\DeclareDocumentCommand{\rpar}{m m}{{#1}\uparrow^{#2}}

\DeclareDocumentCommand{\cwlequiv}{m m}{{#1}\sim{#2}}
\DeclareDocumentCommand{\cwlequivt}{m m m}{{#2}\sim_{#1}{#3}}

\DeclareDocumentCommand{\SensorProcAbt}{ m m m }{\OpABT{\kw{sensor}}{#1;#2[#3]}}
\DeclareDocumentCommand{\SensorProc}{ m m m }{\SensorProcAbt{#1}{#2}{#3}}

\NewDocumentCommand{\TODO}{m m}%
  {{\bfseries\color{#1}[#2]}}%

\usepackage{xspace}
\newcommand{\tillst}{TILLST\xspace} 
\newcommand{\tilst}{TILLST\xspace} 
\newcommand{\tslr}{TSSLR\xspace} 
\newcommand{\tsslr}{TSSLR\xspace} 

\newcommand{\ilst}{ILLST\xspace} 
\newcommand{\illst}{ILLST\xspace} 

\usepackage[nameinlink,noabbrev,capitalise]{cleveref}
\usepackage[nameinlink]{cleveref}
\Crefname{section}{\S\!}{\S}
\Crefname{figure}{Fig.}{Figs.}
\Crefname{tabular}{Tab.}{Tabs.}
\Crefname{theorem}{Thm.}{Thms.}
\Crefname{lemma}{Lem.}{Lems.}
\Crefname{definition}{Def.}{Defs.}
\Crefname{challenge}{Challenge}{Challenge}
\Crefname{mode}{Mode}{Mode}
\Crefname{mode-l}{Mode of Use}{Mode of Use}

\newcommand*{\eg}{\emph{e.g.,}\@\xspace}
\newcommand*{\ie}{\emph{i.e.,}\@\xspace}
\newcommand*{\aka}{\emph{a.k.a.,}\@\xspace}

\newcommand*{\respb}{\emph{resp}\@\xspace}
\newcommand*{\etc}{\emph{etc}\@\xspace}

\newcommand*{\where}{\quad\text{~where~}\quad}
\newcommand*{\andthat}{\quad\text{~and~}\quad}

\newcommand{\millisec}[1]{{#1}~ms}

\appendixtrue

\begin{document}

\title{Semantic Logical Relations for Timed Message-Passing Protocols\ifextended{\,(Extended Paper)}}



\author{Yue Yao}
\affiliation{%
\institution{Carnegie Mellon University}
\city{Pittsburgh}
\country{USA}
}
\email{yueyao@cs.cmu.edu}

\author{Grant Iraci}
\affiliation{%
\institution{University at Buffalo}
\city{Buffalo}
\country{USA}
}
\email{grantira@buffalo.edu}

\author{Cheng-En Chuang}
\affiliation{%
\institution{University at Buffalo}
\city{Buffalo}
\country{USA}
}
\email{chengenc@buffalo.edu}

\author{Stephanie Balzer}
\affiliation{%
\institution{Carnegie Mellon University}
\city{Pittsburgh}
\country{USA}
}
\email{balzers@cs.cmu.edu}

\author{Lukasz Ziarek}
\affiliation{%
\institution{University at Buffalo}
\city{Buffalo}
\country{USA}
}
\email{lziarek@buffalo.edu}

\begin{abstract}
Many of today's message-passing systems not only require
messages to be exchanged in a certain order
but also to happen at a certain \emph{time} or within a certain \emph{time window}. 
Such correctness conditions are particularly prominent in
Internet of Things (IoT) and real-time systems applications,
which interface with hardware devices that come with inherent timing constraints.
Verifying compliance of such systems with the intended \emph{timed protocol} is challenged by their
\emph{heterogeneity}---ruling out any verification method that relies on the system to be implemented in one common language,
let alone in a high-level and typed programming language.
To address this challenge,
this paper contributes a \emph{logical relation}
to verify that its inhabitants (the applications and hardware devices to be proved correct)
comply with the given timed protocol.
To cater to the systems' heterogeneity,
the logical relation is entirely \emph{semantic},
lifting the requirement that its inhabitants are syntactically well-typed.
A semantic approach enables two modes of use of the logical relation for program verification:
\textit{(i)} \emph{once-and-for-all} verification of an \emph{arbitrary} well-typed application, given a type system, and
\textit{(ii)} \emph{per-instance} verification of a specific application / hardware device (\aka foreign code).
To facilitate mode \textit{(i)}, the paper develops
a refinement type system for expressing timed message-passing protocols and
proves that any well-typed program inhabits the logical relation (fundamental theorem).
A type checker for the refinement type system has been implemented in Rust,
using an SMT solver to check satisfiability of timing constraints.
Then, the paper demonstrates both modes of use based on a small case study of
a smart home system for monitoring air quality,
consisting of a controller application and various environment sensors.

\end{abstract}



\maketitle

\section{Introduction}
\label{main:sec:intro}
The computing landscape has gradually been shifting to applications targeting distributed and heterogeneous systems,
including Internet of Things (IoT) and real-time systems applications.
Such applications are predominantly \emph{concurrent}, employ \emph{message-passing},
and often interface with \emph{foreign code}--code inaccessible to the application developer.
For example, consider a smart home system for monitoring air quality.
The controller of the system, an IoT application,
bases its decision on readings it receives from various environment sensors
(\eg BME680~\citep{BME680}),
spread across the home, measuring the surrounding air temperature, humidity, pressure, \etc.
Such sensors are hardware devices that the developer of the controller must interact with 
through a \emph{protocol} defined by the manufacturer in a specification (``datasheet'').
Another characteristic of such applications is their need to comply with
the \emph{timing constraints} dictated by the hardware devices' protocol.
For example, when measuring air quality, the BME680 sensor can only provide readings
after heating up an internal component for 30ms,
after which it requires an additional 20ms to cool down. Any application that uses the sensor must account for its protocol and timing requirements.

This naturally leads to the \emph{research question} that we address in this paper:

\begin{quote}
\emph{How to enable application developers to write timed message-passing programs
that comply with the timing constraints of the underlying hardware devices' protocols?}
\end{quote}

This question raises the following \emph{challenges}; the techniques to overcome them are
the core contribution of our paper:

\begin{enumerate}

\item\label[challenge]{challenge:spec-lang} distillation of a common \emph{specification language}
for \emph{timed message-passing protocols}
to prescribe the sequencing and timing of interactions between applications and devices;

\item\label[challenge]{challenge:op-model} distillation of a common \emph{operational model},
capturing the execution behavior \emph{both} of running applications and devices;

\item\label[challenge]{challenge:verification-framework} development of a \emph{compositional verification framework},
allowing the proof that the operational models of the applications and devices satisfy their specifications.
Compositionality (\aka modularity) allows applications and devices to be verified separately
and guarantees that they can be combined to a verified whole (without the need to re-verify the whole).

\end{enumerate}

To address \Cref{challenge:spec-lang}, we use \emph{types} as a specification language.
In particular, we build on the types developed for process calculi
\citep{KobayashiLICS1997,IgarashiPOPL2001} and
specifically on \emph{session types} \citep{HondaCONCUR1993,HondaESOP1998,HondaPOPL2008}.
Session types are behavioral types that prescribe the protocols of message-passing concurrent programs
and enjoy strong theoretical foundations, 
including a Curry-Howard correspondence between the session-typed $\pi$-calculus and linear logic
\citep{CairesCONCUR2010,WadlerICFP2012,ToninhoESOP2013,ToninhoPhD2015,LindleyMorrisESOP2015,KokkePOPL2019}.
The connection to linear logic endows programming languages developed for logic-based session types
with various desirable properties, such as protocol adherence and deadlock freedom.
The latter ensures global progress and is a result of linearity,
which imposes a tree structure on the runtime configurations of processes.

Among these logic-based session types,
we chose the family based on \emph{intuitionistic linear logic}
\citep{CairesCONCUR2010,ToninhoESOP2013,ToninhoPhD2015},
which distinguishes the \emph{provider} from the \emph{client} side of a protocol.
This distinction naturally accommodates the separation between hardware devices (\ie providers) and application programs (\ie clients),
present in our target domain.
To facilitate expression of timing constraints, we extend intuitionistic linear logic session types (\ilst)
with \emph{temporal predicates},
resulting in \emph{timed intuitionistic linear logic session types} (\tillst).
\tillst adopts an \emph{instant-based} model of time \citep{TemporalLogicSEP},
treating points in time (\ie instants) as primary notions,
and allows temporal predicates to quantify over such points in time
to prescribe at which instant(s) communications must happen relative to a \emph{global clock}.
\tillst sets itself apart from session types based on timed automata
\citep{BocchiCONCUR2014,BartolettiARTICLE2017,BocchiESOP2019} not only in terms of its logical foundation,
but also in the underlying model of time.
Automata-based session types come equipped with local clocks,
which require explicit synchronization (\ie clock resets) to simulate a global notion of time.

To address \Cref{challenge:op-model}, we use a \emph{labelled transition system (LTS)}
\citep{MilnerBook1980,MilnerBook1999,SangiorgiWalkerBook2001}
to express how applications and hardware devices exchange messages and thus compute.
The fundamental notion underlying an LTS is the one of a transition labelled with an \emph{action},
conveying the readiness of a participating entity to engage in an exchange.
Actions range over outputs (\ie sends), inputs (\ie receives), and the empty action.
An empty action denotes an actual computation, \ie a \emph{reduction},
which happens when two entities with complementary, \ie \emph{dual}, actions meet and
exchange a message as a result.
We chose an LTS as our operational model because it is agnostic of
the literal syntax used to represent communicating entities,
catering to the heterogeneity of our target domain.
Moreover, the duality of actions mirrors the provider-client distinction in \ilst:
when a provider is ready to output, its client will (eventually) be ready to input, and vice versa.
To accommodate timing considerations, our LTS denotes the instant at which an entity is ready to engage in an exchange,
in addition to the action.
We refer to the resulting LTS as \emph{timed LTS}.

To address \Cref{challenge:verification-framework}, we use \emph{logical relations}
\citep{TaitARTICLE1967,GirardPhD1972,PlotkinTR1973,StatmanARTICLE1985,PittsStarkHOOTS1998}
as our verification framework.
Logical relations are a device to prescribe the properties of valid programs
in terms of their \emph{computational behavior}
and are defined by considering the types of the underlying language.
As such, a logical relation can be viewed as defining ``inhabitance'' of valid terms in a type.
Logical relations enjoy \emph{compositionality},
ensuring that any two inhabitants can be composed to a compound inhabitant,
as dictated by the type structure.
An important characteristic of logical relations is their \emph{constructive} standpoint:
logical relations define the semantics of a program in terms of how it runs.
This insight fueled the \emph{semantic typing} approach
\citep{LoefARTICLE1982,ConstableBook1986,TimanyJACM2024},
which lifts the requirement that terms inhabiting the logical relation
ought to be (syntactically) well-typed.
Such a semantic approach allows proving inhabitance not only
of well-typed terms---via the ``fundamental theorem'' of the logical relation---but also of untyped terms,
provided they exhibit the computational behavior prescribed by the logical relation.
Because of this property, cross-language logical relations have been successfully employed,
for example,
for compiler correctness proofs
\citep{MinamidePOPL1996,ChlipalaPLDI2007,BentonICFP2009}
and soundness of language interoperability
\citep{PattersonPLDI2022}.

We chose logical relations as our verification framework precisely because of semantic typing.
The logical relation that we contribute enables
semantic typing of \emph{both} the applications and devices in our target domain and
is the core contribution of this paper.
Next, we summarize the \textbf{(1)} key \emph{features} of our logical relation and then
highlight \textbf{(2)} two \emph{modes of use} of our logical relation:
\textbf{(a)} once-and-for-all verification of an arbitrary well-typed application, given a type system, and
\textbf{(b)} per-instance verification of a specific application / hardware device (\aka foreign code).
Because logical relations are compositional, both modes of use work \emph{synergistically},
allowing us to combine components verified by either method to a verified whole.
The subsequent sections should be read as a ``teaser'', conveying the main ideas,
which we will formally develop in \Cref{main:sec:semantics}--\Cref{main:sec:sensor}.

\subsection{A Timed Semantic Session Logical Relation (\tsslr)}
\label{main:sec:intro:tsslr}

Our logical relation is defined by structural induction on
timed intuitionistic linear logic session types (\tillst),
our specification language for timed message-passing protocols (\Cref{challenge:spec-lang}),
and prescribes the computational behavior of its inhabitants
in terms of a timed LTS,
our operational model for applications and devices in our target domain (\Cref{challenge:op-model}).
We refer to our logical relation as
\emph{timed semantic session logical relation} (\tsslr),
to convey its purely semantic nature and grounding in \tillst.
To the best of our knowledge,
\tsslr is the first logical relation facilitating
the verification of timing constraints for message-passing protocols.
It is also the first \emph{entirely} semantic logical relation for session types;
prior logical relations for session types
\citep{PerezESOP2012,CairesESOP2013,PerezARTICLE2014,DeYoungFSCD2020,DerakhshanLICS2021,BalzerARXIV2023,DerakhshanECOOP2024,VanDenHeuvelECOOP2024}
demand inhabitants to be syntactically well-typed.
Besides facilitating \emph{semantic typing} for \emph{timed message-passing protocols}, 
\tsslr has the following unique features:

\subsubsection{Nameless Families of Configurations}
\label{main:sec:intro:nameless-configurations}

The handling of names---identifiers such as channels and locations---can become tedious in formal developments.
All too often, we end up either renaming existing names, for example upon receiving a channel,
or maintaining equivalence classes of names.
These strategies are not only cumbersome but also error-prone.
But even more so, formal developments usually do not even depend on a \emph{particular} choice of a name!
Linear logic can come to help,
because it guarantees that a name is only shared between \emph{two} runtime entities (\eg sensor and controller),
facilitating local reasoning about names.
Intuitionistic linear logic---due to its provider-client distinction--moreover
allows a provider (\eg sensor) to be \emph{polymorphic} in the name it is referred to by its client (\eg controller).
Our development takes advantage of these properties and
introduces the notion of a \emph{nameless family of configurations}.
Configurations are the terms inhabiting our logical relation,
nameless families of configurations are polymorphic in the choice of a name by their clients.
Our \tsslr defines inhabitance in terms of nameless families of configurations to accommodate arbitrary choices of names.

\subsubsection{Computable Trajectories}
\label{main:sec:intro:trajectories}

Labelled transition systems (LTS) \citep{MilnerBook1980,MilnerBook1999,SangiorgiWalkerBook2001}
describe concurrent interactions in a \emph{local} way,
isolating a particular entity that is ready to engage in an action,
while ``framing off'' entities unaffected by the action.
The understanding is that any number of ready entities with mutually complementary actions can reduce concurrently.
For timed message-passing protocols, however, the term ``concurrent'' is too liberal,
because it does not prescribe which reductions among the concurrent ones must happen \emph{simultaneously}.
Transitions in our timed LTS therefore are additionally annotated with the \emph{instant} at which an exchange may happen.
To complement this local description of a potential computation with a \emph{global} perspective,
we introduce the notion of a trajectory.
A \emph{trajectory} is a function that returns for each instant in time
the configuration of all entities at that particular instant.
We can thus think of a trajectory as a description of \emph{how a configuration evolves over time}.
To assert that a trajectory is the result of applying a sequence of timed LTS reductions,
which validates that trajectory,
we introduce the notion of a computable trajectory.
A \emph{computable trajectory} is a pair, consisting of the trajectory and a validating sequence of reductions.
In case of simultaneous reductions, there exists a validating sequence of reductions
for each permutation of simultaneous reductions,
but any of these suffices to assert computability.
The value of the notion of a trajectory is precisely that it ``collapses'' all simultaneous local LTS reductions
to one global reduction step.
Our \tsslr is phrased in terms of computable trajectories
and is thus polymorphic in the equivalence class of instantaneous LTS reduction sequences for a configuration's trajectory.

\subsubsection{Algebra for Computable Trajectories}
\label{main:sec:intro:algebra}

With the definition of computable trajectories in our hands,
it is convenient to define \emph{operations} on computable trajectories,
resulting in an \emph{algebra} for computable trajectories.
To gather an intuition for what operations may be suitable,
it is helpful to remind ourselves that a computable trajectory essentially describes
how a configuration evolves over time.
With that intuition in mind, it is sensible to expect that trajectories can be:
\textbf{(a)} \emph{interleaved}, describing the evolution of the concurrent composition of two configurations;
\textbf{(b)} \emph{partitioned}, to truncate a trajectory relative to a given instant; and
\textbf{(c)} \emph{concatenated}, to sequentially compose two trajectories.
As we will see in \Cref{main:sec:ftlr}, our computable trajectory algebra facilitates
an elegant definition and proof of the usual forward and backwards closure properties of logical relations,
generalized to account for the passage of time. 

\subsection{Two Modes of Use of \tsslr}
\label{main:sec:intro:modes}

Thanks to its semantic formulation, our \tsslr facilitates program verification in two ways:

\begin{enumerate}
\item\label[mode]{mode:all}\label[mode-l]{mode-l:all} \emph{once-and-for-all} verification of an \emph{arbitrary} well-typed application, given a type system;
\item\label[mode]{mode:one}\label[mode-l]{mode-l:one} \emph{per-instance} verification of a \emph{specific} application / hardware device (\aka foreign code).
\end{enumerate}

\Cref{mode:all} requires development of a suitable type system
that is strong enough to ensure that any well-typed term computes as prescribed by the logical relation.
This proof, \ie that any well-typed term inhabits the logical relation,
is referred to as the \emph{fundamental theorem} of the logical relation (FTLR).
The benefit of \Cref{mode:all} is, in a sense, its ``economy of scale'':
By carrying out, once and for all, a difficult proof (proof of the FTLR),
per-program verification reduces to a type checking problem.
If a decidable type checking algorithm exists, then program verification becomes automatic.

Type systems are by design recursively enumerable and
thus never complete with respect to the intended semantics (\ie inhabitance does not necessarily imply well-typedness).
As a result, they may reject perfectly good programs.
This is where \Cref{mode:one} comes to help.
Here, inhabitance of a term is proved directly, without a type system as an intermediary.
The benefit of \Cref{mode:one} is its impartiality:
Any ``computational object'' can be certified, as long as it can be shown to compute as prescribed by the logical relation.
\Cref{mode:one} is therefore indispensable to the verification of systems in our application domain.

\subsubsection{\Cref{mode-l:all}: Refinement Type System for \tillst}
\label{main:sec:intro:mode-all}

To facilitate \Cref{mode-l:all} of our \tsslr,
we contribute a refinement type system for \tillst.
The dependency of \tillst types on temporal predicates
manifests in the typing judgment of our type system:
a term is typed relative to a context of temporal propositions,
in addition to the usual variable context.
Temporal dependencies between communications are expressed using temporal variables,
whose free occurrences are collected in a separate context as well.
To guarantee that any term that has a valid derivation using our type system also inhabits our \tslr,
we prove the corresponding \emph{fundamental theorem}.

We have implemented a \emph{type checker} for our type system.
The type checker is implemented in Rust and uses an SMT solver to check satisfiability of temporal predicates.
Type checking is thus incomplete; our benchmark suite of examples, however, type checks successfully.

\subsubsection{\Cref{mode-l:one}: Per-Device Inhabitance Proof}
\label{main:sec:intro:mode-one}

We illustrate \Cref{mode-l:one} of our \tsslr,
by providing a proof that the BME680 environment sensor inhabits our \tsslr.
To carry out this proof, we translate the BME680's specification~\citep{BME680} given by the manufacturer
to a corresponding \tillst type as well as timed automaton, a term that can stepped using our timed LTS.
As we show in \Cref{main:sec:sensor},
this translation is straightforward because the manufacturer's datasheet effectively defines a timed automaton.

\subsubsection{\Cref{mode-l:all} $+$ \Cref{mode-l:one}: Whole System Verification}
\label{main:sec:intro:mode-both}

A semantic logical relation really comes to shine when combining both modes of use.
For example, given the timed protocol as a \tilst type,
we can develop the controller of our smart home device using our refinement language
and prove it correct using our type checker
(\Cref{mode-l:all}).
Then, we compose our controller with the BME680 sensor,
certified to be correct by \Cref{mode-l:one}.
Compositionality of the logical relations method guarantees correctness of the entire system as a result.
As we will see in \Cref{main:sec:ftlr},
compositionality is guaranteed by the statement of the fundamental theorem,
which takes an open well-typed term (\eg controller),
closes it with values assumed to inhabit the logical relation (\eg sensor),
and asserts inhabitance of the resulting closed term (\eg sensor $+$ controller).
Because the definition of our \tsslr does not require its inhabitants to be well-typed,
the substituted values (\eg sensor) do not have to be well-typed,
accommodating foreign code.

Our example of a smart home device also showcases that semantic logical relations
are \emph{synergistic} with other verification methods.
Although we provide ourselves a proof of inhabitance of the BME680 sensor in \Cref{main:sec:sensor},
a proof of inhabitance carried out by any other method will suffice.
For example, the manufacturer could utilize the UPPAAL tool suite \citep{BengtssonWORKSHOP1995}
to prove that the BME680 sensor complies with the timed automaton specified~\citep{BME680} and
thus assert its inhabitance in our \tsslr.

\subsection{Summary and Contributions}
\label{main:sec:intro:contributions}

In the remainder of this paper, we first give a motivating example (\Cref{main:sec:motivation}),
the air quality monitoring system,
detailing its specification and introducing the reader to
the protocol specification language \tillst, contributed by this paper,
as well as the implementation of the controller in a process language.
While our motivating example is an instance of a \emph{real-time} and (or) \emph{embedded system}, 
our work can see applications beyond real-time and embedded systems; 
our methods and results are not specialized to those domains. 
Communicating systems with timing concerns also include web servers and 
email servers. For instance, the SMTP protocol~\citep{RFC2821} for email exchange 
prescribes timeouts in addition to message format specifications.

In \Cref{main:sec:semantics}, we then develop our semantic logical relation \tslr,
featuring our \emph{timed LTS},
\emph{nameless families of configurations},
\emph{computable trajectories}, and \emph{algebra for computable trajectories}.
The two modes of use of our logical relation
are developed in \Cref{main:sec:type-sys} and \Cref{main:sec:sensor},
which contribute a refinement type system for \tillst to facilitate \Cref{mode-l:all} (\Cref{main:sec:tillst})
and proof of inhabitance of the BME680 sensor in our logical relation,
showcasing \Cref{mode-l:one} (\Cref{main:sec:sensor}).
The proof of correctness of the whole system,
showcasing \Cref{mode-l:all} $+$ \Cref{mode-l:one}, is given in \Cref{main:sec:sensor},
as a consequence of the proof of the fundamental theorem, given in \Cref{main:sec:ftlr}.
\Cref{main:sec:implementation} details our Rust type checker implementation,
\Cref{main:sec:related-work} comments on related work,
and \Cref{main:sec:conclusions-outlook} concludes with an outlook on future work.

\paragraph{Artifact and Technical Report}

Our type checker for the \tillst refinement type system is available as an associated artifact.
An extended version of this paper, which includes an appendix with formalization and proofs,
is available as a technical report \citep{fullpaper}.

\section{Protocol Specification in \tilst}
\label{main:sec:motivation}
This section revisits the air quality monitoring system mentioned in \Cref{main:sec:intro}
and illustrates how to specify the underlying protocol using timed intuitionistic linear logic session types (\tillst)
and how to implement the controller in a process language.
This process language coincides with the term language of our refinement type system,
which we introduce in \Cref{main:sec:tillst} and
prove to inhabit our logical relation in \Cref{main:sec:ftlr}, for well-typed terms.
\Cref{main:sec:sensor} completes the example by giving a representation of the environment sensors
and a proof of their inhabitance in the logical relation.

\subsection{Air Quality Monitoring System}
\label{main:sec:motivation:system-desc}

Let's revisit the air quality monitoring system mentioned in \Cref{main:sec:intro} in more detail.
The system has three components: two environment sensors, $x$ and $y$, 
connected to one central controller.
Sensor $x$ is placed in the bedroom and $y$ is placed in the living room.
The task of the controller is to decide in a timely manner, whether it needs to run 
the air-conditioner, by sending a Boolean representing its decision. To complete the task, 
the controller must configure both sensors and then collect data from them, according to the protocols 
dictated by the sensor specification.
The controller should report $\kw{true}$ if the temperature in either room 
gets too high or the air quality in the bedroom degrades too much. The exact conditions are not important here.
The aggregation of multiple data sources to synthesize a clearer, more informative signal 
is known as \emph{sensor fusion} \citep{SankarDasICRCC2012}. Sensor fusion requires a controller to juggle 
multiple time sensitive communications, challenging to program correctly.

The sensor is modeled against the BME680 4-in-1 environment sensor \cite{BME680}. 
The BME680 sensor measures surrounding temperature, humidity, pressure, and air quality.
Interaction with the BME680 is via a hardware message-passing bus called I2C \cite{nxp-i2c}, and
the sensor operates as a state machine. Initially, the sensor is in a low-power stand-by state. 
In this mode, the controller can send configuration messages to choose the desired measurement type.
Additionally, the sensor can be set to either report continuously and periodically 
(\emph{normal mode}) or to perform just one set of measurements (\emph{forced mode}). 
After configuration, the sensor reports data for each enabled measurement sequentially
by sending corresponding messages. It then returns to the standby state, potentially after a cooldown period.

Let's consider how we can specify the forced mode operation of the sensor in \tillst, required by our controller.
We model two options that have very different timing requirements: (1) temperature only and (2) temperature followed by air quality.
While the sensor measures temperature effectively instantaneously, 
in order to measure air quality, the sensor needs to first heat-up an internal component  
before taking a measurement. The heating takes $\millisec{30}$. 
After that, the sensor must cool down for $\millisec{20}$ before it returns to standby.
If both temperature and air quality data are requested (option 2), heating 
cannot start before the temperature results are read to prevent interference and inaccurate measurements. This means that our controller 
must wait an appropriate duration between temperature and air quality measurements.

The timing constraints involved in these protocols complicates programming even with just a singular sensor. 
The controller in our example, however, interleaves operations on both the bedroom and living room sensors. 
In the following sections we will see how \tillst types and processes allow us to specify and verify protocols cleanly and effectively.

\subsection{Protocol Specification Language and Process Term Language}

Before being in a position to specify the protocol for our example
and implement the controller,
we first must acquaint ourselves with timed intuitionistic linear logic session types (\tillst)
and a suitable process language.
\tillst enrich intuitionistic linear logic session types
\citep{CairesCONCUR2010,ToninhoESOP2013,ToninhoPhD2015}
with \emph{temporal predicates}.
The syntax of \tillst and the process language is given below.
The most distinguishing feature of our language is the superscript $t.p$, 
where $t$ is a time variable and $p$ serves as a predicate on $t$.
\Cref{main:tbl:tillst-types-terms} conveys the protocol semantics.
We first focus on the core process calculus constructs
and discuss support for the exchange of functional values in \Cref{main:sec:motivation:functional-layer}.
\begin{align*}
\Sort{Session Types} \quad A &\bnfdef \tpunitstycst{t}{p} \mid \tptensorstycst{t}{p}{A_1}{A_2} \mid \tpsumstycst{t}{p}{A_1}{A_2} \mid \tpwithstycst{t}{p}{A_1}{A_2} \mid \tplarrstycst{t}{p}{A_1}{A_2} \\
\Sort{Time} \quad T  &\bnfdef t \mid \finitcst \mid \fshiftcst{T}{i} \\
\Sort{Prop} \quad p  &\bnfdef \top \mid \bot \mid \landcst{p_1}{p_2} \mid \lorcst{p_1}{p_2} \mid \limpcst{p_1}{p_2} \mid \eqcst{T_1}{T_2} \mid \leqcst{T_1}{T_2} \\
\Sort{Process} \quad P, Q  &\bnfdef \tpclosecst{t}{p} \mid \tpwaitcst{T}{x}{Q} \mid \tplarrrcvcst{t}{p}{A_1}{x}{P} \mid \tplarrsndcst{x}{T}{P}{Q} \mid \tptensorsndcst{t}{p}{P_1}{P_2} \\ 
                        &\phantom{\bnfdef} \bnfalt \tptensorrcvcst{x}{T}{y}{Q} \mid \tpinlcst{t}{p}{A_1}{A_2}{P} \mid \tpinrcst{t}{p}{A_1}{A_2}{P} \\ 
                        &\phantom{\bnfdef} \bnfalt \tpcasecst{T}{x}{Q_1}{Q_2} \mid \tpoffercst{t}{p}{P_1}{P_2} \mid \tpselectlcst{x}{T}{Q} \mid \tpselectrcst{x}{T}{Q} \\
                        &\phantom{\bnfdef} \bnfalt \tpfwdcst{T}{x} \mid \tpspawnproccst{T}{P}{x}{Q}
\end{align*}

\subsubsection{Timed Process Term Language}
\label{main:sec:motivation:process-language}

Our language is a process language, where run-time processes of the form $\RunProc{a}{P}$ compute by communicating 
with each other over named channels. 
Each process executes some code $P$ called its \emph{process term}. When it is clear 
from the context, we refer to the process executing $P$ as just the process $P$.
Among the channels by which a process $\RunProc{a}{P}$ is connected to other processes,
we designate one channel, $a$, as its \emph{offering channel}. 
Process $P$ is said to \emph{provide for}, or \emph{to be the provider of}, its offering channel $a$.
For all the other channels, $P$ assumes the role of a \emph{client}.
The distinction of provider and client roles of a process
is the hallmark of an intuitionistic system. It has a profound impact on the design (\cref{main:sec:tillst})
and semantics (\cref{main:sec:semantics} and \cref{main:sec:ftlr}) of the system. 
The separation supports a wide range of metatheoretic properties,
including and not limited to deadlock freedom (\cref{main:thm:adequacy}).

We begin with the process terms, denoted by $P$, $Q$.
Process terms can have occurrences of free and bound variables (denoted by $x$, $y$),
for which we adopt the usual conventions to denote binding.
Variables range over names of runtime channels.
%
%
Process terms can be classified into three groups,
depending on the channel they immediately operate on. 
\emph{Provider processes}, such as $\tpclosecst{t}{p}$ and $\tpoffercst{t}{p}{P_1}{P_2}$, act on the offering channel. Because the 
offering channel of a process is distinguished, provider processes do not name the channel explicitly. 
\emph{Client processes}, such as $\tpwaitcst{T}{x}{Q}$ and $\tpcasecst{T}{x}{Q_1}{Q_2}$, act on the channel variable $x$.
In either case, process terms are typically \emph{sequences}, consisting of the next communication action and a continuation process.
For example, provider process $\tpinlcst{t}{p}{A_1}{A_2}{P}$ ($\tpinrcst{t}{p}{A_1}{A_2}{P}$) sends message $\lblleft$ ($\lblright$) and then continues with $P$.
Dually, the client process $\tpcasecst{T}{x}{Q_1}{Q_2}$ receives the message over $x$ and continues with $Q_1$ ($Q_2$). 
An overview of all process terms with description is given in \Cref{main:tbl:tillst-types-terms}.
Processes $\tpfwdcst{T}{x}$ and $\tpspawnproccst{T}{P}{x}{Q}$ have judgmental roles and constitute the third group. The process 
$\tpfwdcst{T}{x}$ forwards any actions from and to channel $x$ to it's offering channel, therefore identifying them. Process 
$\tpspawnproccst{T}{P}{x}{Q}$ spawns $P$ as a separate process and binds its offering channel to $x$ for use in the continuation $Q$.

In addition to specifying what actions a process should take, processes also carry timing related information as superscripts. 
In this regard, providers and clients again take on distinct roles. A provider process has a predicate $t.p$ as a superscript,
indicating at what time 
it is willing to take the action. Here $t$ is a variable ranging over points in time (denoted $T$) 
and $p$ is a proposition involving $t$. The exact definition is inessential and will be clarified shortly.
A predicate $t.p$ constrains a message exchange to only occur at times $T$ 
such that the proposition $p$ holds true for $t$ substituted with $T$ (\ie $\Subst{T}{t}{p}$ is true). 
In other words, provider processes limit the message exchange to a time window, 
while promising to engage in the exchange at any valid choice of time.
Client processes, on the other hand, carry a concrete choice of time $T$ as superscripts. 
The chosen $T$ is expected to satisfy the predicate set up by the provider of the referenced channel, in addition to 
some other requirements guaranteeing that time moves forward.
For example, the process 
$\tpoffercst{t}{p}{P_1}{P_2}$ offers to receive either $\lblleft$ or $\lblright$ at any time satisfying $p$. 
Its client $\tpselectlcst{x}{T}{Q}$ chooses to send $\lblleft$ at exactly time $T$.

\subsubsection{\tillst: Protocol Specification Language}

The sequences of actions that a process takes over its offering channel constitutes its \emph{protocol}.
\emph{Session types} (denoted by $A, B$) are behavioral types that prescribes such protocols.
Generally, the superscript of a session type indicates the time at which a communication action may occur, while the connective 
indicates the nature of the action itself. 
The connectives include \emph{sequencing} and \emph{branching} constructs. 
Sequencing is expressed by operators $\otimes$ and $\multimap$. 
The type $\larrstycst{A_1}{A_2}$ indicates that, after the receipt of a 
channel of type $A_1$, the protocol transitions to behaving as $A_2$.
Conversely, the type $\tensorstycst{A_1}{A_2}$ denotes the sending of a channel of type $A_1$.
Branching is expressed by the types $\bwithstycst{A_1}{A_2}$ and $\bsumstycst{A_1}{A_2}$ ,
offering a choice between the sessions $A_1$ and $A_2$ and making a choice between the sessions $A_1$ and $A_2$, \respb.
The choice is conveyed by receiving and sending labels $\lblleft$ or $\lblright$,
reminiscent of products and sums in functional languages.
The type $\unitstycst$ denotes the end state of a protocol.
The connectives come from intuitionistic linear logic,
given \ilst's foundation.
\Cref{main:tbl:tillst-types-terms} lists all connectives, with corresponding process terms.
Due to the provider-client distinction, each connective has a term for the provider and client.

\begin{table}[t]
\caption{\tilst types and process terms.}
\label{main:tbl:tillst-types-terms}
\small
\bgroup
\DeclareDocumentCommand{\mlt}{+O{l} m}{\begin{tabular}{#1}#2\end{tabular}}
\DeclareDocumentCommand{\tdescp}{}{\mlt[@{}l@{}]{any $T$ \\ s.t. $\Subst{T}{t}{p}$}}
\DeclareDocumentCommand{\tdescc}{}{\mlt[@{}l@{}]{fixed $T$ \\ s.t. $\Subst{T}{t}{p}$}}
\DeclareDocumentCommand{\tdescs}{}{\mlt[@{}l@{}]{fixed $T$}}
\begin{tabular}{@{}lrllll@{}}
\toprule
Role & Type ($A$)  & Process Term ($P$) & Time & Action & Cont. \\
\midrule
\multirow{6}{*}{\mlt[@{}l@{}]{Provider \\ \\ \emph{Acting} on \\ \emph{providing} \\ \emph{channel} \\ $\IsOfProc{P}{T}{A}$}}  
 & $\tpunitstycst{t}{p}$ & $\tpclosecst{t}{p}$                & \multirow{6}{*}{\tdescp} & send closing signal $\kw{cls}$ & none \\
 & $\tplarrstycst{t}{p}{A_1}{A_2}$ & $\tplarrrcvcst{t}{p}{A_1}{x}{P}$   & & accept channel, bind to $x$ & $P$ \\
 & $\tptensorstycst{t}{p}{A_1}{A_2}$ & $\tptensorsndcst{t}{p}{P_1}{P_2}$  & & spawn $P_1$ and send it & $P_2$ \\
 & $\tpsumstycst{t}{p}{A_1}{A_2}$ & $\tpinlcst{t}{p}{A_1}{A_2}{P}$     & & send  $\lblleft$  & $P$ \\
 & $\tpsumstycst{t}{p}{A_1}{A_2}$ & $\tpinrcst{t}{p}{A_1}{A_2}{P}$     & & send $\lblright$ & $P$ \\
 & $\tpwithstycst{t}{p}{A_1}{A_2}$ & $\tpoffercst{t}{p}{P_1}{P_2}$      & & accept $\lblleft$ or $\lblright$ & $P_1$ or $P_2$ \\
\midrule
\multirow{6}{*}{\mlt[@{}l@{}]{Client \\ \\ \emph{Acting on} \\ \emph{chan.} $\IsOf{x}{A}$ }}  
 & ${\tpunitstycst{t}{p}}$ & $\tpwaitcst{T}{x}{Q}$              & \multirow{6}{*}{\tdescc} & wait for closing signal $\kw{cls}$ & $Q$.  \\
 & ${\tplarrstycst{t}{p}{A_1}{A_2}}$ & $\tplarrsndcst{x}{T}{P}{Q}$        & & spawn $P$ and send it & $Q$  \\
 & ${\tptensorstycst{t}{p}{A_1}{A_2}}$ & $\tptensorrcvcst{x}{T}{y}{Q}$      & & receive channel, bind to $y$ & $Q$  \\
 & ${\tpsumstycst{t}{p}{A_1}{A_2}}$ & $\tpcasecst{T}{x}{Q_1}{Q_2}$       & & receive $\lblleft$ or $\lblright$ & $Q_1$ or $Q_2$  \\
 & ${\tpwithstycst{t}{p}{A_1}{A_2}}$ & $\tpselectlcst{x}{T}{Q}$           & & send $\lblleft$  & $Q$  \\
 & ${\tpwithstycst{t}{p}{A_1}{A_2}}$ & $\tpselectrcst{x}{T}{Q}$           & & send $\lblright$ & $Q$  \\
\midrule
 \multicolumn{2}{@{}l@{}}{\multirow{2}{*}{Judgmental rules}} & $\tpfwdcst{T}{x}$                  & \multirow{2}{*}{\tdescs} & \multicolumn{2}{l}{forward $x$ to offerring channel} \\
 &  & $\tpspawnproccst{T}{P}{x}{Q}$      &  & \multicolumn{2}{l}{spawn $P$ and connect to it on $x$}  \\
\bottomrule
\end{tabular}
\egroup
\end{table}

\paragraph{The Logic and Model of Time} The language presented uses a straightforward model for time. Every point in time is essentially 
an integer offset $i$ away from one distinguished \emph{initial} point in time $\finitcst$. This model of time is discrete, linear, and 
unbounded. Points in time can be compared for equality and satisfy less-or-equal-to relations. 
Propositions $p$ include the whole quantifier-free fragment of first order logic. This choice of logic is expressive enough for our 
examples and many real-world applications. 

\subsubsection{Exchange of Functional Values}
\label{main:sec:motivation:functional-layer}

Our language may be easily extended to support evaluating, 
sending, and receiving functional expressions. Let $e$ be the syntactic sort of expressions in a (typed) functional language of your choice. 
For the sake of simplicity, we assume $e$ is at least terminating and pure. Lifting those restrictions presents difficulties largely 
orthogonal to the development of this paper. Let $\tau$ be the sort of types in said language. 
For the sake of our sensor example, let us 
assume that the language includes at least Booleans ($\booltycst$) and integer numbers ($\inttycst$). 
\begin{align*}
\Sort{Functional Types} \quad \tau &\bnfdef \booltycst \mid \inttycst \mid \dots \\
\Sort{Functional Expressions} \quad e &\bnfdef x \mid \trexcst \mid \faexcst \mid \numcst{n} \mid \dots \\
\Sort{Session Types} \quad A &\bnfdef \dots \mid \tpbangstycst{t}{p}{\tau}{A} \mid \tpquerystycst{t}{p}{\tau}{A} \\
\Sort{Process} \quad P  &\bnfdef \dots \mid \tpproduceproccst{t}{p}{e}{P} \mid \tpconsumeproccst{T}{y}{x}{Q} \\
                                       & \mid \tpqueryproccst{t}{p}{x}{P} \mid \tpsupplyproccst{T}{x}{v}{Q}
\end{align*}

We extend the session types and process terms accordingly.
Sessions types $\tpbangstycst{t}{p}{\tau}{A}$ and $\tpquerystycst{t}{p}{\tau}{A}$ allow
sending a functional value of type $\tau$ and receiving a value of type $\tau$, \respb. 
Provider process $\tpproduceproccst{t}{p}{e}{P}$ evaluates $e$ to a value and then sends it over the offering 
channel. The message is intended for the client process $\tpconsumeproccst{T}{y}{x}{Q}$, which binds the message to a variable $x$. Be aware that 
we are overloading variable names $x, y$ for both channel and functional variables. The pair $\tpqueryproccst{t}{p}{x}{P}$
and $\tpsupplyproccst{T}{x}{v}{Q}$ perform analogous actions, with provider/client roles switched.

\subsection{Code for the Controller}

We now have the tools to prescribe the protocols for the sensor ($A_\kw{BME680}$) 
and the controller ($A_\kw{Hub}$):
\begin{align*}
A_\kw{BME680} &\triangleq \tpwithstycst{t_1}{t_0 \leq t_1}{A_\kw{T}}{A_\kw{TG}}\\
A_\kw{T} &\triangleq \tpbangstycst{t_2}{t_1 \leq t_2}{\tau_\kw{Temp}}{\tpunitstycst{t_3}{t_2 \leq t_3}} \\
A_\kw{TG} &\triangleq \tpbangstycst{t_2}{t_1 \leq t_2}{\tau_\kw{Temp}}{ 
    \tpbangstycst{t_3}{t_2 + \millisec{30} \leq t_3 }{\tau_\kw{Gas}}{\tpunitstycst{t_4}{t_3 + \millisec{20} \leq t_4}}} \\
A_\kw{Hub} &\triangleq
   \tplarrstycst{t_1}{t_0 \leq t_1}{A_\kw{BME680}}{
   \tplarrstycst{t_2}{t_2 = t_1}{A_\kw{BME680}}{
        \tpbangstycst{t_3}{t_1 + \millisec{50} \leq t_3}{\booltycst}{\tpunitstycst{t_4}{t_4 = t_3}}}}
\end{align*}

Suppose we start the protocol at time $t_0$. Notice that $t_0$ is free everywhere. 
The protocol of the sensor starts by accepting a configuration message of either $\lblleft$ or $\lblright$, choosing between measuring
just the temperature ($A_\kw{T}$) or both temperature and air quality~\footnote{The BME680 datasheet \citep{BME680} refers to the feature that measures 
air quality as ``gas'' measurement, hence the letter ``\kw{G}''.} ($A_\kw{TG}$). 
The message simultaneously puts the sensor into (forced) operational mode~\footnote{We combine the configuration of measurement and mode selection, which have no timing requirements between them.}. The message 
can happen at anytime $t_1$ after initial time $t_0$ ($t_1. t_0 \leq t_1$). 
If $A_\kw{T}$ is selected, the temperature result (a functional value of type $\tau_\kw{temp}$)
is immediately available for collection. This is conveyed by the predicate $t_2. t_1 \leq t_2$, licensing the client to receive the functional 
value at any time $t_2$ in the future of $t_1$. This illustrates an important feature of the language, we are now in a position to appreciate:
\begin{quote}
\tilst supports \emph{binding} the actual time of communication for future reference. 
\end{quote}
A more precise reading of the type $A_\kw{BME680}$ is that a process may accept a $\lblleft$ or $\lblright$ at any time that follows 
$t_0$, and \emph{let $t_1$ be the actual time at which the exchange occurs}.
This now-bound point in 
time $t_1$ may be used both for specifying protocols in types, which we are seeing now,
and for choosing the time of future actions, 
which we will see when we examine the process term for the controller. 
Wrapping up this branch, because only temperature is collected, the 
sensor may be shut-down any time after the temperature is received, a point in time now-bound at $t_2$, conveyed by the predicate $t_2 \leq t_3$.
On the other hand, if $A_\kw{TG}$ is selected, then the sensor temperature result is immediately available, as before. 
After the temperature result is collected at $t_2$, the subsequent air quality result ($\tau_\kw{gas}$) is only available after $\millisec{30}$, 
allowing enough time for the sensor to warm up. This is expressed by $t_2 + \millisec{30} \leq t_3$. Finally, after the collection of both 
results at time $t_3$, the sensor takes $\millisec{20}$ to cool down before it can be closed at $t_4$.

The controller type $A_\kw{Hub}$ uses \emph{higher-order channels}, which are channels whose messages are names of other 
channels. At some time $t_1$ (or $t_2$, since $t_1 = t_2$) after $t_0$, the controller simultaneously gets hold of two channels, each connected 
to a sensor. Upon receiving the pair of sensors, the controller has $\millisec{50}$ to compute and make available a $\booltycst$ response, 
ready for collection at $t_3$.
Then, the controller is immediately terminated at $t_3$ ($t_4 = t_3$).

The ability to prescribe the relative timing between actions is critical for this specification. 
Many protocols, including this one, require a client to ``\kw{wait}'' 
or ``\kw{sleep}'' for a fixed or variable amount of time between consecutive actions.
Existing modeling tools and languages based on \emph{timed automata} (\eg \cite{BocchiESOP2019})
address this via imperative \emph{clock resets} (as shown in \cref{main:sec:sensor}). 
However, these are not the only kinds of timing requirements. Other common requirements include
the action takes place ``no-later'' or ``no earlier'' than some other action, 
or that the action ``leaves enough room'' for some other action. These timing requirements can be logically complicated to express indirectly via clock resets. 
\tillst accommodates these and introduces a rich declarative language for timing specification, 
allowing for intuitive and natural timing specifications.

We turn now to the process term for the controller, using \Cref{main:tbl:tillst-types-terms} as a guide:
\begin{align*}
\IsOfProc{P}{t_0}{A_\kw{Hub}} &\triangleq 
  \tplarrrcvcst{t_1}{t_0 \leq t_1}{A_\kw{BME680}}{x}{\,} 
  \tplarrrcvcst{t_2}{t_2 = t_1}{A_\kw{BME680}}{y}{P_\kw{Hub}}  \\
&\phantom{\triangleq~~} \tpselectrcst{x}{t_1}{\,}  u_1 \leftarrow \kw{consume}^{t_1}\,x; \\
&\phantom{\triangleq~~} \tpselectlcst{y}{t_1}{\,}
  u_2 \leftarrow \kw{consume}^{t_1}\,y; \tpwaitcst{t_1}{y}{\,} \\
&\phantom{\triangleq~~} v_1 \leftarrow \kw{consume}^{t_1 + \millisec{30}}\,x; \tpwaitcst{t_1 + \millisec{50}}{x}{\,} \\
&\phantom{\triangleq~~} \tpproduceproccst{t_3}{t_1 + 50 \leq t_3}{\kw{NeedAC}(u_1, u_2, v_1)}{\tpclosecst{t_4}{t_4=t_3}}
\end{align*}
As dictated by its protocol, the controller initializes at $t_0$, then simultaneously accepts two channels $x$ and $y$, 
each connected to a sensor.
Recall that $y$ connects to 
the living room sensor, from which we only require temperature, and $x$ connects to the bedroom sensor, from which we require both the temperature and air quality data.
After receiving the channels, the controller \emph{interleaves} actions to the sensors. 
It first sends $\lblright$ to $x$, selecting both measurements and receives the temperature results, all done at $t_1$ at which the channels 
where received. Instead of waiting on the subsequent air quality results, it turns its attention to $y$. It configures, collects from, and then 
closes $y$. It then reads the air quality result from $x$ at $t_1 + \millisec{30}$ and closes it at $t_2 + \millisec{50}$, namely $\millisec{20}$ 
after the previous step. Finally, it computes the $\booltycst$ through some defined algorithm $\kw{NeedAC}$ in the functional language and then 
closes itself.

The ability of a process to interleave actions on different channels as a client,
without revealing this interleaving as part of its offering protocol,
preserves abstraction and is one of the key benefits of intuitionism.
To meet timing requirements, applications need to be able to put multiple processes in motion. If we were to modify the example to 
collect temperature and air quality data from both $x$ and $y$, then both sensors must be configured to warm up in parallel to meet the 
output deadline. Intuitionism again helps here, as the session type only prescribes what happens over the offering channel, processes have 
the freedom to re-order client actions as they see fit.

\section{Semantics through Timed Semantic Session Logical Relation (\tslr)}
\label{main:sec:semantics}
The heterogeneity of our target domain
requires the semantics of programs, \ie how programs run (\aka communicate), to be the front and centerpiece of the development. 
In this section, we begin this exploration by looking at the \emph{dynamic semantics}, which concerns the stepping of 
programs through time using a timed labelled transition system. 
We will then distill the meaning of \emph{session types} based on their role as classifiers for program 
behavior through the lens of the \emph{logical relations} method.
In \Cref{main:sec:sensor}, we build on this foundation
to show that the implementation of the BME680 environment sensor discussed in \Cref{main:sec:motivation}
inhabits our logical relation.
Thanks to semantic typing and the use of a timed labelled transition system,
these ``foreign objects'' can be easily accommodated
by simply extending the list of transition rules while
maintaining an almost identical proof structure.

\subsection{Dynamic Semantics} 
\label{main:sec:dynamics}
The dynamics describes a transition system that models computation 
as evolving with time in a way that is consistent with our instant-based model of time. 
In this section, we will set up a \emph{timed labeled transition system}, 
where transitions are labeled by both the action (if any) taken and the time at which it takes place.

At runtime, a program amounts to a configuration $\Omega$ of processes, defined as:
\begin{align*}
\ConfSort{} \quad \Omega & \bnfdef \StopConf \mid \RunProc{a}{P} \mid \FwdProc{a}{b} \mid \conc{\Omega_1}{\Omega_2} 
\end{align*}

The nullary configuration is written as ``$\StopConf$''. The runtime incarnation of a process term $P$ is $\RunProc{a}{P}$,
where $a$ is the providing channel of the process. The forwarding configuration $\FwdProc{a}{b}$ identifies runtime channels $a$ and $b$ 
by forwarding all messages between them. Finally, $\conc{\Omega_1}{\Omega_2}$ denotes the concurrent composition of two configurations.

As usual, structural congruence rules identify configurations up to reordering. Nullary configurations may be dropped silently, 
and forwarding configurations may be dropped after renaming the channels referenced accordingly. 
Structural congruence is set up to be a congruent equivalence relation. Below we are showing the critical rules. 
For the complete set of rules see \refapx[apx:lang:statics:cong].
\ifdefined\InApx
\begin{mathpar}
\defruler[C-Stop][rule:cong:Stop]
{\strut}
{\conc{\StopConf}{\Omega} \equiv \Omega}

\defruler[C-Comm][rule:cong:comm]
{\strut}
{\conc{\Omega_1}{\Omega_2} \equiv \conc{\Omega_2}{\Omega_1}}

\defruler[C-Fwd][rule:cong:fwd]
{\strut}
{\conc{\RunProc{a}{P}}{\FwdProc{b}{a}} \equiv \RunProc{b}{P}}

\defruler[C-Ctr][rule:cong:contr]
{\strut}
{\conc{\FwdProc{c}{b}}{\FwdProc{b}{a}} \equiv \FwdProc{c}{a}}

\defruler[C-Refl][rule:cong:refl]
{\strut}
{\Omega \equiv \Omega}

\defruler[C-Trans][rule:cong:trans]
{\Omega \equiv \Omega' \\ 
 \Omega' \equiv \Omega''}
{\Omega \equiv \Omega''}

\defruler[C-Cong][rule:cong:cong]
{\Omega_1 \equiv \Omega_1'}
{\conc{\Omega_1}{\Omega_2} \equiv \conc{\Omega_1'}{\Omega_2}}
\end{mathpar}

\else
\begin{mathpar}
\defruler[C-Stop][rule:cong:Stop]
{\strut}
{\conc{\StopConf}{\Omega} \equiv \Omega}

\defruler[C-Fwd][rule:cong:fwd]
{\strut}
{\conc{\RunProc{a}{P}}{\FwdProc{b}{a}} \equiv \RunProc{b}{P}}

\defruler[C-Comm][rule:cong:comm]
{\strut}
{\conc{\Omega_1}{\Omega_2} \equiv \conc{\Omega_2}{\Omega_1}}

\defruler[C-Ctr][rule:cong:contr]
{\strut}
{\conc{\FwdProc{c}{b}}{\FwdProc{b}{a}} \equiv \FwdProc{c}{a}}
\end{mathpar}
\fi

Next, we define the \emph{timed labelled transition system}. 
The judgment
$\Omega \LLStepsTo{\alpha}{T} \Omega'$ asserts that
at time $T$ the configuration $\Omega$ is ready to step to $\Omega'$ taking action $\alpha$. 
Actions $\alpha$ are defined as follows:
%
%
\begin{align*}
\Sort{Action} \quad \alpha & \bnfdef  \actsil  \mid \actsndchancst{a}{b} \mid \actrcvchancst{a}{b} \mid \actsndlblcst{a}{\lblleft} \mid \actsndlblcst{a}{\lblright} 
                           \mid \actrcvlblcst{a}{\lblleft} \mid \actrcvlblcst{a}{\lblright} \mid \actsndclosecst{a} \mid \actrcvclosecst{a} 
\end{align*}

The action $\actsil$ is the \emph{nullary}, \emph{silent}
\footnote{Commonly known as $\tau$ transitions in the $\pi$-calculus literature. We choose $\varepsilon$ to avoid conflicting with functional types.} 
action, corresponding to an actual reduction step of the configuration. Silent actions are usually omitted.
Non-silent actions always carry a message, a channel over which the message propagates, 
and a direction. The symbol ``$!$'' means send and the symbol ``$?$'' means receive. Messages may be labels, channel names, or 
the closing signal.
For example, $\Omega \LLStepsTo{\actsndlblcst{a}{\lblleft}}{T} \Omega'$ means
that configuration $\Omega$ at time $T$ may send the label $\lblleft$ over channel $a$ and become $\Omega'$. 

Our transition system allows any pair of compatible processes to communicate anywhere within the configuration. This is achieved by 
structural congruence in conjunction with the \emph{framing} and communication rules.

\begin{mathpar}
\defruler[D-Frame][rule:dyn:frame]
{\Omega_1 \LLStepsTo{\alpha}{T} \Omega_1'}
{\conc{\Omega_1}{\Omega_2} \LLStepsTo{\alpha}{T} \conc{\Omega_1'}{\Omega_2}}

\defruler[D-Comm][rule:dyn:comm]
{\Omega_1 \LLStepsTo{\alpha}{T} \Omega_1' \\ 
 \Omega_2 \LLStepsTo{\overline{\alpha}}{T} \Omega_2' \\
}
{\conc{\Omega_1}{\Omega_2} \LLStepsTo{}{T} \conc{\Omega_1'}{\Omega_2'}}
\end{mathpar}

The premise of rule \ruleref{rule:dyn:frame} ``frames off'' the surrounding configuration $\Omega_2$, allowing 
us to isolate $\Omega_1$ for a local transition. 

Two processes are compatible to communicate if they are willing to take \emph{complementary} (\aka dual) actions.
Complementary actions are defined as follows, with $\overline{\overline{\alpha}} = \alpha$:

\begin{mathpar}
\overline{\actrcvlblcst{a}{\lblleft}} \triangleq \actsndlblcst{a}{\lblleft} 

\overline{\actrcvlblcst{a}{\lblright}} \triangleq \actsndlblcst{a}{\lblright} 

\overline{\actsndlblcst{a}{\lblleft}} \triangleq \actrcvlblcst{a}{\lblleft} 

\overline{\actsndlblcst{a}{\lblright}} \triangleq \actrcvlblcst{a}{\lblright} 

\\
\overline{\actsil} \triangleq \actsil 

\overline{\actrcvchancst{a}{b}} \triangleq \actsndchancst{a}{b} 

\overline{\actsndchancst{a}{b}} \triangleq \actrcvchancst{a}{b} 

\overline{\actrcvclosecst{a}} \triangleq \actsndclosecst{a}

\overline{\actsndclosecst{a}} \triangleq \actrcvclosecst{a}

\end{mathpar}

In \ruleref{rule:dyn:comm}, if a pair of processes is willing to take complementary actions, then they communicate and both transition. 
The overall step involving both processes is silent because the communication happens ``internally''.

%
%




\begin{figure}
\raggedright
\ifdefined\InApx 

$\boxed{\Omega \LLStepsTo{\alpha}{T} \Omega'}$
\begin{mathpar}
\inferrule[]
{\strut}
{
 \RunProc{b}{P} \opconc \RunProc{a}{\tpfwdcst{T}{b}} 
 \LLStepsTo{}{T}
 \RunProc{a}{P}
}

\inferrule[]
{\strut}
{\RunProc{a}{\tpspawnproccst{T}{P}{x}{Q}} 
 \LLStepsTo{}{T} 
 \RunProc{b}{P} \opconc \RunProc{a}{\Subst{b}{x}{Q}}
}

\inferrule[]{\strut}{
 \RunProc{a}{\tpwaitcst{T}{b}{Q}}
 \LLStepsTo{\actrcvclosecst{b}}{T}
 \RunProc{a}{Q}
}

\inferrule[]{\strut}{
 {\RunProc{a}{\tpselectlcst{b}{T}{Q}}}
 \LLStepsTo{\actsndlblcst{b}{\lblleft}}{T}
 {\RunProc{a}{Q}}
}

\inferrule[]{\strut}{
 {\RunProc{a}{\tpselectrcst{b}{T}{Q}}}
 \LLStepsTo{\actsndlblcst{b}{\lblright}}{T}
 {\RunProc{a}{Q}}
}

\inferrule[]{\strut}{
 {\RunProc{a}{\tpcasecst{T}{b}{Q_1}{Q_2}}}
 \LLStepsTo{\actrcvlblcst{b}{\lblleft}}{T}
 {\RunProc{a}{Q_1}}
}

\inferrule[]{\strut}{
 {\RunProc{a}{\tpcasecst{T}{b}{Q_1}{Q_2}}}
 \LLStepsTo{\actrcvlblcst{b}{\lblright}}{T}
 {\RunProc{a}{Q_2}}
}

\inferrule[]{\strut}{
 \RunProc{a}{\tptensorrcvcst{b}{T}{y}{Q}} 
 \LLStepsTo{\actrcvchancst{b}{c}}{T}
 \RunProc{a}{\Subst{c}{y}{Q}}
}

\inferrule[]{\strut}{
 {\RunProc{a}{\tplarrsndcst{b}{T}{P}{Q}} }
 \LLStepsTo{\actsndchancst{b}{c}}{T}
 {\RunProc{a}{Q} \opconc \RunProc{c}{P}}
}

\inferrule[]{\Subst{T}{t}{p}}{
 {\RunProc{a}{\tpclosecst{t}{p}}}
 \LLStepsTo{\actsndclosecst{a}}{T}
 {\StopConf}
}

\inferrule[]{\Subst{T}{t}{p}}{
 {\RunProc{a}{\tpoffercst{t}{p}{P_1}{P_2}}}
 \LLStepsTo{\actrcvlblcst{a}{\lblleft}}{T}
 {\RunProc{a}{\Subst{T}{t}{P_1}}}
}

\inferrule[]{\Subst{T}{t}{p}}{
 {\RunProc{a}{\tpoffercst{t}{p}{P_1}{P_2}}}
 \LLStepsTo{\actrcvlblcst{a}{\lblright}}{T}
 {\RunProc{a}{\Subst{T}{t}{P_2}}} 
}

\inferrule[]{\Subst{T}{t}{p}}{
 {\RunProc{a}{\tpinlcst{t}{p}{A_1}{A_2}{P}}}
 \LLStepsTo{\actsndlblcst{a}{\lblleft}}{T}
 {\RunProc{a}{\Subst{T}{t}{P}}}
}

\inferrule[]{\Subst{T}{t}{p}}{
 {\RunProc{a}{\tpinrcst{t}{p}{A_1}{A_2}{P}}}
 \LLStepsTo{\actsndlblcst{a}{\lblright}}{T}
 {\RunProc{a}{\Subst{T}{t}{P}}}
}

\inferrule[]{\Subst{T}{t}{p}}{
 {\RunProc{a}{\tptensorsndcst{t}{p}{P_1}{P_2}}}
 \LLStepsTo{\actsndchancst{a}{c}}{T}
 {\RunProc{a}{\Subst{T}{t}{P_2}} \opconc \RunProc{c}{\Subst{T}{t}{P_1}}}
}

\inferrule[]{\Subst{T}{t}{p}}{
 {\RunProc{a}{\tplarrrcvcst{t}{p}{A}{x}{P}}}
 \LLStepsTo{\actrcvchancst{a}{c}}{T}
 {\RunProc{a}{\Subst{T, c}{t, x}{P}}}
}
\end{mathpar}

\else

\begin{tabular}{l}
	
\defsteprule[D-\kw{fwd}][rule:dyn:fwd]
	{\RunProc{a}{\tpfwdcst{T}{b}}}
	{}{T}
	{\FwdProc{a}{b}}\\

\defsteprule[D-\kw{spawn}][rule:dyn:cut]
	{\RunProc{a}{\tpspawnproccst{T}{P}{x}{Q}}}
	{}{T} 
	{\RunProc{b}{P} \opconc \RunProc{a}{\Subst{b}{x}{Q}}}\\

\\

\defsteprule[D-$\unitstycst$ L][rule:dyn:one:l]
 {\RunProc{a}{\tpwaitcst{T}{b}{Q}}}
 {\actrcvclosecst{b}}{T}
 {\RunProc{a}{Q}}\\

\defsteprule[D-$\&$ L1][rule:dyn:with:l1]
 {\RunProc{a}{\tpselectlcst{b}{T}{Q}}}
 {\actsndlblcst{b}{\lblleft}}{T}
 {\RunProc{a}{Q}}\\

\defsteprule[D-$\&$ L2][rule:dyn:with:l2]
 {\RunProc{a}{\tpselectrcst{b}{T}{Q}}}
 {\actsndlblcst{b}{\lblright}}{T}
 {\RunProc{a}{Q}} \\

\defsteprule[D-$\oplus$ L1][rule:dyn:oplus:l1]
 {\RunProc{a}{\tpcasecst{T}{b}{Q_1}{Q_2}}}
 {\actrcvlblcst{b}{\lblleft}}{T}
 {\RunProc{a}{Q_1}}\\

\defsteprule[D-$\oplus$ L2][rule:dyn:oplus:l2]
 {\RunProc{a}{\tpcasecst{T}{b}{Q_1}{Q_2}}}
 {\actrcvlblcst{b}{\lblright}}{T}
 {\RunProc{a}{Q_2}}\\

\defsteprule[D-$\otimes$ L][rule:dyn:tensor:l]
 {\RunProc{a}{\tptensorrcvcst{b}{T}{y}{Q}} }
 {\actrcvchancst{b}{c}}{T}
 {\RunProc{a}{\Subst{c}{y}{Q}}}\\

\defsteprule[D-$\multimap$ L][rule:dyn:lolli:l]
 {\RunProc{a}{\tplarrsndcst{b}{T}{P}{Q}} }
 {\actsndchancst{b}{c}}{T}
 {\RunProc{a}{Q} \opconc \RunProc{c}{P}}\\

\\

\textsc{With premise $\Subst{T}{t}{p}$:} \\

\defsteprule[D-$\unitstycst$][rule:dyn:one:r]
 {\RunProc{a}{\tpclosecst{t}{p}}}
 {\actsndclosecst{a}}{T}
 {\StopConf}\\

\defsteprule[D-$\&$ 1][rule:dyn:with:r1]
 {\RunProc{a}{\tpoffercst{t}{p}{P_1}{P_2}}}
 {\actrcvlblcst{a}{\lblleft}}{T}
 {\RunProc{a}{\Subst{T}{t}{P_1}}}\\

\defsteprule[D-$\&$ 2][rule:dyn:with:r2]
 {\RunProc{a}{\tpoffercst{t}{p}{P_1}{P_2}}}
 {\actrcvlblcst{a}{\lblright}}{T}
 {\RunProc{a}{\Subst{T}{t}{P_2}}} \\

\defsteprule[D-$\oplus$ R1][rule:dyn:oplus:r1]
 {\RunProc{a}{\tpinlcst{t}{p}{A_1}{A_2}{P}}}
 {\actsndlblcst{a}{\lblleft}}{T}
 {\RunProc{a}{\Subst{T}{t}{P}}}\\

\defsteprule[D-$\oplus$ R2][rule:dyn:oplus:r2]
 {\RunProc{a}{\tpinrcst{t}{p}{A_1}{A_2}{P}}}
 {\actsndlblcst{a}{\lblright}}{T}
 {\RunProc{a}{\Subst{T}{t}{P}}}\\

\defsteprule[D-$\otimes$][rule:dyn:tensor]
 {\RunProc{a}{\tptensorsndcst{t}{p}{P_1}{P_2}}}
 {\actsndchancst{a}{c}}{T}
 {\RunProc{a}{\Subst{T}{t}{P_2}} \opconc \RunProc{c}{\Subst{T}{t}{P_1}}}\\

\defsteprule[D-$\multimap$][rule:dyn:lolli]
 {\RunProc{a}{\tplarrrcvcst{t}{p}{A}{x}{P}}}
 {\actrcvchancst{a}{c}}{T}
 {\RunProc{a}{\Subst{T, c}{t, x}{P}}}\\
\end{tabular}

\fi

\caption{Timed labelled transitions $\Omega \LLStepsTo{\alpha}{T} \Omega'$.}
\label{rules:inststep}
\end{figure}

\cref{rules:inststep} lists the remaining timed labelled transition rules of our system. 
Clients can only step at a time $T$, dictated by the client, or in the 
case of \ruleref{rule:dyn:fwd} and \ruleref{rule:dyn:cut}, specified by the process term.
All provider rules require $\Subst{T}{t}{p}$, meaning that the time of interaction must 
satisfy the demand imposed. 
The rules of \cref{rules:inststep} give our instant-based model of time computational meaning:
computation happens \emph{at an instant}.
Messages between processes are sent and received at exactly the same instant.
In particular, there is no notion of an in-flight message.

To describe the computation of a configuration over a period of time,
silent labelled transitions can be ``chained'' together, each happening at progressing times.
For example, the reductions $\Omega \LLStepsTo{}{T_1} \Omega_1 \cdots \LLStepsTo{}{T_n} \Omega_n \LLStepsTo{}{T_{n+1}} \Omega'$
take an initial configuration $\Omega$, starting at $T$,
to a configuration $\Omega'$ at $T'$,
such that $T \leq T_i \leq T_{i+1} \leq T'$ for all $i$.
This chaining is expressed by the judgment
$$\ClockProc{T}, \Omega \MultiStepsTo{} \ClockProc{T'}, \Omega'$$
asserting that an initial configuration $\Omega$ at $T$ computes and reaches $\Omega'$ at $T'$.
Configurations $\Omega$ and $\Omega'$ are called the \emph{initial and terminal configurations}, \respb.
The judgment is a generalization of the usual multistep reduction relation,
which not only requires performing an \emph{instantaneous} computation step
but also \emph{advancing the clock}.
To convey which of these steps are performed, we annotate the judgment with \emph{proof terms} $\sigma$.
The resulting multistep reduction judgment
${\sigma}~:~{\ClockProc{T}, \Omega \MultiStepsTo{} \ClockProc{T'}, \Omega'}$
is defined in \cref{rules:mstep}.
These rules define both the judgment and syntax of its proof terms $\sigma$. 
Rule~\ruleref{rule:dyn:stepc} steps the configuration while maintaining time and 
rule~\ruleref{rule:dyn:stept} explicitly advances the time to some time in the future. 
Proof terms $\sigma$ syntactically represent sequences of silent transition steps 
with monotonically increasing time, and are referred to simply as \emph{sequences}.

\begin{figure}
\footnotesize
\input{rules/mstep}
\caption{Timed multistep reduction judgment}
\label{rules:mstep}
\end{figure}

The aforementioned dynamics is inherently non-deterministic. At each step, the judgment allows 
the choice between progressing time or progressing the configuration. Not all such choices 
are fruitful.
For example, choosing to advance time from $T$ to $T'$ for a configuration $\Omega$ at $T$,
will rule out any pending silent step $\Omega \LLStepsTo{}{T} \Omega'$ at $T$, 
possibly resulting in a stuck configuration.

\subsection{Timed Semantic Session Logical Relation}
\label{main:sec:tslr}
As emphasized earlier, session types prescribe and classify process behavior on their offering channel.
Given a session type $A$, we are interested in capturing precisely the runtime behavior that a configuration of processes must exhibit in order to comply with the protocol prescribed by $A$.
For our logical relation, we thus define a family of sets $\LInt{A}{T}$,
indexed by session type $A$,
that characterizes configurations that are \emph{providers} of protocol $A$ at time $T$.
To account for the provider-client distinction inherent to intuitionism,
we must complement this characterization with
a family of sets $\LSInt{A}{T}$,
indexed by session type $A$,
that characterizes configurations that a \emph{client may use} as prescribed by type $A$ at time $T$.
These two characterizations account for the dual roles providers and clients assume with regard to timing considerations,
(see \Cref{main:sec:motivation:process-language}):
whereas providers specify temporal predicates and assert availability at any valid choice of time,
clients satisfy temporal predicates and may choose a suitable valid time.

We define the sets $\LInt{A}{T}$ and $\LSInt{A}{T}$ in \Cref{main:sec:tsslr:lr}.
These definitions rely on our timed labelled transition system and multistep reductions
introduced in \Cref{main:sec:dynamics}.
For the latter we find it convenient to adopt a more global perspective
and work instead with the notion of \emph{computable trajectories},
which we define precisely in \Cref{main:sec:tsslr:traj}.
Trajectories (denoted by $r$, $s$) are functions from \emph{points in time} to configurations,
characterizing the program execution state at each point in time.
We are especially interested in the subset of trajectories that can be realized by some sequence $\sigma$. 
Intuitively, $\sigma$ is a discrete but computable counterpart for the more convenient notion of 
trajectories. Towards this end we define a relation $\Rel{r}{\sigma}$ asserting $r$ is computed by the 
sequence $\sigma$. An element of the relation is termed a \emph{computable trajectory}, or just \emph{trajectory}
for short.
Section \cref{main:sec:tsslr:traj} defines three operations on computable trajectories, a \emph{computable trajectory algebra}, each 
shown to respect the computability relation and shown to commute and cancel properly. 
\begin{itemize}
  \item An \emph{interleaving} operator $\il{r_1}{r_2}$ combines two computable trajectories into one describing the 
  execution of their concurrent composition.
  \item The \emph{partitioning} operators $\lpar{r_1}{T}$ and $\rpar{r_2}{T}$ access the before-$T$ and after-$T$ component 
  of the trajectory, \respb.
  \item Dually, the \emph{concatenation} operator $\concat{r_1}{r_2}$ pieces two trajectories of connected domains together.
\end{itemize}

\subsubsection{Computable Trajectories}
\label{main:sec:tsslr:traj}

In the following sections, unless explicitly stated or deducible from the context, we 
assume that processes, configurations, and syntactic elements of the temporal logic are all 
closed.

As discussed, the execution of a program through time can be understood 
in terms of functions from points in time to configurations, namely \emph{trajectories}. 
 
\begin{definition}[Temporal intervals]
  A temporal interval $I$ is either \emph{bounded} or \emph{unbounded}: 
  \begin{itemize}
    \item (Bounded) $\TInt{T_1}{T_2} \triangleq \SetCmp{T}{T_1 \leq T < T_2}$
    \item (Unbounded) $\TIntInf{T} \triangleq \SetCmp{T'}{T \leq T}$
  \end{itemize}
\end{definition}
  
\begin{definition}[Trajectories, lines, segments]
A \emph{trajectory} is a function $I \to \mathsf{Conf}$. 
If $I$ is unbounded, then it is said to be a \emph{line}, otherwise it is said to be a \emph{segment}. 
The set of trajectories over interval $I$ is denoted by $\Trj{I}$.
\end{definition}

Not all trajectories are computational by nature. For example, take the 
reals as a model, the function that sends all rational numbers to 
$\RunProc{a}{\tpinlcst{t}{p}{A_1}{A_2}{P}}$ and irrational numbers to 
$\RunProc{a}{\tpinrcst{t}{p}{A_1}{A_2}{P}}$ certainly does not result from our dynamics. We are interested in functions that result from our dynamics. This motivates the following set of definitions.

\begin{definition}[$\trjConst{I}{\Omega}$ ]
$\trjConst{I}{\Omega}(-)$ is the constant function sending $I$ to $\Omega$;
\end{definition}
\begin{definition}[$\trjExt{T}{\Omega}{r}$]
Function $\trjExt{T}{\Omega}{-}$ extends the domain leftward to $T$ of its argument trajectory by sending the 
additional inputs to $\Omega$. Full definitions are available in \refapx[apx:semantics:trj].
\end{definition}

\begin{definition}[Computable trajectory]\label{main:defn:ctrj}
Let $\Rel{r}{\sigma}$ be the strongest relation satisfying: 
\begin{itemize}
  \item $\Rel{\trjConst{\TInt{T}{T'}}{\Omega}}{\StepRefl{T}{\Omega}}$, where $T'$ may be $\infty$.
  \item $\Rel{r}{\StepC{T}{\Omega}{\Omega'}{\sigma}}$ if $\Rel{r}{\sigma}$.
  \item $\Rel{\trjExt{T}{\Omega}{r}}{\StepT{T}{T'}{\Omega}{\sigma}}$ if $\Rel{r}{\sigma}$.
\end{itemize}

An element $\pairexcst{r}{\sigma}$ of the relation is called a \emph{computable trajectory},
or just \emph{trajectory}.
Computable trajectories are denoted by $w$.
The trajectory $\pairexcst{r}{\sigma}$ is more precisely called a
\emph{computable line}, if $r$ is a line, and a \emph{computable segment}, 
if $r$ is a segment.
\end{definition}

\noindent Intuitively, $\Rel{r}{\sigma}$ takes a discrete description of computation over time, given by 
$\sigma$,
and fills in the gaps by (1) assuming that the configuration stays unchanged until the next silent 
transition step and (2) maintaining the last configuration 
indefinitely into the future, if the trajectory is a line.

\begin{definition}
Let $w = \pairexcst{r}{\sigma}$ be a computable trajectory. The domain of $w$, write $\Dom{w}$, 
is the domain of $r$. For all $T \in \Dom{w}$, define $w(T) \triangleq r(T)$.
\end{definition}

\begin{definition}[$\ctrj{T_1}{T_2}{\Omega}{\Omega'}$ and $\cwl{T}{\Omega}$]
Let $\ctrj{T_1}{T_2}{\Omega}{\Omega'}$ denote the set of computable trajectories 
$\pairexcst{r}{\sigma}$ such that (1) $r \in \Trj{\TInt{T_1}{T_2}}$, 
(2) initial configuration of $\sigma$ is $\Omega$, and (3) 
if $r$ is a segment then the terminal configuration of $\sigma$ is $\Omega'$, 
otherwise formally write $\Omega' = \cdot$.
Further define $\cwl{T}{\Omega} \triangleq \ctrj{T}{\infty}{\Omega}{\cdot}$.
\end{definition}

From a computable segment we can recover a sequence
(proof in \refapx[apx:semantics:ctrj]):
\begin{lemma}
If $w \in \ctrj{T_1}{T_2}{\Omega_1}{\Omega_2}$, then 
$\IsOf{\sigma}{\ClockProc{T_1}, \Omega_1 \MultiStepsTo{} \ClockProc{T_2}, \Omega_2}$.
\end{lemma}

Computable trajectories are functions with computability receipts. Therefore, it is 
natural to consider and define common operations on functions for computable trajectories. 
The subtlety of the operations lies in the handling of the receipts, that is the operations 
on trajectories must be justified by operations on sequences. These operations on the 
sequences are constructive, therefore proofs carried out with computable trajectories 
yield effective ways to determine schedules.

In this section, we describe the operations and the necessary properties they 
satisfy, for the exact construction please refer to \refapx[apx:semantics:ctrj].

\begin{definition}[Equivalence]
Trajectories $w_1 \in \ctrjI{I_1}{\Omega_1}{\Omega_1'}$ and $w_2 \in \ctrjI{I_2}{\Omega_2}{\Omega_2'}$
are said to be \emph{equivalent} on interval $I$, $\cwlequivt{I}{w_1}{w_2}$, 
iff for all $T \in I$, $w_1(T) = w_2(T)$. 

In particular, if $I = I_1 = I_2$ and $\cwlequivt{I}{w_1}{w_2}$, write just $\cwlequiv{w_1}{w_2}$.
\end{definition}

The equivalence relations on computable trajectories treat them as functions. We remark that 
$\cwlequivt{I}{w_1}{w_2}$ is reflexive, symmetric, and transitive by definition. 

We will now define a pair of operators that are dual to each other. 

\begin{definition}[Concatenation]
The \emph{concatenation} operator $\concat{w_1}{w_2}$ concatenates trajectories with 
connected domains:
$$
\concat{\cdot}{\cdot} : 
\ctrj{T_1}{T_2}{\Omega_1}{\Omega_2}
\to \ctrj{T_2}{T_3}{\Omega_2}{\Omega_3}
\to \ctrj{T_1}{T_3}{\Omega_1}{\Omega_3}
$$
\end{definition}
Concatenation of trajectories coincides with the constituents on the domain 
each is defined:

\begin{corollary}
$\forall w_1, w_2$, 
$\cwlequivt{\Dom{w_1}}{w_1}{(\concat{w_1}{w_2})}$ and
$\cwlequivt{\Dom{w_2}}{w_2}{(\concat{w_1}{w_2})}$. 
\end{corollary}

Proof in \refapx[apx:semantics:ctrj]. Additionally, given a trajectory and a point in its domain, it is possible to partition the domain 
by this point, motivating the following pairs of operators.

\begin{definition}[Partition]
The left and right partition of $w$ by time $T$ s.t. $T \in \TInt{T_1}{T_2}$ are: 
\begin{align*}
  \lpar{\cdot}{T} &: \ctrj{T_1}{T_2}{\Omega_1}{\Omega_2} \to \ctrj{T_1}{T}{\Omega}{w(T)} \\
  \rpar{\cdot}{T} &: \ctrj{T_1}{T_2}{\Omega_1}{\Omega_2} \to \ctrj{T}{T_2}{w(T)}{\Omega_2} 
\end{align*}
\end{definition}

A trajectory coincides with its parts on each domain (proof in \refapx[apx:semantics:ctrj]):
\begin{corollary} 
For all $w$ and $T \in \Dom{w}$, 
$\cwlequivt{\Dom{\lpar{w}{T}}}{\lpar{w}{T}}{w}$ and
$\cwlequivt{\Dom{\rpar{w}{T}}}{\rpar{w}{T}}{w}$.
\end{corollary}

The operators are therefore dual in the following sense:
\begin{lemma}
For all $w$ and $T \in \Dom{w}$, $\cwlequiv{w}{\concat{(\lpar{w}{T})}{(\rpar{w}{T})}}$
\end{lemma}
\begin{proof}
Take arbitrary $T' \in \Dom{w}$. If $T' \geq T$, then 
$(\concat{(\lpar{w}{T})}{(\rpar{w}{T})})(T') = (\rpar{w}{T})(T') = w(T')$. Otherwise,
$(\concat{(\lpar{w}{T})}{(\rpar{w}{T})})(T') = (\lpar{w}{T})(T') = w(T')$.
\end{proof}

Partitioning and concatenation both operate on the trajectory of 
a monolithic configuration. 
The computation of a \emph{compound} configuration through time consists of multiple 
configurations computing through time concurrently. This motivates us to define 
the interleaving operator:

\begin{definition}[Interleaving]
The \emph{interleaving} operator has the following signature:
$$
\il{\cdot}{\cdot} :
\ctrj{T}{T'}{\Omega_1}{\Omega_1'} \to 
\ctrj{T}{T'}{\Omega_2}{\Omega_2'} \to
\ctrj{T}{T'}{\conc{\Omega_1}{\Omega_1'}}{\conc{\Omega_2}{\Omega_2'}}
$$
\end{definition}

The operator is constructed such that (proof in \refapx[apx:semantics:ctrj]): 
\begin{corollary}
For all $w_1$ and $w_2$ and $T$, $\il{w_1}{w_2}(T) = \conc{(w_1(T))}{(w_2(T))}$. 
\end{corollary}

As a further corollary, partitioning and concatenation distribute over interleaving.

\begin{corollary}
For all $w_1 \in \ctrjI{I}{\Omega_1}{\Omega_1'}$ and $w_2 \in \ctrjI{I}{\Omega_2}{\Omega_2'}$ 
and for all $T \in I$, 
\begin{itemize}
  \item $\cwlequiv{\lpar{(\il{w_1}{w_2})}{T}}{\il{(\lpar{w_1}{T})}{(\lpar{w_2}{T})}}$ and $\cwlequiv{\rpar{(\il{w_1}{w_2})}{T}}{\il{(\rpar{w_1}{T})}{(\rpar{w_2}{T})}}$.
  \item $\cwlequiv{\concat{(\il{\lpar{w_1}{T}}{\lpar{w_2}{T}})}{(\il{\rpar{w_1}{T}}{\rpar{w_2}{T}})}}
                  {\il{w_1}{w_2}}$
\end{itemize}
\end{corollary}
\begin{proof}
Fix arbitrary $T' \in I$ and case split between $T' \geq T$ and $T' < T$, 
in either case proceed by straightforward computation.
\end{proof}

The interleaving operator is named after the critical setup in its construction: to 
obtain a singular sequence from two sequences by interleaving instantaneous 
steps from both sequences while maintaining monotonic ordering in time. In other
words, the computational content of the proof constitutes an \emph{algorithm} for scheduling
process execution.

Computable trajectories not only facilitate a succinct definition of our logical relation,
but also greatly benefit the proof of the fundamental theorem of our logical relation.

\subsubsection{Logical Relation}
\label{main:sec:tsslr:lr}


One difficulty presented by session-typed languages is the handling of
names. Channel names are runtime values that clients use to distinguish between providers. Although
the name of the providing channel is recorded as part of the provider's syntax $\RunProc{a}{P}$, 
the semantic of the provider does not, and should not, meaningfully depend on the channel name. 
We introduce \emph{nameless families} to semantically capture this.
Nameless families are the actual inhabitants of our logical relation. 

\begin{definition}
A \emph{nameless family of configurations} $\bfOmega$ is a family of configurations differing only in the choice of a single 
(root) process offering channel name. 
Formally that is for some fixed $P$ and $\Omega$, $\bfOmega[a] = \conc{\RunProc{a}{P}}{\Omega}$ for all indices $a$.
\end{definition}

The definition can be point-wise extended to computable trajectories.
When it is clear from the context, we may speak of nameless families of trajectories and 
configurations by just ``trajectories'' and ``configurations'', \respb.
%

Our logical relation necessitates defining an
auxiliary sort $\mathcal{A}$ of \emph{urgent types}, inspired by \cite{BocchiESOP2019}: 
$$ \calA \bnfdef \unitstycst \mid \tensorstycst{A_1}{A_2} \mid \larrstycst{A_1}{A_2} \mid \bsumstycst{A_1}{A_2} \mid \bwithstycst{A_1}{A_2}$$
Urgent types represent a session at a client-instantiated time, 
right before the communication has occurred. The definition is \emph{not} recursive: the 
component types (if any) are regular timed session types.
We can instantiate a type $A$ at some point in time $T$, rendering it urgent.

\begin{definition}[Urgency instantiation $\expose{A}{T}$]
Let $\expose{A}{T}$ be the urgent type of $A$ instantiated at $T$:
\begin{align*}
  \expose{\tpunitstycst{t}{p}}{T}               &\triangleq \unitstycst \\
  \expose{(\tptensorstycst{t}{p}{A_1}{A_2})}{T} &\triangleq \tensorstycst{(\Subst{T}{t}{A_1})}{(\Subst{T}{t}{A_2})}\\
  \expose{(\tplarrstycst{t}{p}{A_1}{A_2})}{T}   &\triangleq \larrstycst{(\Subst{T}{t}{A_1})}{(\Subst{T}{t}{A_2})}\\
  \expose{(\tpsumstycst{t}{p}{A_1}{A_2})}{T}    &\triangleq \bsumstycst{(\Subst{T}{t}{A_1})}{(\Subst{T}{t}{A_2})}\\
  \expose{(\tpwithstycst{t}{p}{A_1}{A_2})}{T}   &\triangleq \bwithstycst{(\Subst{T}{t}{A_1})}{(\Subst{T}{t}{A_2})}
\end{align*}
\end{definition}


Our logical relation distinguishes itself from tradition in that it is defined 
for pairs $\gl{\bfw}{\bfOmega}$, called \emph{temporal computability pairs},
where the components are nameless families of computable trajectories and configurations, 
satisfying $\bfw \in \cwl{T}{\bfOmega}$.
Here $\bfOmega$ is the initial configuration and $\bfw$ is the evidence that it carries to 
substantiate its semantic property. Intuitively, the logical relation classifies initial 
configurations by examining their proposed trajectories at various points in time according 
to the types they advertise.

\begin{figure}[htb]
\small
\newcolumntype{L}{>$l<$}
\newcolumntype{C}{>$c<$}
\begin{tabular}{lcp{.9\textwidth}}
$\gl{\bfw}{\bfOmega} \in \LSInt{\wp{A}{t}{p}}{T}$ & iff &
  $\forall T'. \Subst{T'}{t}{p} \land (T' \geq T) \implies \bfw(T') \in \VSInt{\expose{A}{T'}}{T'}$ \\
$\gl{\bfw}{\bfOmega} \in \LInt{\wp{A}{t}{p}}{T}$ & iff & 
  $\forall T'. \Subst{T'}{t}{p} \implies (T' \geq T) \land \bfw(T') \in \VInt{\expose{A}{T'}}{T'}$ \\
\\
$\bfOmega \in \VAnyInt{\unitstycst}{T}$ & iff & 
  $\forall a. \bfOmega[a] \LLStepsTo{\actsndclosecst{a}}{T} \StopConf$. \\
$\bfOmega \in \VAnyInt{\bwithstycst{A_1}{A_2}}{T}$ & iff & 
  $\forall a$ $\exists{\bfw_1} \exists \bfOmega_1$ s.t. $\bfOmega[a] \LLStepsTo{\actrcvlblcst{a}{\lblleft}}{T} \bfOmega_1[a]$ and 
  $\gl{\bfw_1}{\bfOmega_1} \in \LAnyInt{A_1}{T}$, and \newline
  $\exists{\bfw_2} \exists \bfOmega_2$ s.t. $\bfOmega[a] \LLStepsTo{\actrcvlblcst{a}{\lblright}}{T} \bfOmega_2[a]$ and
  $\gl{\bfw_2}{\bfOmega_2} \in \LAnyInt{A_2}{T}$.
    \\
$\bfOmega \in \VAnyInt{\bsumstycst{A_2}{A_2}}{T}$ & iff & $\forall a$ either \newline 
  $\exists \bfw_1 \exists \bfOmega_1$ s.t. $\bfOmega[a] \LLStepsTo{\actsndlblcst{a}{\lblleft}}{T} \bfOmega_1[a]$ and 
  $\gl{\bfw_1}{\bfOmega_1} \in \LAnyInt{A_1}{T}$, or
  \newline
  $\exists \bfw_2 \exists \bfOmega_2$ s.t. $\bfOmega[a] \LLStepsTo{\actsndlblcst{a}{\lblright}}{T} \bfOmega_2[a]$ and 
  $\gl{\bfw_2}{\bfOmega_1} \in \LAnyInt{A_2}{T}$.
  \\
$\bfOmega \in \VAnyInt{\tensorstycst{A_1}{A_2}}{T}$ & iff &
  $\forall a \exists c \exists \bfw_1 \bfw_2 \exists \bfOmega_1 \bfOmega_2$ s.t. 
  $\bfOmega[a] \LLStepsTo{\actsndchancst{a}{c}}{T} \conc{\bfOmega_1[c]}{\bfOmega_2[a]}$, \newline
  $\gl{\bfw_1}{\bfOmega_1} \in \LAnyInt{A_1}{T}$ and 
  $\gl{\bfw_2}{\bfOmega_2} \in \LAnyInt{A_2}{T}$.
  \\
$\bfOmega \in \VInt{\larrstycst{A_1}{A_2}}{T}$ & iff &
  $\forall a \forall \bfw_1 \forall \bfOmega_1$ s.t. 
  if $\gl{\bfw_1}{\bfOmega_1} \in \LSInt{A_1}{T}$ then \newline
  $\forall c \exists \bfw_2$ s.t.
  $\bfOmega[a] \LLStepsTo{\actrcvchancst{a}{c}}{T} \bfOmega_2[a]$ and
  $\gl{\bfw_2}{\conc{\bfOmega_1[c]}{\bfOmega_2}} \in \LInt{A_2}{T}$. 
  \\
$\bfOmega \in \VSInt{\larrstycst{A_1}{A_2}}{T}$ & iff &
  $\forall a \forall \bfw_1 \forall \bfOmega_1$ s.t. 
  if $\gl{\bfw_1}{\bfOmega_1} \in \LInt{A_1}{T}$ then \newline
  $\forall c \exists \bfw_2$ s.t. 
  $\bfOmega[a] \LLStepsTo{\actrcvchancst{a}{c}}{T} \bfOmega_2[a]$ and
  $\gl{\bfw_2}{\conc{\bfOmega_1[c]}{\bfOmega_2}} \in \LSInt{A_2}{T}$. 
  \\
\end{tabular}
\caption{Timed semantic session logical relation.}
\label{fig:lr}
\end{figure}

We define four (unary) relations:

\begin{tabular}{rl}
  $\gl{\bfw}{\bfOmega} \in \LSInt{A}{T}$  & \quad \text{Interpreting \emph{latent} configurations to be used with evidence $\bfw$.} \\ 
  $\bfOmega \in \VSInt{\calA}{T}$  & \quad \text{Interpreting \emph{urgent} configuration to be used.} \\
  $\gl{\bfw}{\bfOmega} \in \LInt{A}{T}$  & \quad \text{Interpreting \emph{latent} configuration as provider with evidence $\bfw$.} \\
  $\bfOmega \in \VInt{\calA}{T}$  & \quad \text{Interpreting \emph{urgent} configuration as provider.} \\
\end{tabular}\\

All four relations take the current time $T$ as an input, indicating that meaning of a type is time-dependent.
Two \emph{latent relations} classify initial configurations whose advertised services 
are yet to occur, therefore they require a proposed computable trajectory from $\bfOmega$ 
to justify their semantic inhabitance. They are akin to expression interpretations 
in traditional logical relations. On the other hand, the \emph{urgent relations} 
classify configurations after a time of interaction has been picked and fixed. They 
are akin to value interpretations in traditional logical relations. 
The logical relations are defined inductively over $A$ in \cref{fig:lr}.

\subsubsection{Support for Functional Value Exchange} 
We briefly sketch modifications required to support communicating functional values. First, actions $\alpha$ are enriched with 
a complementary pair of actions: $\actsndvalcst{a}{v}$ and $\actrcvvalcst{a}{v}$ for sending and receiving value $v$ over $a$.
Let $e \evalsto v$ be a suitably defined big-step evaluation judgment for functional expressions. 
The transition rules now include transitions engendering these actions:

\begin{small}
\raggedright
\begin{tabular}{l}
\defsteprule[D-$!$ L][rule:dyn:produce:l]
 {\RunProc{a}{\tpconsumeproccst{T}{b}{x}{Q}} }
 {\actrcvvalcst{b}{v}}{T}
 {\RunProc{a}{\Subst{v}{x}{Q}}}\\

\defsteprule[D-$?$ L][rule:dyn:query:l]
 {\RunProc{a}{\tpsupplyproccst{T}{b}{v}{Q}} }
 {\actsndvalcst{b}{e}}{T}
 {\RunProc{a}{Q}} where $e \evalsto v$ \\

\textsc{With premise $\Subst{T}{t}{p}$:} \\

\defsteprule[D-$!$][rule:dyn:produce:r]
 {\RunProc{a}{\tpproduceproccst{t}{p}{v}{P}}}
 {\actsndvalcst{a}{e}}{T}
 {\RunProc{a}{P}} where $e \evalsto v$ \\

\defsteprule[D-$?$][rule:dyn:query:r1]
 {\RunProc{a}{\tpqueryproccst{t}{p}{x}{P}}}
 {\actrcvvalcst{a}{v}}{T}
 {\RunProc{a}{\Subst{v}{x}{P}}}\\
\end{tabular}
\end{small}

This completes the necessary changes to the transition system. On the logical relation side, we start by defining 
$\expose{\tpbangstycst{t}{p}{\tau}{A}}{T} \triangleq \bangstycst{\tau}{\Subst{T}{t}{A}}$ and 
$\expose{\tpquerystycst{t}{p}{\tau}{A}}{T} \triangleq \querystycst{\tau}{\Subst{T}{t}{A}}$.

The logical relation is then enriched to account for the two additional session types by adding to the $\VAnyInt{-}{T}$ cases: 
\begin{small}
\begin{align*}
\bfOmega \in \VAnyInt{\bangstycst{\tau}{A}}{T} & \triangleq
  \forall a \exists \bfw_1 \exists \bfOmega_1. (\bfOmega[a] \LLStepsTo{\actsndvalcst{a}{v}}{T} \bfOmega_1[a]) \land 
  (\gl{\bfw_1}{\bfOmega_1} \in \LAnyInt{A}{T}) \land (v \in \Interp{\tau}) \\
\bfOmega \in \VAnyInt{\querystycst{\tau}{A}}{T} & \triangleq
  \forall a \exists\bfw_1 \exists \bfOmega_1. (\bfOmega[a] \LLStepsTo{\actrcvvalcst{a}{v}}{T} \bfOmega_1[a]) \land 
  \forall (v \in \Interp{\tau}). \gl{\bfw_1}{\bfOmega_1} \in \LAnyInt{A}{T}
\end{align*}
\end{small}

Here $\Interp{\tau}$ denote some suitable logical relation defined for the functional language. The exact detail of the definition
depends on the language features available in the functional layer. However, for a wide range of choices it is well studied.
These definitions here says that the process must be willing to send (receive) a functional value, and that 
it continues to behave according to $A$ after the message exchange. If the process sends then the value sent must be well-behaving. 
Otherwise, it may assume that the value it received is well-behaving.

\section{Automatic Verification Through Refinement Type System for \tilst}
\label{main:sec:type-sys}
This section explores \Cref{mode-l:all} of our logical relation,
once-and-for-all verification using a type system.
To facilitate this mode, we first develop a refinement type system for \tillst (\Cref{main:sec:tillst})
and then show that well-typed terms inhabit the logical relation (\Cref{main:sec:ftlr})

\subsection{Refinement Type System for \tilst}
\label{main:sec:tillst}

Our refinement type system considers the process language introduced in \Cref{main:sec:motivation:process-language}
and assigns \tillst types to them.
The resulting typing rules are shown in \Cref{main:fig:tillst-typing} and
employ the judgment
$$\calG; \calF \mid \Delta \entails{\IsOfProc{P}{T}{A}}.$$
The judgment differs from the usual judgment found in \illst systems
by its dependence on the temporal variables in $\calG$ and propositions in $\calF$.
It reads as \textit{``at time $T$, process term $P$ provides a session of type $A$,
given the typing of channels in $\Delta$ and
assuming truth of the propositions in $\calF$''}.
We call attention to the dependencies on the temporal variable context $\calG$:
the linear channel context $\Delta$, propositional context $\calF$, 
the time of assertions $T$,
and the type $A$ are all scoped under context $\calG$.


\begin{figure}[ht]
\centering
\begin{small}
\input{rules/typing.tex}
\end{small}
\caption{Process term typing rules of \tilst.}
\label{main:fig:tillst-typing}
\end{figure}

The time $T$ is the time \emph{at which the judgment is asserted}.
The validity and meaning of typing thus depends on $T$. 
As usual, we give the typing rules in a \emph{sequent calculus},
where the conclusion denotes the protocol state before the message exchange
and the premises the protocol states after the message exchange. 
Therefore, the time \emph{at} the conclusion must \emph{precede} that of the premises.

Judgmentally, a type $\wp{A}{t}{p}$ at time $T$ internalizes a family of derivations indexed by 
time instances $t \geq T$ such that $p$ is true. This is enforced in the right rules 
by asserting the premises at time variable $t$ along with the premise 
$p \entails{\leqcst{T}{t}}$.
Use of the type $\wp{A}{t}{p}$, on the other hand, is required to occur at some concrete time $T'$ that is accessible from ``now'' 
and satisfies the predicate $p$. 
This is enforced in the left rules by typing the premises at a fixed future $T'$ ($\leqcst{T}{T'}$) satisfying 
$p(T)$.

Since right rules \emph{bind} the time of communication $t$ for typing the providing process' continuation,
premises of right rules extend $\calG$ with the variable $t$ and $\calF$ with the proposition $p$,
allowing temporal predicates of future interactions to refer to the times of past interactions.
Left rules, on the other hand, update the type of the channel variable interacted with for the continuation,
substituting $T'$ for $t$ in $A$.

\tilst adopts a \emph{global} notion of time, in the sense that time expressions reference points in time that are 
commonly known and agreed upon. In particular, a closed 
time expression means the same for all processes.
Furthermore, time passes at the same pace for every process. When a process 
advances time to communicate over some channel $x$ in $\Delta$, the same amount of time passes for the remaining 
channels in $\Delta$. To see this consider the two types:
$$A \triangleq \tplarrstycst{t_1}{t_0 \leq t_1 \leq t_0 + 15}{X}{(\tptensorstycst{t_2}{t_2 = t_1 + 10}{X}{C})}, \quad 
B \triangleq \tplarrstycst{t_1}{t_0 \leq t_1 \leq t_0 + 10}{X}{(\tptensorstycst{t_2}{t_2 = t_1 + 10}{X}{C})}$$
Both $A$ and $B$ accept a process of some generic type $X$, work on it, and then return 
it after 10 units of time, counting from reception of $X$. The difference is that $B$ needs it within 10 units of time from now, 
but $A$ can wait a bit longer. Suppose we are programming a client $P$:
$$\calG; \calF \mid \IsOf{x}{A}, \IsOf{y}{B}, \IsOf{z}{X} \entails{\IsOfProc{P}{t_0}{D}}.$$
Process $P$ must send the process $\IsOf{z}{X}$ to $x$ and $y$ in some order. Here are four possible implementations $P_i$ where $i = \{1, 2, 3, 4\}$: 
\begin{align*}
P_1 &\triangleq \tplarrsndcst{x}{t_0}{\fwdcst{z}}{\tptensorrcvcst{x}{t_0 + 10}{z}{\tplarrsndcst{y}{t_0 + 10}{\fwdcst{z}}{\tptensorrcvcst{y}{t_0 + 20}{z}{\cdots}}}} \\
P_2 &\triangleq \tplarrsndcst{y}{t_0+3}{\fwdcst{z}}{\tptensorrcvcst{y}{t_0 + 13}{z}{\tplarrsndcst{x}{t_0 + 15}{\fwdcst{z}}{\tptensorrcvcst{x}{t_0 + 25}{z}{\cdots}}}} \\
P_3 &\triangleq \tplarrsndcst{x}{t_0+3}{\fwdcst{z}}{\tptensorrcvcst{x}{t_0 + 13}{z}{\tplarrsndcst{y}{t_0 + 13}{\fwdcst{z}}{\tptensorrcvcst{y}{t_0 + 23}{z}{\cdots}}}} \\
P_4 &\triangleq \tplarrsndcst{y}{t_0+6}{\fwdcst{z}}{\tptensorrcvcst{y}{t_0 + 16}{z}{\tplarrsndcst{x}{t_0 + 16}{\fwdcst{z}}{\tptensorrcvcst{x}{t_0 + 25}{z}{\cdots}}}} 
\end{align*}

$P_1$ and $P_2$ will type-check while $P_3$ and $P_4$ will not. Process $P_1$ immediately sends $z$ to $x$, and sends $z$ \emph{again immediately}
to $y$ once it receives $z$ back. It barely makes the deadline imposed by $y$ in the second send. Process $P_3$, in contrast to $P_1$, waits 
3 units of time in the first send, causing it to miss the deadline in the second send. Process $P_2$ switches the order between 
$x$ and $y$, giving it a bit of slack. It chooses to wait for 3 units for the first send and an additional 2 units 
for the second send. Process $P_4$, in contrast to $P_2$, waits 6 units for the first send, causing the second send to miss the deadline. In all 
cases, as a process $P_i$ spends time with either participant $x$ or $y$, time also progresses for the other participant, 
reducing the window-of-communication.

The above example provides a context for us to discuss an important \emph{asymmetry} between types as antecedents versus succedents.
When an antecedent $\IsOf{x}{\wp{A}{t}{p}}$ moves from $T$ to a future $T'$, part of the internalized derivations at $T$, specifically 
those between $T$ and $T'$, are no longer internalizable at this new time $T'$, as connectives only internalize derivations concerning the future. 
At the same time the client loses access to these derivations because time progresses equally for both parties. 
For antecedents, we impose semantic requirements only for the reachable times;
for succedents, we require all times under quantification to be reachable and well-behaving.

As a consequence of this asymmetry, the identity rule \ruleref{rule:fwd} and the cut-rule \ruleref{rule:cut} both carry
an extra premise. The definitions are available in \cref{main:fig:tillst-retyping}. 

\begin{figure}[ht]
\centering
\input{rules/retyping.tex}
\caption{Retyping rules of \tilst.}
\label{main:fig:tillst-retyping}
\end{figure}

The premise $\FwdComp{A}{A'}{T}$ (termed \emph{forward-retyping}) in \ruleref{rule:fwd} picks out the 
part of $A$ that remains reachable at $T$ and rewrites the type accordingly to $A'$. Its effect is to 
ensure that a forward does not happen ``too late'',
\ie that the process providing along $x$ must be available at least for the entire period that
the forwarding process promises to be available to its client.
Without it the system would be unsound. 
To see what goes wrong otherwise, consider the following process:
$$
\IsOf{x}{\tpunitstycst{t}{\fbetween{t_0}{t}{\fshiftcst{t_0}{5}}}}, 
\IsOf{y}{\tpunitstycst{t}{\fbetween{t_0}{t}{\fshiftcst{t_0}{5}}}} 
\entails{\IsOfProc{\tpwaitcst{\fshiftcst{t_0}{2}}{x}{\tpfwdcst{\fshiftcst{t_0}{2}}{y}}}{t_0}{\tpunitstycst{t}{\fbetween{t_0}{t}{\fshiftcst{t_0}{5}}}}}
$$
The process starts at $t_0$ and waits until $\fshiftcst{t_0}{2}$ to close $x$, then 
it starts forwarding $y$. Itself advertises to be available $\fbetween{t_0}{t}{\fshiftcst{t_0}{5}}$ along its providing channel,
which includes $\fshiftcst{t_0}{1}$.
At this time, however, it will not be available as it is waiting to communicate with $x$.
This is a soundness problem pertaining specifically to \ruleref{rule:fwd}. 

A dual problem of \emph{completeness} arises in the \ruleref{rule:cut}~rule.
The premise $\CutComp{A}{A'}{T}$ (termed \emph{cut-retyping}) in \ruleref{rule:cut} allows the cut as long 
as the reachable parts of $A'$ at $T$ are covered by $A$, effectively allowing $A$ to be cut against an $A'$ with a 
broader quantification. To see why it is useful, let us consider another example.
Suppose at time $\fshiftcst{t_0}{2}$ we 
have a provider $P$ and client $Q$ satisfying: 
$$\Delta_1 \entails{\IsOfProc{P}{\fshiftcst{t_0}{2}}{\tpunitstycst{t}{\leqcst{\fshiftcst{t_0}{2}}{t}}}},
\quad 
\Delta_2, \IsOf{x}{\tpunitstycst{t}{t_0 \leq t}} 
\entails{\IsOfProc{Q}{\fshiftcst{t_0}{2}}{C}}$$
Process $Q$ expects to be able to close $x$ at any time following $t_0$, and $P$ is a process 
that can be closed at any time after $\fshiftcst{t_0}{2}$. However, since the typing is 
derived at $\fshiftcst{t_0}{2}$, the earliest possible time that $Q$ can communicate through 
$x$ is the current time $\fshiftcst{t_0}{2}$.
Therefore, this cut should be permitted, despite the fact that the types provided and used are syntactically different.

\subsection{Fundamental Theorem}
\label{main:sec:ftlr}
To prove that all process terms with valid derivations
using the rules in \Cref{main:fig:tillst-typing}
inhabit our \tslr,
and thus are timely,
we prove the fundamental theorem of the logical relation.
The theorem is stated for \emph{open terms},
allowing our program to be composed with other objects,
as long as they inhabit the logical relation.
We first introduce auxiliary definitions to account for open process terms,
then state the fundamental theorem.
Because process terms contain 
both free channel variables and temporal variables, both contexts need to be accounted for.

\begin{definition}[$\delta \in \LSInt{\Delta}{T}$]
A \emph{sub-forest} for context $\Delta$ is a map $\delta$ from $\Delta$ to temporal computability pairs.
We say $\delta \in \LSInt{\Delta}{T}$ iff for all $(\IsOf{x}{A}) \in \Delta$, $\delta_x \in \LSInt{A}{T}$
\end{definition}

\begin{definition}[$\Delta \sementails{\IsOfProc{P}{T}{A}}$]
A \emph{(runtime) channel substitution} for $\Delta$ is a map $\gamma$ from $\Delta$ to channel names.
We say $\Delta \sementails{\IsOfProc{P}{T}{A}}$ iff 
for all $\delta \in \LSInt{\Delta}{T}$ and for all substitutions $\gamma$,
$\exists \bfw$ s.t. $\gl{\bfw}{\conc{(\ilfam{x \in \Delta}{\bfOmega_x[\gamma(x)]})}{\RunProc{-}{\hat{\gamma}(P)}}} \in \LInt{A}{T}$, 
where $\bfOmega_x$ is initial configuration of $\delta_x$.
\end{definition}

\begin{definition}[$\calG; \calF \mid \Delta \sementails{\IsOfProc{P}{T}{A}}$]\label{main:defn:subst}
Let $\varphi$ be any assignment of time variables $t \in \calG$ such that $\varphi$ satisfies $\calF$. Let $\hat{\varphi}$ 
be the substitution function induced by the assignment. Then
$\hat{\varphi}(\Delta) \sementails{\IsOfProc{\hat{\varphi}(P)}{\hat{\varphi}(T)}{\hat{\varphi}(A)}}$
\end{definition}

\begin{theorem}[FTLR]\label{main:thm:ftlr}
If $\calG; \calF \mid \Delta \entails{\IsOfProc{P}{T}{A}}$ then $\calG; \calF \mid \Delta \sementails{\IsOfProc{P}{T}{A}}$.
\end{theorem}

Proof in \refapx[apx:semantics:ftlr]. Additionally, the usual forward and backward closure property is now generalized to account for 
the passage of time. For the forward direction, given $\gl{\bfw}{\bfOmega} \in \LSInt{A}{T}$, the 
right partition of the computable trajectories at any future time $T'$ inhabits 
the same type at $T'$. 
Dually, for the backwards direction, inhabitance in the type is preserved for any past 
time as long as there is a way to extend the trajectories to that past time.

\begin{lemma}[Forward and backward closure] (Proof in \refapx[apx:semantics:fbc])\label{main:lem:closure}
\begin{itemize}
\item If $\gl{\bfw}{\bfOmega} \in \LSInt{A}{T}$ then 
$\forall T'. (T' \geq T) \implies \gl{\rpar{\bfw}{T'}}{\bfw(T')} \in \LSInt{A}{T'}$

\item If $\bfw_1 \in \ctrj{T'}{T}{\bfOmega_1}{\bfOmega_2}$ and $\gl{\bfw_2}{\bfOmega_2} \in \LInt{A}{T}$,
then $\gl{(\concat{\bfw_1}{\bfw_2})}{\bfOmega_1} \in \LInt{A}{T'}$.
\end{itemize}
\end{lemma}

Semantically, the retyping relations $\FwdComp{A}{B}{T}$ and $\CutComp{A}{B}{T}$ allow us to translate 
between $\LSInt{-}{T}$ and $\LInt{-}{T}$, captured in the following two lemmas:

\begin{lemma}[Semantic retyping]\label{lem:sem-retyping} (Proof of its generalization in \refapx[apx:semantics:sem-retyping])\label{main:lem:retyping}
\begin{itemize}
\item If $\gl{\bfw}{\bfOmega} \in \LSInt{A}{T}$ and $\FwdComp{A}{B}{T}$ then $\gl{\bfw}{\bfOmega} \in \LInt{B}{T}$.
\item If $\gl{\bfw}{\bfOmega} \in \LInt{A}{T}$ and $\CutComp{A}{B}{T}$ then $\gl{\bfw}{\bfOmega} \in \LSInt{B}{T}$.
\end{itemize}
\end{lemma}

The proofs of \Cref{main:lem:closure} and \Cref{lem:sem-retyping} greatly benefit from the abstractions
afforded by computable trajectories and are essentially carried out by equational reasoning in terms of the trajectory algebra.

As an immediate result of \Cref{main:thm:ftlr}, we can prove the following adequacy theorem for 
closed terms of type $\tpunitstycst{t}{p}$:

\begin{theorem}[Adequacy]\label{main:thm:adequacy}
If $\IsOfProc{P}{\finitcst}{\tpunitstycst{t}{\eqcst{t}{\fshiftcst{\finitcst}{\overline{n}}}}}$, then \\
$\exists\sigma. \IsOf{\sigma}{\ClockProc{\finitcst}, \RunProc{a}{P} \MultiStepsTo{} \ClockProc{\fshiftcst{\finitcst}{\overline{n}}}, \Omega}$, 
and for some $\Omega$ s.t. $\Omega \LLStepsTo{\actsndclosecst{a}}{\fshiftcst{\finitcst}{\overline{n}}}{\StopConf}$, \\
\end{theorem}
\begin{proof}
By the fundamental theorem (\Cref{main:thm:ftlr}) we have $\emptyset \sementails{\IsOfProc{P}{T}{A}}$ because temporal context is empty. 
There exists $\bfw$ s.t. 
$\gl{\bfw}{\bfOmega} \in \LInt{\tpunitstycst{t}{\eqcst{t}{\fshiftcst{\finitcst}{\overline{n}}}}}{\finitcst}$.
where $\bfOmega = \RunProc{-}{P}$. Let $T \triangleq \fshiftcst{\finitcst}{\overline{n}}$. Therefore, 
 $\bfw(T) \in \VInt{\unitstycst}{T}$. That is for any $a$, $\bfw(T)[a] \LLStepsTo{\actsndclosecst{a}}{T} \StopConf$. 
Obtain the desired sequence by additionally consulting 
$\lpar{\bfw[a]}{T} \in \ctrj{\finitcst}{\fshiftcst{\finitcst}{\overline{n}}}{\RunProc{a}{P}}{w(T)}$.
\end{proof}

Because silent transitions can only occur at a time satisfying their respective predicate as imposed by the process term, 
existence of a sequence $\sigma$ suffices to ensure that no process missed its ``deadline'' during the computation. 
In particular, \cref{main:thm:adequacy} entails deadlock-freedom as well as termination for the process 
term $P$ in question. 

\subsubsection{Support for Functional Value Exchange.} We briefly sketch the necessary changes to support functional value exchange. 
For the type system, we will introduce another context $\Gamma$ containing assumptions of form $\IsOf{x}{\tau}$, asserting $x$ 
holds a functional value of type $\tau$. The context $\Gamma$ is structural, and assumptions within may be contracted and 
weakened at will. This modification will be propagated throughout the system. On the fundamental theorem side, \cref{main:defn:subst}
will be modified to include substitutions for functional variables. This change propagates throughout the proofs. 

\section{Whole System Manual Verification: \tslr In Action}
\label{main:sec:sensor}
In this section, we will once more revisit the smart home example introduced in \cref{main:sec:intro} and \cref{main:sec:motivation:system-desc}. We start by reviewing
our progress so far:
In \cref{main:sec:motivation}, we distilled \tillst types representing the protocols of the sensor ($A_\kw{BME680}$) 
and controller ($A_\kw{Hub}$) and provided a process implementation ($P_\kw{Hub}$).
With the rules in \cref{main:sec:tillst}, we can easily check (\eg using our type checker in \cref{main:sec:implementation}) 
that our implementation inhabits the proposed type 
($\IsOfProc{P_\kw{Hub}}{t_0}{A_\kw{Hub}}$). By the fundamental theorem (\cref{main:thm:ftlr}), proved in \cref{main:sec:ftlr}, we
conclude that $P_\kw{Hub}$ adheres to the protocol prescribed by $A_\kw{Hub}$. That is: 
$$
t_0; \cdot \mid \cdot \sementails{\IsOfProc{P_\kw{Hub}}{t_0}{A_\kw{Hub}}}
$$
This is where we are right now.

However, clients of the hub device would appreciate an end-to-end \emph{whole system} guarantee: when the controller is connected to the  
sensors, the entire, heterogeneous system will still behave according to the protocol. Concretely, suppose that we have  
protocol-adhering sensors connected on channels $a$ and $b$, represented by some configuration $\Omega_a$ and $\Omega_b$. 
We would like to ensure that the configuration 
$$
\bfOmega_0 \triangleq \Omega_a \opconc \Omega_b \opconc \RunProc{-}{\tpspawnproccst{T}{P_\kw{Hub}}{x}{
    \tplarrsndcst{x}{T}{\tpfwdcst{T}{a}}{\tplarrsndcst{x}{T}{\tpfwdcst{T}{b}}{\tpfwdcst{T}{x}}}}}
$$
adheres to the type $\tpbangstycst{t_3}{T + \millisec{50} \leq t_3}{\booltycst}{\tpunitstycst{t_4}{t_4 = t_3}}$. Formally, that is: 
$$
\exists \bfw. \gl{\bfw}{\bfOmega_0} \in \LInt{\tpbangstycst{t_3}{T + \millisec{50} \leq t_3}{\booltycst}{\tpunitstycst{t_4}{t_4 = t_3}}}{T}
$$

The goal of this section is to provide such a whole-system guarantee.

Our immediate obstacle is that we do not have a representation for these sensors. The sensors are unlikely to be programmed in the 
process language proposed in \Cref{main:sec:motivation:process-language}. Additionally, we do not have, and will not have, access to the code inside the sensors. On top of this, 
the origins of the timing requirements are often non-computational (\eg warming up an internal component), therefore it is likely 
infeasible to model the sensor using terms in our calculus.

What we are given is an operational manual (\emph{datasheet}) for the sensor. 
These documents often model the operation of the sensor as a \emph{state machine}, either directly with a figure 
or indirectly with textual descriptions.  In the case where timing is relevant, state machines can be enriched with 
timing information, resulting in Timed Automata~\cite{AlurDillARTICLE1994}.
\cref{fig:sensor-tm} shows the state machine that we extracted from the BME680 specification \cite{BME680}.

\tikzset{->, node distance=2cm}

\begin{figure}[ht] 
\centering 
\begin{tikzpicture}[scale=0.8]

\node[state, initial] (s0) {$S_0$};
\node[state, xshift=1.5cm, yshift=0.7cm] (s1) {$S_1$};
\node[state, xshift=1.5cm, yshift=-0.7cm] (s2) {$S_2$};
\node[state, right of=s1] (s3) {$S_3$};
\node[state, right of=s2] (s4) {$S_4$};
\node[state, right of=s4] (s5) {$S_5$};
\node[state, accepting, right of=s3] (s6) {$\checkmark$};

\draw 
(s0) edge[bend left, above] node{$\actrcv{}{\lblleft}\phantom{\qquad}$} (s1)
(s0) edge[bend right, below] node{$\actrcv{}{\lblright}\phantom{qquad}$} (s2)
(s1) edge[bend left, above] node{$\actsnd{}{\kw{val}(v_\kw{temp})}$} (s3)
(s3) edge[bend left, above] node{$\actrcvclosecst{}$} (s6)
(s2) edge[bend left, above] node{$\actsnd{}{\kw{val}(v_\kw{temp})}$, $t$} (s4)
(s4) edge[bend right, below] node{$t \geq 30$, $\actsnd{}{\kw{val}(v_\kw{gas})}$, $t$} (s5)
(s5) edge[bend right, right] node{$t \geq 20$, $\actrcvclosecst{}$} (s6);

\end{tikzpicture}
\caption{BME680 Sensor specified using Timed Automaton}
\label{fig:sensor-tm}
\end{figure}

We provide a quick introduction on timed automata. Timed automaton come with a sequence of named timers, termed \emph{clocks}. 
The initial reading of the clock is always zero. As one steps through the states, the clocks constantly tick up. 
Here, we just have one clock $t$. Transitions in a timed automaton have three parts: a clock condition, 
an action, and a set of clocks to ``reset''. 
For example, for the transition $S_4 \LLStepsTo{}{t \geq 30, \actsndvalcst{}{v_\kw{gas}}, t} S_5$, the condition $t \geq 30$ specifies that the transition 
is only enabled if the clock reads above or equal to $30$; the action of this transition is to send a functional value for air quality; finally clock $t$ is reset (becomes zero) after taking this transition. If the condition part of the transition is missing, 
then the transition is always enabled. If clock resets are omitted, then no clock will be reset.
For this transition, because the previous transition from $S_3$ to $S_4$ resets the clock $t$, 
this effectively means that this transition must wait for $\millisec{30}$. As one can see, this style of modeling is imperative in the 
treatment of time. Clocks in this setting can be viewed as shared variables, incremented and accessed periodically and concurrently with 
the executing process.

We can faithfully represent this automaton in our system by making slight adjustments. First, we enrich our configuration syntax 
with a new process form $\SensorProc{a}{S_i}{T}$, representing an instance of the automaton that is currently at state $S_i$ 
and entered $S_i$ at time $T$. This induces other necessary, but inessential changes to structural congruence. The definition of 
nameless configurations also needs to be extended to include this new process form. To represent the transitions, we introduce the new transition 
rules in \cref{rules:sensor}. Each transition corresponds to exactly one rule.

\begin{figure}[ht]
\raggedright

\ifdefined\InApx
\begin{mathpar}
\defrule[$S_0-L$][rule:sensor:s0l]
{T \leq T_1}
{\SensorProc{a}{S_0}{T} 
 \LLStepsTo{\actrcvlblcst{a}{\lblleft}}{T_1} 
 \SensorProc{a}{S_1}{T_1}}

\defrule[$S_0-R$][rule:sensor:s0r]
{T \leq T_1}
{\SensorProc{a}{S_0}{T} 
 \LLStepsTo{\actrcvlblcst{a}{\lblright}}{T_1} 
 \SensorProc{a}{S_2}{T_1}}

\defrule[$S_1$][rule:sensor:s1]
{T \leq T_1 \\ \IsOf{v}{\tau_\kw{temp}}}
{\SensorProc{a}{S_1}{T} 
 \LLStepsTo{\actrcvlblcst{a}{\kw{val}(v_\kw{temp})}}{T_1} 
 \SensorProc{a}{S_3}{T_1}}

\defrule[$S_3$][rule:sensor:s3]
{T \leq T_1}
{\SensorProc{a}{S_3}{T} 
 \LLStepsTo{\actrcvclosecst{a}}{T_1} 
 \StopConf}

\defrule[$S_2$][rule:sensor:s2]
{T \leq T_1 \\ \IsOf{v}{\tau_\kw{temp}}}
{\SensorProc{a}{S_2}{T} 
 \LLStepsTo{\actsndlblcst{a}{\kw{val}(v_\kw{temp})}}{T_1} 
 \SensorProc{a}{S_4}{T_1}}

\defrule[$S_4$][rule:sensor:s4]
{T + 30 \leq T_1 \\ \IsOf{v}{\tau_\kw{gas}}}
{\SensorProc{a}{S_3}{T} 
 \LLStepsTo{\actsndlblcst{a}{\kw{val}(v_\kw{gas})}}{T_1} 
 \SensorProc{a}{S_5}{T_1}}\\

\defrule[$S_5$][rule:sensor:s5]
{T + 20 \leq T_1}
{\SensorProc{a}{S_5}{T} 
 \LLStepsTo{\actrcvclosecst{a}}{T_1} 
 \StopConf}
\end{mathpar}

\else

\begin{tabular}{l}
\defsensorrule[$S_0-L$][rule:sensor:s0l]
{T \leq T_1} 
{\SensorProc{a}{S_0}{T} 
 \LLStepsTo{\actrcvlblcst{a}{\lblleft}}{T_1} 
 \SensorProc{a}{S_1}{T_1}} \\

\defsensorrule[$S_0-R$][rule:sensor:s0r]
{T \leq T_1}
{\SensorProc{a}{S_0}{T} 
 \LLStepsTo{\actrcvlblcst{a}{\lblright}}{T_1} 
 \SensorProc{a}{S_2}{T_1}} \\

\defsensorrule[$S_1$][rule:sensor:s1]
{T \leq T_1 \quad \IsOf{v}{\tau_\kw{temp}}}
{\SensorProc{a}{S_1}{T} 
 \LLStepsTo{\actrcvlblcst{a}{\kw{val}(v_\kw{temp})}}{T_1} 
 \SensorProc{a}{S_3}{T_1}} \\

\defsensorrule[$S_3$][rule:sensor:s3]
{T \leq T_1} 
{\SensorProc{a}{S_3}{T} 
 \LLStepsTo{\actrcvclosecst{a}}{T_1} 
 \StopConf} \\

\defsensorrule[$S_2$][rule:sensor:s2]
{T \leq T_1 \quad \IsOf{v}{\tau_\kw{temp}}}
{\SensorProc{a}{S_2}{T} 
 \LLStepsTo{\actsndlblcst{a}{\kw{val}(v_\kw{temp})}}{T_1} 
 \SensorProc{a}{S_4}{T_1}} \\

\defsensorrule[$S_4$][rule:sensor:s4]
{T + 30 \leq T_1 \quad \IsOf{v}{\tau_\kw{gas}}}
{\SensorProc{a}{S_3}{T} 
 \LLStepsTo{\actsndlblcst{a}{\kw{val}(v_\kw{gas})}}{T_1} 
 \SensorProc{a}{S_5}{T_1}}\\

\defsensorrule[$S_5$][rule:sensor:s5]
{T + 20 \leq T_1}
{\SensorProc{a}{S_5}{T} 
 \LLStepsTo{\actrcvclosecst{a}}{T_1} 
 \StopConf}
\end{tabular}

\fi
\caption{Rules for representing BME680 sensor}
\label{rules:sensor}
\end{figure}

The remainder of the argument goes as follows: 
\begin{enumerate}
\item Show that the  $\SensorProc{-}{S_0}{T}$ can be semantically assigned the type $A_\kw{BME680}$. 
$$\exists \bfw ~\text{s.t.}~ \gl{\bfw}{\SensorProc{-}{S_0}{T}} \in \LInt{A_\kw{BME680}}{T}$$
\item Observe that by typing rules (recall that temporal context is empty, both $x, y$ has type $A_\kw{BME680}$): 
\begin{small}
$$
\dots \entails{\IsOfProc{\tpspawnproccst{T}{P_\kw{Hub}}{z}{
    \tplarrsndcst{z}{T}{\tpfwdcst{T}{x}}{\tplarrsndcst{z}{T}{\tpfwdcst{T}{y}}{\tpfwdcst{T}{z}}}}}{T}{\tpbangstycst{t_3}{T + \millisec{50} \leq t_3}{\booltycst}{\tpunitstycst{t_4}{t_4 = t_3}}}}
$$
\end{small}
\item Appeal to the result of \cref{main:thm:ftlr} with channel substitution $[x \mapsto a, y \mapsto b]$ and 
subforest $[x \mapsto \SensorProc{a}{S_0}{T}, y \mapsto \SensorProc{b}{S_0}{T}]$ and conclude what we want to show.
\end{enumerate}

The challenges of the proof concentrate in the first step. Towards it, we provide the following proof sketch.
The idea is that we build up our proof by consecutively analyzing each state, by proving the following sub-goals in order: 
\begin{enumerate}
    \item $\exists \bfw$ s.t. $\gl{\bfw}{\SensorProc{-}{S_5}{T}} \in \LInt{\tpunitstycst{t_4}{T + 20 \leq t_4}}{T}$
    \item $\exists \bfw$ s.t. $\gl{\bfw}{\SensorProc{-}{S_4}{T}} \in \LInt{\tpbangstycst{t_3}{T + 30 \leq t_3}{\tau_\kw{Gas}}{\tpunitstycst{t_4}{t_3 + 20 \leq t_4}}}{T}$
    \item $\exists \bfw$ s.t. $\gl{\bfw}{\SensorProc{-}{S_2}{T}} \in \LInt{\tpbangstycst{t_2}{T \leq t_2}{\tau_\kw{Temp}}{ \tpbangstycst{t_3}{t_2 + 30 \leq t_3 }{\tau_\kw{Gas}}{\tpunitstycst{t_4}{t_3 + 20\kw{ms} \leq t_4}}}}{T}$
    \item $\exists \bfw$ s.t. $\gl{\bfw}{\SensorProc{-}{S_3}{T}} \in \LInt{\tpunitstycst{t_3}{T \leq _3}}{T}$
    \item $\exists \bfw$ s.t. $\gl{\bfw}{\SensorProc{-}{S_1}{T}} \in \LInt{\tpbangstycst{t_2}{T \leq t_2}{\tau_\kw{Temp}}{\tpunitstycst{t_3}{t_2 \leq t_3}}}{T}$
\end{enumerate}

These goals analyze the automaton in reverse topological order, each building up from before. 
In each case, the proof mostly constitutes unfolding definitions and making observations. Two exemplary proof cases 
can be found in \refapx[apx:pf-sensor].


\section{Rust Implementation}
\label{main:sec:implementation}
We implemented our refinement type system for \tillst as a DSL for Rust,
whose syntax can be found in \refapx[apx:dsl].
Our implementation includes a type checker, but no code generation.
We chose Rust for its strong support for systems applications in both language design and tooling,
witnessed by several prior session type encodings \cite{ChenECOOP2022,JespersenWGP2015}.
While those encodings remain within the Rust type system, we opted for a DSL to support temporal predicates.
\cref{fig:dsl-type} shows the DSL type of sensor hub from \cref{main:sec:intro:mode-one}.
Macro \code{!rtsm\{...\}} delimits DSL blocks that the Rust compiler passes to our parser and type checker.
Types generally take the form of \code{TyOp < t where p, ...>}, where \code{TyOp} is the name 
of the type operator, \code{t where p} represents predicate $t.p$, and the remaining arguments are 
continuation session types. Type operators \code{ExChoice}, \code{Lolli}, \code{Produce} and \code{Unit}, 
stands for $\oplus$, $\multimap$, $!$, and $\mathbf{1}$ \respb.
We apply the typing rules from \cref{main:sec:tillst} in a syntax-directed traversal of the process
and use Rust for the functional layer.

\begin{figure}[h]
    \centering
    \footnotesize
    \begin{dslblock}
!rtsm { type BME680 = ExChoice <t1 where Geq<t1, t0>, TEMP, TEMP_AIR> 
        type HUB    = Lolli <t1 where Leq<t0, t1>, BME680,
                      Lolli <t2 where Eq<t2, t1>, BME680,
                      Produce <sort_bool, t3 where Leq<Shift<t1, 50>, t3>,
                      Unit <t4 where Eq<t4, t3>>>>>                        ...}
    \end{dslblock}
    \caption{The sensor hub type in the \tilst Rust DSL}
    \label{fig:dsl-type}
\end{figure}



%
%


The challenge in implementing \tilst is the temporal judgment $\calG; \calF \entails p$.
In \refapx[apx:dsl] we establish an encoding of our temporal predicates into FOL.
Our implementation uses this to generate queries to an SMT solver.
We encode the temporal model via sorts for times and durations with assertions of the axioms.
On the provider side, we check the judgment $\calG, t; \calF, p(t) \entails{\leqcst{T}{t}}$ 
to ensure the provider is not too late.
On the client side, we must check two judgments. 
First, we examine if the client communicates at the right time: $\calG; \calF \entails{\leqcst{T}{T'}}$. 
Then, we ensure the communication can go forward in time: $\calG; \calF \entails{p(T')}$.
We thus encode $\calG; \calF \entails p$ as the question:
\emph{is there an assignment of temporal variables in which $\calF$ and not $p$?}
An \kw{unsat} result is then interpreted as validating the judgment, otherwise a type error is generated.
For each judgment, the type checker writes out a query in the solver-agnostic SMT-LIB2 format
and invokes the solver with a timeout.
This yields a sound decision, but one that may be incomplete.

In practice \textsf{cvc5}\cite{BarbosaTACAS22}, our choice of SMT solver,
is capable of answering the queries needed to type check the range of examples
we have implemented.
Examples include the keyless entry protocol on modern automobiles~\cite{keyless} and 
a radar collision detector for airplane traffic control~\cite{CDx2009}.
These examples, including the running smart home example, can be found in  
\refapx[apx:rust:example].


\section{Related Work}
\label{main:sec:related-work}

\paragraph{Logical Relations for Session Types}

Prior work on logical relations for session types is relatively young,
starting out with unary logical relations for proving termination
\citep{PerezESOP2012, PerezARTICLE2014,DeYoungFSCD2020}
and then tackling binary logical relations for proving
parametricity \citep{CairesESOP2013} and
noninterference \citep{DerakhshanLICS2021,BalzerARXIV2023,DerakhshanECOOP2024,VanDenHeuvelECOOP2024}.
Except for the work by \citet{VanDenHeuvelECOOP2024},
which targets cyclic process networks based on classical linear logic session types,
all the remaining logical relations are developed for intuitionistic linear logic session types,
like ours.
Our logical relation is most closely related to the
unary logical relations for termination;
\tslr asserts not only termination, and thus deadlock freedom, but also \emph{timeliness}.
In contrast to any existing logical relations for session types,
\tslr does not require its inhabitants to be syntactically well-typed.
As result, our work facilitates \emph{semantic typing} and
enables both once-and-for-all verification, given a type system, and
per-instance verification of foreign code;
both modes are indispensable in our target domain.

Being entirely semantic, our logical relation builds on the foundations laid by
\citet{LoefARTICLE1982,ConstableBook1986,TimanyJACM2024},
brought to scale in the context of the Iris framework \citep{JungJFP2018}.
Iris has fueled a multitude of verification efforts,
contributing Iris-based program logics targeting per-instance verification of functional program correctness.
In this context, we highlight RustBelt by \citet{JungPOPL2018},
which combines both modes of use of logical relations
to prove the Rust core language and selected libraries
memory safe and race free.
%
%

\paragraph{Intuitionistic Metric Temporal Logic (IMTL)}

Our refinement type system for \tillst is related to work by \citet{SaPPDP2023}
on Intuitionistic Metric Temporal Logic (IMTL),
an intuitionistic account of Metric Temporal Logic (MTL) \citep{KoymansARTICLE1990,OuaknineWorrellLICS2005}.
Metric Temporal Logic (MTL) \citep{KoymansARTICLE1990,OuaknineWorrellLICS2005}
extends linear temporal logic (LTL) \citep{PnueliFOCS1977} with temporal intervals.
Rather than interpreting propositions over models, as is done in prior work on MTL,
\citeauthor{SaPPDP2023} view temporal logic through the lens of the \emph{propositions-as-types paradigm},
focusing on how propositions are proved.
A similar endeavor has been undertaken prior by \citet{KojimaIgarashiARTICLE2011},
albeit for linear temporal logic (LTL) \citep{PnueliFOCS1977} and a reduced set of temporal modalities.
The technical contributions by \citeauthor{SaPPDP2023} comprise
a syntactic proof of cut elimination, entailing not only consistency of the logic,
but also temporal causality (``future events cannot affect the present'') and
temporal monotonicity (``a proof can never move backwards in time'').
%
Similarly to \tillst,
the authors assume an \emph{instant-based} model of time,
witnessed by the fact that cut reductions happen at the judgmental present time.
However, IMTL was conceived as a logic,
with cut reductions as the primary notion of computation.
Our refinement type system for \tillst, in contrast,
relies on an actual execution dynamics,
which really infuses meaning to an instant-based model of time.
Our dynamics is also fundamental to our semantic typing approach,
allowing us to show inhabitance of terms with or without a typing derivation.
%

\paragraph{Temporal Session Types}

Our refinement type system for \tillst is also more distantly related to
extension of Honda-style session type systems,
both binary \citep{HondaCONCUR1993,HondaESOP1998} and multiparty \citep{HondaPOPL2008},
to support timing constraints.
Adding the notion of a delay, they connect session types to communicating timed automata
\citep{AlurDillARTICLE1994,BengtssonWORKSHOP1995,KrcalCAV2006,LampkaEMSOFT2009}.
%

In the multiparty session types setting, 
\citet{BocchiCONCUR2014} and \citet{BartolettiARTICLE2017}
consider temporal guards on communications.
These systems assume access only to local clocks and a fixed view of durations as rational numbers.
Our type system is defined with respect to a single global clock abstracted over a model of time.
Temporal predicates may reference the time of any prior event in the protocol.
This matches our domain, 
where IoT and wireless device hardware maintains synchronized clocks with known bounds on drift, \eg \citep{bluetooth}.
Work by \citet{NeykovaARTICLE2017} uses a corresponding extension of the Scribble
protocol description language \citep{HondaICDCIT2011} to build runtime monitoring of protocol adherence.
Multiparty reactive sessions (MRS) by \citet{CanoTR2019} take a global view of time as discretized instants.
This synchronous reactive programming model of time has found applications to embedded systems problems
via languages such as Esterel \citep{BerryGonthierARTICLE1992,BerryBook1999} and Lustre \citep{HalbwachsARTICLE1991}.
MRS connects logical instants to the external world via reaction to events.
This means constraints cannot be directly specified using physical notions of time.
Work by \citet{LeBrunDardhaESOP2023} uses the concept of timeouts to model message delivery failure,
but similarly leaves correspondence between timeout and physical time intentionally unspecified.

\citet{BocchiESOP2019} extend temporal guards to asynchronous binary session types.
This too gives access only to local views of time.
The choice of asynchrony also leads to a subtyping relation 
that is covariant on output and contravariant on input times.
In a synchronous system, the time both participants communicate must coincide.
Safely substituting types in our synchronous system requires
retyping relations, which distinguish the client and provider roles.
For systems applications such as in \cref{main:sec:motivation},
the distinction of which side may be more permissive is critical to correctness.

Rate-based session types (RBST) \citep{IraciOOPSLA2023} introduces a periodic construct 
to binary session types that specifies parts of the protocol repeat at a fixed interval.
Timing in this system is attached only to control flow, with no association to specific communications.
This is insufficient to express narrower constraints on specific events,
such as the exchange in \cref{main:sec:motivation}.
\tilst is also able to express dynamic change to the connectivity in protocols,
i.e. spawning of new processes, and supports higher-order channels.
It does this while still maintaining desirable properties of \ilst, such as deadlock freedom, which RBST lacks.

\paragraph{Verification and Modeling of Embedded, Control, or Hybrid Systems} 

Generally, when it comes to verification of computational systems, approaches can be 
broadly divided into external and internal/integrated methods. 
External methods employed by tools such as UPPAAL~\cite{BengtssonWORKSHOP1995} separate the implementation of the system 
from the modeling, specification, and proof. 
The modeling may be approached by timed-automata, and the logic for specification 
can be a domain specific logic such as (Differential) Dynamic Logic~\cite{dDL, dyn-logic}. 
Our type theoretic method offers a different, internal and integrated approach, where 
the implementation and the verification conditions are designed, expressed, and 
proved in unison in a singular language. Concretely, through logical relations, 
type checking serves the role of program verification.
Our methods enjoy compositionality and support higher-order features 
(\eg sending channel names).

While our approach is internal, the technical development supplements and embraces external methods; 
it is synergistic with external verification methods thanks to the use of a logical relation. 
As the example in \cref{main:sec:sensor} demonstrates, the modeling of the sensor and the proof 
of the verification conditions as dictated by the logical relation can be approached with 
any number of external verification methods. Therefore, our approach takes an 
inclusive and constructive stance when it comes to existing verification works.


\section{Conclusions}
\label{main:sec:conclusions-outlook}
This paper contributes a compositional framework to enable the \emph{verification} of
\emph{timed message-passing} systems,
such as IoT applications and real-time systems.
The framework consist of a \textit{(a)} language to specify timed protocols,
\tillst, rooted in intuitionistic linear logic session types,
a \textit{(b)} timed labelled transition system
to characterize how programs run,
and a \textit{(c)} logical relation, \tslr, to prove programs compliant with their specifications.
To cater to the heterogeneity of its application domain,
the paper adopts a \emph{semantic typing} approach,
freeing programs to be proved correct from any well-typedness constraint.
As a result, the \tslr can be used in two modes:
once-and-for-all verification, given a type system, and
per-instance verification of foreign code.
The paper illustrates both modes based on the example of an IoT application,
using a prototype implementation for the type-based verification.


There exist various avenues to be explored as part of future work.
Most immediate is support of recursive behavior in terms of coinductive types,
which have been shown to integrate smoothly with a Curry-Howard interpretation of linear session types
\citep{LindleyMorrisICFP2016,DerakhshanPfenningLMCS2022},
yet must be given a semantic typing interpretation.
Another interesting future research direction is to integrate the framework with a cost model
to bound the execution time of internal computation.
Existing work \citep{DasLICS2018,DasICFP2018} in the context of intuitionistic linear session types
may serve as a valuable starting point,
but again, will have to be endowed with a semantic typing interpretation.

\begin{acks}

This material is based upon work supported by the
\grantsponsor{NSF}{National Science Foundation}{https://www.nsf.gov/}
under Grant No. (\grantnum{NSF}{2211996} and \grantnum{NSF}{2211997})
and upon work supported by the
\grantsponsor{AFOSR}{Air Force Office of Scientific Research}{https://www.afrl.af.mil/AFOSR/}
under award number \grantnum{AFOSR}{FA9550-21-1-0385} (Tristan Nguyen, program manager).
Any opinions, findings, and conclusions or recommendations expressed in this material are those of the author(s) and do not necessarily reflect the views of
the National Science Foundation or
the U.S. Department of Defense.
\end{acks}

\clearpage

\bibliographystyle{ACM-Reference-Format}
\bibliography{ref}

\ifappendix

\clearpage

\def\InApx{}

\appendix





\section{Language Typing and Dynamics}
\subsection{Syntax of Temporal Logic}
\label{apx:lang:syntax}
\ifdefined\InApx

\begin{synchartmini}{tsyn}
\Sort{Time} &  T  & \bnfdef & t, \cdots & \text{Frame variables} \\
            &     & \bnfalt & \fshiftcst{T}{i} & \text{Shifting by integer duration} \\
            &     & \bnfalt & \finitcst        & \text{Time of initialization} \\
            \\
\Sort{Prop}  & p & \bnfdef & \topcst   & \text{Truth}  \\
             &   & \bnfalt & \botcst   & \text{Falsehood}           \\
             &   & \bnfalt & \landcst{p_1}{p_2} & \text{Conjunction}   \\
             &   & \bnfalt & \lorcst{p_1}{p_2}  & \text{Disjunction}   \\
             &   & \bnfalt & \limpcst{p_1}{p_2} & \text{Implication}   \\
             &   & \bnfalt & \eqcst{T_1}{T_2}   & \text{Equality}   \\
             &   & \bnfalt & \leqcst{T_1}{T_2}  & \text{Less than or equal}   \\
\\
  & \calG & \bnfdef & t_1, \cdots, t_n & \text{Temporal variables context} \\
  & \calF & \bnfdef & p_1, \cdots, p_n & \text{Proposition context} \\
\end{synchartmini}

Derived connectives:
\begin{align*}
	\lnotcst{p} & \triangleq \limpcst{p}{\bot} \\
	T_1 \neq T_2 & \triangleq \lnotcst{(\eqcst{T_1}{T_2})} \\
	T_1 < T_2 & \triangleq \landcst{(\leqcst{T_1}{T_2})}{(T_1 \neq T_2)} \\
	T_1 \geq T_2 & \triangleq \leqcst{T_2}{T_1} \\
	T_1 > T_2 & \triangleq T_2 < T_1
\end{align*}

\else


$$
\Sort{Time} \quad T  \bnfdef t \mid \fshiftcst{T}{D} \mid \finitcst, \quad
\Sort{Duration} \quad D  \bnfdef \dzerocst \mid \dunitcst \mid \dsumcst{D_1}{D_2} \mid \ddiffcst{T_1}{T_2}, \quad
\Sort{Prop} \quad p  \bnfdef \dots \mid \leqcst{T_1}{T_2} 
$$

\fi
\subsection{Syntax of Process Language}
\label{apx:tlogic:syntax}
\ifdefined\InApx

\begin{synchartmini}{ilsill}
\Sort{Chan} &  x  & \bnfdef & x, y, z, \cdots & \text{channel variables (Static)} \\
            &     & \bnfalt & a, b, c, \cdots & \text{channel symbols (Runtime)} \\
            \\
\TypeSort{} &  A, B, C  & \bnfdef & \tpunitstycst{t}{p}                  & \text{closing} \\
            &     & \bnfalt & \tptensorstycst{t}{p}{A_1}{A_2}   & \text{tensor} \\
            &     & \bnfalt & \tplarrstycst{t}{p}{A_1}{A_2}     & \text{lolli} \\
            &     & \bnfalt & \tpsumstycst{t}{p}{A_1}{A_2}      & \text{internal choice} \\
            &     & \bnfalt & \tpwithstycst{t}{p}{A_1}{A_2}     & \text{external choice} \\
	    \\
\ProcSort{} &  P  & \bnfdef & \tpfwdcst{T}{x}                  & \text{forward} \\
            &     & \bnfalt & \tpspawnproccst{T}{P}{x}{Q}      & \text{spawn} \\
            &     & \bnfalt & \tpclosecst{t}{p}                & \text{provider of $\unitstycst$} \\
            &     & \bnfalt & \tpwaitcst{T}{x}{Q}              & \text{client of $\unitstycst$} \\
            &     & \bnfalt & \tplarrrcvcst{t}{p}{A_t}{x}{P}   & \text{provider of $\multimap$} \\
            &     & \bnfalt & \tplarrsndcst{x}{T}{P}{Q}        & \text{client of $\multimap$} \\
            &     & \bnfalt & \tptensorsndcst{t}{p}{P_1}{P_2}  & \text{provider of $\otimes$} \\
            &     & \bnfalt & \tptensorrcvcst{x}{T}{y}{Q}      & \text{client of $\otimes$} \\
            &     & \bnfalt & \tpinlcst{t}{p}{A_1}{A_2}{P}     & \text{provider of $\oplus$} \\
            &     & \bnfalt & \tpinrcst{t}{p}{A_1}{A_2}{P}     & \text{provider of $\oplus$} \\
            &     & \bnfalt & \tpcasecst{T}{x}{Q_1}{Q_2}       & \text{client of $\oplus$} \\
            &     & \bnfalt & \tpoffercst{t}{p}{P_1}{P_2}      & \text{provider of $\&$} \\
            &     & \bnfalt & \tpselectlcst{x}{T}{Q}           & \text{client of $\&$} \\
            &     & \bnfalt & \tpselectrcst{x}{T}{Q}           & \text{client of $\&$} \\
\\

\ConfSort{} & \Omega & \bnfdef & \StopConf & \text{Stopped (nullary) configuration} \\
            &        & \bnfalt & \RunProc{a}{P} & \text{Process running $P$ on channel $a$} \\
            &        & \bnfalt & \FwdProc{a}{b} & \text{Forwarding from $a$ (as provider) to $b$ (the client)} \\
            &        & \bnfalt & \conc{\Omega_1}{\Omega_2} & \text{Concurrent composition} \\
\end{synchartmini}

\else 

\fi


\clearpage

\subsection{Typing - Process Typing}
\label{apx:lang:statics}
\subsubsection{Retyping Relations}
\label{apx:lang:statics:retyping}
\ifdefined\InApx
$\boxed{\calG; \calF \entails{\FwdComp{A}{B}{T}}}$ 
$\boxed{\calG; \calF \entails{\CutComp{A}{B}{T}}}$ 

\begin{mathpar}
\inferrule
  {\calG, t; \calF, q \entails{p} \\\\
   \calG, t; \calF, q \entails{\leqcst{T}{t}} }
  {\calG, t; \calF \entails{\FwdComp{\tpunitstycst{t}{p}}{\tpunitstycst{t}{q}}{T}}}

\inferrule
  {\calG, t; \calF, q \entails{p} \\
   \calG, t; \calF, q \entails{\leqcst{T}{t}} \\\\
   \calG, t; \calF, q \entails{\FwdComp{A_1}{B_1}{t}} \\
   \calG, t; \calF, q \entails{\FwdComp{A_2}{B_2}{t}}
  }
  {\calG; \calF \entails{\FwdComp{\tpwithstycst{t}{p}{A_1}{A_2}}{\tpwithstycst{t}{q}{B_1}{B_2}}{T}}}

\inferrule
  {\calG, t; \calF, q \entails{p} \\
   \calG, t; \calF, q \entails{\leqcst{T}{t}} \\\\
   \calG, t; \calF, q \entails{\FwdComp{A_1}{B_1}{t}} \\\\
   \calG, t; \calF, q \entails{\FwdComp{A_2}{B_2}{t}}
  }
  {\calG; \calF \entails{\FwdComp{\tpsumstycst{t}{p}{A_1}{A_2}}{\tpsumstycst{t}{q}{B_1}{B_2}}{T}}}

\inferrule
  {\calG, t; \calF, q \entails{p} \\
   \calG, t; \calF, q \entails{\leqcst{T}{t}} \\\\
   \calG, t; \calF, q \entails{\FwdComp{A_1}{B_1}{t}} \\\\
   \calG, t; \calF, q \entails{\FwdComp{A_2}{B_2}{t}}
  }
  {\calG; \calF \entails{\FwdComp{\tptensorstycst{t}{p}{A_1}{A_2}}{\tptensorstycst{t}{q}{B_1}{B_2}}{T}}}

\inferrule
  {\calG, t; \calF, q \entails{p} \\
   \calG, t; \calF, q \entails{\leqcst{T}{t}} \\\\
   \calG, t; \calF, q \entails{\FwdComp{B_1}{A_1}{t}} \\\\
   \calG, t; \calF, q \entails{\FwdComp{A_2}{B_2}{t}}
  }
  {\calG; \calF \entails{\FwdComp{\tplarrstycst{t}{p}{A_1}{A_2}}{\tplarrstycst{t}{q}{B_1}{B_2}}{T}}}

\inferrule
  {\calG, t; \calF, \leqcst{T}{t}, q \entails{p}}
  {\calG; \calF \entails{\CutComp{\tpunitstycst{t}{p}}{\tpunitstycst{t}{q}}{T}}}

\inferrule
  {\calG, t; \calF, \leqcst{T}{t}, q \entails{p} \\\\
   \calG, t; \calF, \leqcst{T}{t}, q \entails{\CutComp{A_1}{B_1}{t}} \\\\
   \calG, t; \calF, \leqcst{T}{t}, q \entails{\CutComp{A_2}{B_2}{t}}}
  {\calG; \calF \entails{\CutComp{\tpwithstycst{t}{p}{A_1}{A_2}}{\tpwithstycst{t}{q}{B_1}{B_2}}{T}}}

\inferrule
  {\calG, t; \calF, \leqcst{T}{t}, q \entails{p} \\\\
   \calG, t; \calF, \leqcst{T}{t}, q \entails{\CutComp{A_1}{B_1}{t}} \\\\
   \calG, t; \calF, \leqcst{T}{t}, q \entails{\CutComp{A_2}{B_2}{t}}}
  {\calG; \calF \entails{\CutComp{\tpsumstycst{t}{p}{A_1}{A_2}}{\tpsumstycst{t}{q}{B_1}{B_2}}{T}}}

\inferrule
  {\calG, t; \calF, \leqcst{T}{t}, q \entails{p} \\\\
   \calG, t; \calF, \leqcst{T}{t}, q \entails{\CutComp{A_1}{B_1}{t}} \\\\
   \calG, t; \calF, \leqcst{T}{t}, q \entails{\CutComp{A_2}{B_2}{t}}}
  {\calG; \calF \entails{\CutComp{\tptensorstycst{t}{p}{A_1}{A_2}}{\tptensorstycst{t}{q}{B_1}{B_2}}{T}}}

\inferrule
  {\calG, t; \calF, \leqcst{T}{t}, q \entails{p} \\\\
   \calG, t; \calF, \leqcst{T}{t}, q \entails{\CutComp{B_1}{A_1}{t}} \\\\
   \calG, t; \calF, \leqcst{T}{t}, q \entails{\CutComp{A_2}{B_2}{t}}}
  {\calG; \calF \entails{\CutComp{\tplarrstycst{t}{p}{A_1}{A_2}}{\tplarrstycst{t}{q}{B_1}{B_2}}{T}}}
\end{mathpar}

\else 
\begin{mathpar}
\inferrule
  {\calG, t; \calF, q \entails{p} \\\\
   \calG, t; \calF, q \entails{\leqcst{T}{t}} }
  {\calG, t; \calF \entails{\FwdComp{\tpunitstycst{t}{p}}{\tpunitstycst{t}{q}}{T}}}

\inferrule
  {\calG, t; \calF, q \entails{p} \\
   \calG, t; \calF, q \entails{\leqcst{T}{t}} \\\\
   \calG, t; \calF, q \entails{\FwdComp{A_i}{A_i'}{t}}
  }
  {\calG; \calF \entails{\FwdComp{\tpwithstycst{t}{p}{A_1}{A_2}}{\tpwithstycst{t}{q}{B_1}{B_2}}{T}}}

\inferrule
  {\calG, t; \calF, q \entails{p} \\
   \calG, t; \calF, q \entails{\leqcst{T}{t}} \\\\
   \calG, t; \calF, q \entails{\FwdComp{A_i}{A_i'}{t}}
  }
  {\calG; \calF \entails{\FwdComp{\tpsumstycst{t}{p}{A_1}{A_2}}{\tpsumstycst{t}{q}{B_1}{B_2}}{T}}}

\inferrule
  {\calG, t; \calF, q \entails{p} \\
   \calG, t; \calF, q \entails{\leqcst{T}{t}} \\\\
   \calG, t; \calF, q \entails{\FwdComp{A_i}{A_i'}{t}}
  }
  {\calG; \calF \entails{\FwdComp{\tptensorstycst{t}{p}{A_1}{A_2}}{\tptensorstycst{t}{q}{B_1}{B_2}}{T}}}

\inferrule
  {\calG, t; \calF, q \entails{p} \\
   \calG, t; \calF, q \entails{\leqcst{T}{t}} \\\\
   \calG, t; \calF, q \entails{\FwdComp{A_i}{A_i'}{t}}
  }
  {\calG; \calF \entails{\FwdComp{\tplarrstycst{t}{p}{A_1}{A_2}}{\tplarrstycst{t}{q}{B_1}{B_2}}{T}}}

\inferrule
  {\calG, t; \calF, \leqcst{T}{t}, q \entails{p}}
  {\calG; \calF \entails{\CutComp{\tpunitstycst{t}{p}}{\tpunitstycst{t}{q}}{T}}}

\inferrule
  {\calG, t; \calF, \leqcst{T}{t}, q \entails{p} \\\\
   \calG, t; \calF, \leqcst{T}{t}, q \entails{\CutComp{A_i}{A_i'}{t}}}
  {\calG; \calF \entails{\CutComp{\tpwithstycst{t}{p}{A_1}{A_2}}{\tpwithstycst{t}{q}{B_1}{B_2}}{T}}}

\inferrule
  {\calG, t; \calF, \leqcst{T}{t}, q \entails{p} \\\\
   \calG, t; \calF, \leqcst{T}{t}, q \entails{\CutComp{A_i}{A_i'}{t}}}
  {\calG; \calF \entails{\CutComp{\tpsumstycst{t}{p}{A_1}{A_2}}{\tpsumstycst{t}{q}{B_1}{B_2}}{T}}}

\inferrule
  {\calG, t; \calF, \leqcst{T}{t}, q \entails{p} \\\\
   \calG, t; \calF, \leqcst{T}{t}, q \entails{\CutComp{A_i}{A_i'}{t}}
   }
  {\calG; \calF \entails{\CutComp{\tptensorstycst{t}{p}{A_1}{A_2}}{\tptensorstycst{t}{q}{B_1}{B_2}}{T}}}

\inferrule
  {\calG, t; \calF, \leqcst{T}{t}, q \entails{p} \\\\
   \calG, t; \calF, \leqcst{T}{t}, q \entails{\CutComp{A_i}{A_i'}{t}}
  }
  {\calG; \calF \entails{\CutComp{\tplarrstycst{t}{p}{A_1}{A_2}}{\tplarrstycst{t}{q}{B_1}{B_2}}{T}}}
\end{mathpar}

\fi

\clearpage
\subsubsection{Process Term Typing}
\label{apx:lang:statics:proctyping}
\ifdefined\InApx
$\boxed{\calF \mid \Delta \entails{\IsOfProc{P}{T}{A}}}$
\else
\def \MathparLineskip {\lineskip=0.2cm}
\fi

\begin{footnotesize}
\begin{mathpar}
\defrule[Fwd][rule:fwd]
  {\calG; \calF \entails{\FwdComp{A}{A'}{T}}}
  {\calG; \calF \mid \IsOf{x}{A} \entails{\IsOfProc{\tpfwdcst{T}{x}}{T}{A'}}}

\defrule[Cut][rule:cut]
  {\calG; \calF \mid \Delta_1 \entails{\IsOfProc{P}{T}{A}} \\\\
   \calG; \calF \mid \Delta_2, \IsOf{x}{A'} \entails{\IsOfProc{Q}{T}{C}} \\\\
   \calG; \calF \entails{\CutComp{A}{A'}{T}}
  }
  {\calG; \calF \mid \Delta_1, \Delta_2 
   \entails{\IsOfProc{\tpspawnproccst{T}{P}{x}{Q}}{T}{C}}} \\

\defrule[$\unitstycst$ R][rule:unit-r]
  {\calG, t; \calF, p(t) \entails{\leqcst{T}{t}}}
  {\calG; \calF \mid \emptyset \entails{\IsOfProc{\tpclosecst{t}{p}}{T}{\tpunitstycst{t}{p}}}}

\defrule[$\unitstycst$ L][rule:unit-l]
  {\calG; \calF \mid \Delta \entails{\IsOfProc{Q}{T'}{C}} \\
   \calG; \calF \entails{\leqcst{T}{T'}} \\
   \calG; \calF \entails{p(T')}
   }
  {\calG; \calF \mid \Delta, \IsOf{x}{\tpunitstycst{t}{p}} \entails{\IsOfProc{\tpwaitcst{T'}{x}{Q}}{T}{C}}}

\defrule[$\otimes$ R][rule:tensor-r]
  {\calG, t; \calF, p(t) \mid \Delta_1 \entails{\IsOfProc{P_1}{t}{A_1}} \\\\
   \calG, t; \calF, p(t) \mid \Delta_2 \entails{\IsOfProc{P_2}{t}{A_2}} \\\\
   \calG, t; \calF, p(t) \entails{\leqcst{T}{t}}
   }
  {\calG; \calF \mid \Delta_1, \Delta_2 \entails{\IsOfProc{\tptensorsndcst{t}{p}{P_1}{P_2}}{T}{\tptensorstycst{t}{p}{A_1}{A_2}}}}

\defrule[$\otimes$ L][rule:tensor-l]
  {\calG; \calF \mid \Delta, \IsOf{x}{\Subst{T'}{t}{A_1}}, \IsOf{y}{\Subst{T'}{t}{A_2}} \entails{\IsOfProc{Q}{T'}{C}} \\\\
   \calG; \calF \entails{\leqcst{T}{T'}} \\
   \calG; \calF \entails{p(T')} }
  {\calG; \calF \mid \Delta, \IsOf{x}{\tptensorstycst{t}{p}{A_1}{A_2}} \entails{\IsOfProc{\tptensorrcvcst{x}{T'}{y}{Q}}{T}{C}}}\\

\defrule[$\multimap$ R][rule:lolli-r]
  {\calG, t; \calF, p(t) \mid \Delta, \IsOf{x}{A_1} \entails{\IsOfProc{P_2}{t}{A_2}} \\\\
   \calG, t; \calF, p(t) \entails{\leqcst{T}{t}}
  }
  {\calG; \calF \mid \Delta \entails{\IsOfProc{\tplarrrcvcst{t}{p}{A_1}{x}{P_2}}{T}{\tplarrstycst{t}{p}{A_1}{A_2}}}}

\defrule[$\multimap$ L][rule:lolli-l]
  {\calG; \calF \mid \Delta_1 \entails{\IsOfProc{P}{T'}{\Subst{T'}{t}{A_1}}} \\\\
   \calG; \calF \mid \Delta_2, \IsOf{x}{\Subst{T'}{t}{A_2}} \entails{\IsOfProc{Q}{T'}{C}} \\\\
   \calG; \calF \entails{\leqcst{T}{T'}} \\
   \calG; \calF \entails{p(T')}
   }
  {\calG; \calF \mid \Delta_1, \Delta_2, \IsOf{x}{\tplarrstycst{t}{p}{A_1}{A_2}} 
   \entails{\IsOfProc{\tplarrsndcst{x}{T'}{P}{Q}}{T}{C}}}\\

\defrule[$\oplus$ R1][rule:plus-r1]
  {\calG, t; \calF, p(t) \mid \Delta \entails{\IsOfProc{P}{t}{A_1}} \\\\
   \calG, t; \calF, p(t) \entails{\leqcst{T}{t}}
  }
  {\calG; \calF \mid \Delta \entails{\IsOfProc{\tpinlcst{t}{p}{A_1}{x}{P}}{T}{\tpsumstycst{t}{p}{A_1}{A_2}}}}

\defrule[$\oplus$ R2][rule:plus-r2]
  {\calG, t; \calF, p(t) \mid \Delta \entails{\IsOfProc{P}{t}{A_2}} \\\\
   \calG, t; \calF, p(t) \entails{\leqcst{T}{t}}
  }
  {\calG; \calF \mid \Delta \entails{\IsOfProc{\tpinrcst{t}{p}{A_1}{x}{P}}{T}{\tpsumstycst{t}{p}{A_1}{A_2}}}}

\defrule[$\oplus$ L][rule:plus-l]
  {\calG; \calF \mid \Delta, \IsOf{x}{\Subst{T'}{t}{A_1}} \entails{\IsOfProc{Q_1}{T'}{C}} \\\\
   \calG; \calF \mid \Delta, \IsOf{x}{\Subst{T'}{t}{A_2}} \entails{\IsOfProc{Q_2}{T'}{C}} \\\\
   \calG; \calF \entails{\leqcst{T}{T'}} \\
   \calG; \calF \entails{p(T')}
   }
  {\calG; \calF \mid \Delta, \IsOf{x}{\tpsumstycst{t}{p}{A_1}{A_2}} 
   \entails{\IsOfProc{\tpcasecst{T'}{x}{Q_1}{Q_2}}{T}{C}}}

\defrule[$\&$ R][rule:with-r]
  {\calG, t; \calF, p(t) \mid \Delta \entails{\IsOfProc{P_1}{t}{A_1}} \\\\
   \calG, t; \calF, p(t) \mid \Delta \entails{\IsOfProc{P_2}{t}{A_2}} \\\\
   \calG, t; \calF, p(t) \entails{\leqcst{T}{t}}
   }
  {\calG; \calF \mid \Delta \entails{\IsOfProc{\tpoffercst{t}{p}{P_1}{P_2}}{T}{\tpwithstycst{t}{p}{A_1}{A_2}}}}\\

\defrule[$\&$ L1][rule:with-l1]
  {\calG; \calF \mid \Delta, \IsOf{x}{\Subst{T'}{t}{A_1}} \entails{\IsOfProc{Q}{T'}{C}} \\\\
   \calG; \calF \entails{\leqcst{T}{T'}} \\
   \calG; \calF \entails{p(T')}
   }
  {\calG; \calF \mid \Delta, \IsOf{x}{\tpwithstycst{t}{p}{A_1}{A_2}} \entails{\IsOfProc{\tpselectlcst{x}{T'}{Q}}{T}{C}}}

\defrule[$\&$ L2][rule:with-l2]
  {\calG; \calF \mid \Delta, \IsOf{x}{\Subst{T'}{t}{A_2}} \entails{\IsOfProc{Q}{T'}{C}} \\\\
   \calG; \calF \entails{\leqcst{T}{T'}} \\
   \calG; \calF \entails{p(T')}
   }
  {\calG; \calF \mid \Delta, \IsOf{x}{\tpwithstycst{t}{p}{A_1}{A_2}} \entails{\IsOfProc{\tpselectrcst{x}{T'}{Q}}{T}{C}}}
\end{mathpar}
\end{footnotesize}

\subsubsection{Complementary Actions}
\label{apx:lang:statics:complement-action}

\begin{mathpar}
\overline{\actrcvlblcst{a}{\lblleft}} \triangleq \actsndlblcst{a}{\lblleft} 

\overline{\actrcvlblcst{a}{\lblright}} \triangleq \actsndlblcst{a}{\lblright} 

\overline{\actsndlblcst{a}{\lblleft}} \triangleq \actrcvlblcst{a}{\lblleft} 

\overline{\actsndlblcst{a}{\lblright}} \triangleq \actrcvlblcst{a}{\lblright} 

\\
\overline{\actsil} \triangleq \actsil 

\overline{\actrcvchancst{a}{b}} \triangleq \actsndchancst{a}{b} 

\overline{\actsndchancst{a}{b}} \triangleq \actrcvchancst{a}{b} 

\overline{\actrcvclosecst{a}} \triangleq \actsndclosecst{a}

\overline{\actsndclosecst{a}} \triangleq \actrcvclosecst{a}

\end{mathpar}

\subsubsection{Structural Congruence}
\label{apx:lang:statics:cong}
\ifdefined\InApx
\begin{mathpar}
\defruler[C-Stop][rule:cong:Stop]
{\strut}
{\conc{\StopConf}{\Omega} \equiv \Omega}

\defruler[C-Comm][rule:cong:comm]
{\strut}
{\conc{\Omega_1}{\Omega_2} \equiv \conc{\Omega_2}{\Omega_1}}

\defruler[C-Fwd][rule:cong:fwd]
{\strut}
{\conc{\RunProc{a}{P}}{\FwdProc{b}{a}} \equiv \RunProc{b}{P}}

\defruler[C-Ctr][rule:cong:contr]
{\strut}
{\conc{\FwdProc{c}{b}}{\FwdProc{b}{a}} \equiv \FwdProc{c}{a}}

\defruler[C-Refl][rule:cong:refl]
{\strut}
{\Omega \equiv \Omega}

\defruler[C-Trans][rule:cong:trans]
{\Omega \equiv \Omega' \\ 
 \Omega' \equiv \Omega''}
{\Omega \equiv \Omega''}

\defruler[C-Cong][rule:cong:cong]
{\Omega_1 \equiv \Omega_1'}
{\conc{\Omega_1}{\Omega_2} \equiv \conc{\Omega_1'}{\Omega_2}}
\end{mathpar}

\else
\begin{mathpar}
\defruler[C-Stop][rule:cong:Stop]
{\strut}
{\conc{\StopConf}{\Omega} \equiv \Omega}

\defruler[C-Fwd][rule:cong:fwd]
{\strut}
{\conc{\RunProc{a}{P}}{\FwdProc{b}{a}} \equiv \RunProc{b}{P}}

\defruler[C-Comm][rule:cong:comm]
{\strut}
{\conc{\Omega_1}{\Omega_2} \equiv \conc{\Omega_2}{\Omega_1}}

\defruler[C-Ctr][rule:cong:contr]
{\strut}
{\conc{\FwdProc{c}{b}}{\FwdProc{b}{a}} \equiv \FwdProc{c}{a}}
\end{mathpar}
\fi

\subsection{Dynamics}
\label{apx:lang:dynamics}
\ifdefined\InApx
$\boxed{\IsOf{\sigma}{\ClockProc{T_1}, \Omega_1 \MultiStepsTo{} \ClockProc{T_2}, \Omega_2}}$
\else
\fi

\begin{mathpar}
\defruler[G-\kw{refl}][rule:dyn:refl]
{\strut}
{\IsOf{\StepRefl{T}{\Omega}}{\ClockProc{T}, \Omega \MultiStepsTo{} \ClockProc{T}, \Omega}}

\defruler[G-\kw{stepT}][rule:dyn:stept]
{\leqcst{T_1}{T_2} \\
  \IsOf{\sigma}{\ClockProc{T_2}, \Omega \MultiStepsTo{} \ClockProc{T_3}, \Omega'}}
{\IsOf{\StepT{T_1}{T_2}{\Omega}{\sigma}}{\ClockProc{T_1}, \Omega \MultiStepsTo{} \ClockProc{T_3}, \Omega'}}

\defruler[G-\kw{stepC}][rule:dyn:stepc]
{ \Omega \LLStepsTo{}{T_1} \Omega' \\
  \IsOf{\sigma}{\ClockProc{T_1}, \Omega' \MultiStepsTo{} \ClockProc{T_2}, \Omega''}}
{\IsOf{\StepC{T_1}{\Omega}{\Omega'}{\sigma}}{\ClockProc{T_1}, \Omega \MultiStepsTo{} \ClockProc{T_2}, \Omega''}}
\end{mathpar}
\begin{small}

\begin{mathpar}
\defruler[D-Frame][rule:dyn:frame]
{\Omega_1 \LLStepsTo{\alpha}{T} \Omega_1'}
{\conc{\Omega_1}{\Omega_2} \LLStepsTo{\alpha}{T} \conc{\Omega_1'}{\Omega_2}}

\defruler[D-Comm][rule:dyn:comm]
{\Omega_1 \LLStepsTo{\alpha}{T} \Omega_1' \\ 
 \Omega_2 \LLStepsTo{\overline{\alpha}}{T} \Omega_2' \\
}
{\conc{\Omega_1}{\Omega_2} \LLStepsTo{}{T} \conc{\Omega_1'}{\Omega_2'}}
\end{mathpar}
\end{small}



\clearpage
\section{Semantics}
\label{apx:semantics}

\subsection{Operators on Sequences}
\label{apx:semantics:seq}
\begin{definition}[Concatenation of Sequences]
We define the \emph{concatenation} of $\sigma_1$ and $\sigma_2$:
$$ 
\concat{\cdot}{\cdot} : 
(\ClockProc{T_0}, \Omega_0 \MultiStepsTo{} \ClockProc{T_1}, \Omega_1) \to 
(\ClockProc{T_1}, \Omega_1 \MultiStepsTo{} \ClockProc{T_2}, \Omega_2) \to 
(\ClockProc{T_0}, \Omega_0 \MultiStepsTo{} \ClockProc{T_2}, \Omega_2)
$$

The sequence $\concat{\sigma_1}{\sigma_2}$ is defined inductively over structure of $\sigma_1$ as follows:
\begin{align*}
  \concat{\StepRefl{T}{\Omega}}{\sigma_2}                  & \triangleq \sigma_2 \\
  \concat{\StepC{T}{\Omega}{\Omega'}{\sigma_1}}{\sigma_2}  & \triangleq \StepC{T}{\Omega}{\Omega'}{\concat{\sigma_1}{\sigma_2}} \\
  \concat{\StepT{T}{T'}{\Omega}{\sigma_1}}{\sigma_2}       & \triangleq \StepT{T}{T'}{\Omega}{\concat{\sigma_1}{\sigma_2}}
\end{align*}
\end{definition}

The definition $\concat{\sigma_1}{\sigma_2}$ can then be extended to 
$\IsOf{\sigma_1}{\ClockProc{T_0}, \Omega_0 \MultiStepsTo{} \ClockProc{T_1}, \Omega_1}$ \newline
and $\IsOf{\sigma_2}{\ClockProc{T_1'}, \Omega_1 \MultiStepsTo{} \ClockProc{T_2}, \Omega_2}$
with $T_1 \leq T_1'$, by setting 
$$\concat{\sigma_1}{\sigma_2} \triangleq \concat{(\concat{\sigma_1}{\StepT{T_1}{T_1'}{\Omega_1}{\StepRefl{T_1'}{\Omega_1}}})}{\sigma_2}$$

\begin{definition}[Interleaving of Sequences]
We define the \emph{interleaving} of $\sigma_1$ and $\sigma_2$:
\begin{small}
$$ 
\il{\cdot}{\cdot} : 
(\ClockProc{T}, \Omega_1 \MultiStepsTo{} \ClockProc{T_1}, \Omega_1') \to 
(\ClockProc{T}, \Omega_2 \MultiStepsTo{} \ClockProc{T_2}, \Omega_2') \to 
(\ClockProc{T}, \conc{\Omega_1}{\Omega_2} \MultiStepsTo{} \ClockProc{\max(T_1, T_2)}, \conc{\Omega_1'}{\Omega_2'})
$$
\end{small}

The sequence $\il{\sigma_1}{\sigma_2}$ is defined inductively over the $\sigma_1$ and $\sigma_2$, 
with the induction metric being the total length of both sequences as follows (symmetric cases are omitted):
\begin{align*}
  \il{\StepRefl{T}{\Omega_1}}{\StepRefl{T}{\Omega_2}} 
      & \triangleq \StepRefl{T}{\conc{\Omega_1}{\Omega_2}} \\
  \il{\StepRefl{T}{\Omega_1}}{\StepC{T}{\Omega_2}{\Omega_2'}{\sigma_2}} 
      & \triangleq \StepC{T}{(\Omega_1 \opconc \Omega_2)}{(\Omega_1 \opconc \Omega_2')}{\il{\StepRefl{T}{\Omega_1}}{\sigma_2}} \\
  \il{\StepRefl{T}{\Omega_1}}{\StepT{T}{T'}{\Omega_2}{\sigma_2}} 
      & \triangleq \StepT{T}{T'}{\conc{\Omega_1}{\Omega_2}}{\il{\StepRefl{T'}{\Omega_1}}{\sigma_2}} \\
  \il{\StepC{T}{\Omega_1}{\Omega_1'}{\sigma_1}}{\StepC{T}{\Omega_2}{\Omega_2'}{\sigma_2}} 
      & \triangleq 
        \StepC{T}{\conc{\Omega_1}{\Omega_2}}{\conc{\Omega_1'}{\Omega_2}}{
              \StepC{T}{\conc{\Omega_1'}{\Omega_2}}{\conc{\Omega_1'}{\Omega_2'}}{
                \il{\sigma_1}{\sigma_2}}}\\
  \il{\StepC{T}{\Omega_1}{\Omega_1'}{\sigma_1}}{\StepT{T}{T'}{\Omega_2}{\sigma_2}} 
      & \triangleq 
          \StepC{T}{\conc{\Omega_1}{\Omega_2}}{\conc{\Omega_1'}{\Omega_2}}{
            \StepT{T}{T'}{\conc{\Omega_1'}{\Omega_2}}{
              \il{\sigma_1}{\sigma_2}
            }} \\
  \il{\StepT{T}{T_1}{\Omega_1}{\sigma_1}}{\StepT{T}{T_2}{\Omega_2}{\sigma_2}} 
      & \triangleq 
      \begin{cases*}
              \StepT{T}{T_1}{\Omega_1, \Omega_2}{\il{\sigma_1}{\StepT{T_1}{T_2}{\Omega_2}{\sigma_2}}} 
              & if $T_1 \leq T_2$ \\ 
              \StepT{T}{T_2}{\Omega_1, \Omega_2}{\il{\StepT{T_2}{T_1}{\Omega_1}{\sigma_1}}{\sigma_2}} 
              & if otherwise
      \end{cases*}
\end{align*}
\end{definition}

This definition can be extended to 
$\IsOf{\sigma_1}{\ClockProc{T_1}, \Omega_1 \MultiStepsTo{} \ClockProc{T_1'}, \Omega_1'}$
and $ \IsOf{\sigma_2}{\ClockProc{T_2}, \Omega_2 \MultiStepsTo{} \ClockProc{T_2'}, \Omega_2'}$:
\begin{align*}
  \il{\sigma_1}{\sigma_2} &: 
    \ClockProc{\min(T_1, T_2)}, \conc{\Omega_1}{\Omega_2} \MultiStepsTo{} \ClockProc{\max(T_1', T_2')}, \conc{\Omega_1'}{\Omega_2'} \\ 
  \il{\sigma_1}{\sigma_2} &\triangleq 
    \il{(\concat{\StepRefl{\min(T_1, T_2)}{\Omega_1}}{\concat{\sigma_1}{\StepRefl{\max(T_1', T_2')}{\Omega_2}}})}
       {(\concat{\StepRefl{\min(T_1, T_2)}{\Omega_2}}{\concat{\sigma_2}{\StepRefl{\max(T_1', T_2')}{\Omega_2'}}})}
\end{align*}

\subsection{Operators on trajectories}
\label{apx:semantics:trj}


\begin{notation}
Given a fixed $I$, we will use $I = \TInt{T_1}{T_2}$ to extract the end points of an interval, 
with the understanding that only $T_1$ is 
bound to a point in time if the $I$ is unbounded. In this case we may informally say $T_2 = \infty$.
\end{notation}

\begin{definition}
Given an empty interval $I$ (that is, $I = \emptyset$). Let $\trjTriv{I}$ denote the trivial function over it. 
In particular, let $\trjTriv{T}$ denote the trivial function on the (empty) interval $\TInt{T}{T}$.
\end{definition}

\begin{definition}
Given an interval $I$ and some configuration $\Omega$. Let $\trjConst{I}{\Omega}$ denote the constant function on $I$: 
$$\trjConst{I}{\Omega} : I \to \mathsf{Conf} \triangleq t \mapsto \Omega$$
\end{definition}

\begin{definition}
The \emph{concatenation} operation stitches two trajectories with touching domains: 
\begin{align*}
  \trjCat{\cdot}{\cdot} &: \Trj{\TInt{T_1}{T_2}} \times \Trj{\TInt{T_2}{T_3}} \to \Trj{\TInt{T_1}{T_3}} \\
  \trjCat{r}{s} &\triangleq t \mapsto
        \begin{cases*}
          r(t) & $t \in \TInt{T_1}{T_2}$ \\
          s(t) & $t \in \TInt{T_2}{T_3}$ 
        \end{cases*}
\end{align*}
\end{definition}

\begin{definition}
The following operator extends the domain of $r : \Trj{\TInt{T_1}{T_2}}$ to $\TInt{T_0}{T_2}$ by a constant.
$$\trjExt{T_0}{\Omega}{r} \triangleq \trjCat{\trjConst{\TInt{T_0}{T_1}}{\Omega}}{r}$$
\end{definition}

\begin{definition}
The \emph{interleaving} of two trajectories point-wise parallel composes configurations:
\begin{align*}
  \trjConc{\cdot}{\cdot} &: \Trj{\TInt{T_1}{T_2}} \times \Trj{\TInt{T_1}{T_2}} \to \Trj{\TInt{T_1}{T_2}} \\
  \trjConc{r}{s} &\triangleq t \mapsto \conc{r(t)}{s(t)}
\end{align*}
\end{definition}

\subsection{Operators on Computable Trajectories}
\label{apx:semantics:ctrj}
\begin{lemma}
If $w \in \ctrj{T_1}{T_2}{\Omega_1}{\Omega_2}$ then 
$\IsOf{\sigma}{\ClockProc{T_1}, \Omega_1 \MultiStepsTo{} \ClockProc{T_2}, \Omega_2}$.
\end{lemma}
\begin{proof}
Let $\pairexcst{r}{\sigma} = w$. Induction over $\sigma$ while keeping the relation $\calR$ in mind. 
The inductive cases are straightforward. In the base case we have $\sigma = \StepRefl{T_1}{\Omega_1}$ 
and by the relation conclude $\Omega_2 = \Omega_1$. Take $\StepT{T_1}{T_2}{\Omega_1}{\StepRefl{T_2}{\Omega_1}}$ 
to satisfy our proof obligation.
\end{proof}

\begin{definition}
For all configurations $\Omega$, there is a trivial trajectory on interval $\TInt{T_1}{T_2}$, defined as
$\constcwl{\TInt{T_1}{T_2}}{\Omega} \triangleq \pairexcst{\trjConst{\TInt{T_1}{T_2}}{\Omega}}{\StepRefl{T_1}{\Omega}}$.
\end{definition}

\begin{definition}[Concatenation]
The concatenation of computable trajectories is defined by:
$$\concat{\pairexcst{r}{\sigma}}{\pairexcst{s}{\sigma'}} \triangleq \pairexcst{\trjCat{r}{s}}{\concat{\sigma}{\sigma'}}$$
\end{definition}
\begin{proof}
The goal of the proof is show that the construction preserves the relation $\calR$. Go by induction on $\sigma$ and 
simplify the concatenation. In each case, verify that the relation is preserved.
\end{proof}

\begin{corollary}
$\forall w_1, w_2$, 
$\cwlequivt{\Dom{w_1}}{w_1}{(\concat{w_1}{w_2})}$ and
$\cwlequivt{\Dom{w_2}}{w_2}{(\concat{w_1}{w_2})}$. 
\end{corollary}
\begin{proof}
Directly by definition. Use the fact that concatenation of trajectories glues together their domain, then case split
the input for its relation to the point where domains touch.
\end{proof}

\begin{definition}[Partition]
Partitions of a computation trajectory is defined inductively over its sequence:
\begin{small}
\begin{align*}
  \lpar{\pairexcst{\trjConst{\TInt{T_1}{T_2}}{\Omega}}{\StepRefl{T_1}{\Omega}}}{T} 
    &\triangleq \pairexcst{\trjConst{\TInt{T_1}{T}}{\Omega}}{\StepRefl{T_1}{\Omega}} \\
  \rpar{\pairexcst{\trjConst{\TInt{T_1}{T_2}}{\Omega}}{\StepRefl{T}{\Omega}}}{T} 
    &\triangleq \pairexcst{\trjConst{\TInt{T}{T_2}}{\Omega}}{\StepRefl{T}{\Omega}} \\
  \lpar{\pairexcst{r}{\StepC{T_1}{\Omega}{\Omega'}{\sigma}}}{T} 
    &\triangleq \pairexcst{r'}{\StepC{T_1}{\Omega}{\Omega'}{\sigma'}} 
      \text{ where } \pairexcst{r'}{\sigma'} = \lpar{\pairexcst{r}{\sigma}}{T} \\
  \rpar{\pairexcst{r}{\StepC{T_1}{\Omega}{\Omega'}{\sigma}}}{T} &\triangleq \rpar{\pairexcst{r}{\sigma}}{T}\\
  \lpar{\pairexcst{\trjExt{T_1}{\Omega}{r}}{\StepT{T_1}{T_1'}{\Omega}{\sigma}}}{T}  
    &\triangleq 
    \begin{cases*}
              \pairexcst{\trjConst{\TInt{T_1}{T}}{\Omega}}{\StepT{T_1}{T}{\Omega}{\StepRefl{T}{\Omega}}}
              & if $T \in \TInt{T_1}{T_1'}$ \\
              \pairexcst{\trjExt{T_1}{\Omega}{r'}}{\StepT{T}{T_1'}{\Omega}{\sigma'}}
              \text{ where } \pairexcst{r'}{\sigma'} = \lpar{\pairexcst{r}{\sigma}}{T}
              & if $T \in \TInt{T_1'}{T_2}$ 
    \end{cases*} \\
  \rpar{\pairexcst{\trjExt{T_1}{\Omega}{r}}{\StepT{T_1}{T_1'}{\Omega}{\sigma}}}{T}  
    &\triangleq 
    \begin{cases*}
              \pairexcst{\trjExt{T_1}{\Omega}{r'}}{\StepT{T_1}{T_1'}{\Omega}{\sigma'}}
              \text{ where } \pairexcst{r'}{\sigma'} = \rpar{\pairexcst{r}{\sigma}}{T}
              & if $T \in \TInt{T_1}{T_1}$ \\
              \rpar{\pairexcst{r}{\sigma}}{T}   
              & if $T \in \TInt{T_1'}{T_2}$
    \end{cases*}
\end{align*}
\end{small}
\end{definition}

\begin{corollary} 
For all $w$ and $T \in \Dom{w}$, 
$\cwlequivt{\Dom{\lpar{w}{T}}}{\lpar{w}{T}}{w}$ and
$\cwlequivt{\Dom{\rpar{w}{T}}}{\rpar{w}{T}}{w}$.
\end{corollary}
\begin{proof}
Above results are better shown simultaneously. First the goal is again to show that the relations are maintained. 
Go by induction on the structure of $\sigma$. In each case 
use how partitioning is defined, and straightforward from there. 
To show the latter statement, pay attention to the action on the first component during induction.
\end{proof}

\begin{definition}[Interleaving]
The \emph{interleaving} on trajectories is defined as follows:
\begin{align*}
  \il{\cdot}{\cdot} &:
\ctrj{T}{T'}{\Omega_1}{\Omega_1'} \to 
\ctrj{T}{T'}{\Omega_2}{\Omega_2'} \to
\ctrj{T}{T'}{\conc{\Omega_1}{\Omega_1'}}{\conc{\Omega_2}{\Omega_2'}} \\
\il{\pairexcst{r_1}{\sigma_1}}{\pairexcst{r_2}{\sigma_2}} &\triangleq
\pairexcst{\trjConc{r_1}{r_2}}{\il{\sigma_1}{\sigma_2}}
\end{align*}
\end{definition}
\begin{proof}
The goal is to show that relation is preserved.
Go by induction on the total length of sequences in the input. Expand the definition for interleaving sequences. 
Check conditions of the relation.
\end{proof}
 
\begin{corollary}
For all $w_1$ and $w_2$ and $T$, $\il{w_1}{w_2}(T) = \conc{(w_1(T))}{(w_2(T))}$.
\end{corollary}
\begin{proof}
By definition of interleaving trajectory.
\end{proof}

\begin{lemma}
The equivalence relation is congruence w.r.t. the operators defined.
\end{lemma}
\begin{proof}
Go by definition. Fix arbitrary	time $T$ in the domain, then try to compute the configuration at that time. 
Use the results we have proved before to simplify.
\end{proof}

\clearpage
\subsection{Semantic Retyping}
\label{apx:semantics:sem-retyping}
\begin{theorem}[Forward Retyping]\label{apx:thm:fwdcomp}
If $\calG; \calF \entails{\FwdComp{\wp{A}{t}{p}}{\wp{B}{t}{q}}{T}}$ then for all $\calF$ satisfying temporal variable assignment $\varphi$,
if $\gl{\bfw}{\bfOmega} \in \LSInt{\wp{\hatA}{t}{\hatp}}{\hatT}$ then $\gl{\bfw}{\bfOmega} \in \LInt{\wp{\hatB}{t}{\hatq}}{\hatT}$.
\end{theorem}
\begin{proof}
Proof proceeds by induction over the derivation $\calG; \calF \entails{\FwdComp{\wp{A}{t}{p}}{\wp{B}{t}{q}}{T}}$.

In every case, let $\varphi$ be fixed. Our goal is to show that, 
$$
\forall T'. \Subst{T'}{t}{\hatq} \implies (T' \geq \hatT) \land (\bfw(T') \in \VInt{\expose{\hatB}{T'}}{T'})
$$

Let $T'$ be fixed so that $\Subst{T'}{t}{\hatq}$ holds. It remains to be shown that 
$$
T' \geq \hatT \andthat \bfw(T') \in \VInt{\expose{\hatB}{T'}}{T'}
$$

In every one of the cases we have as premise
$$ \calG, t; \calF, q \entails{p} \andthat \calG, t; \calF, q \entails{T \leq t}$$

With an extended assignment $\varphi[t \mapsto T']$, we see $\Subst{T'}{t}{\hatq} \implies \Subst{T'}{t}{\hatp}$. 

Therefore, we simply have $\Subst{T'}{t}{\hatp}$. 

On the other hand, our assumption is
$$
\forall T'. \Subst{T'}{t}{\hatp} \land (\hatT \leq T') \implies \bfw(T') \in \VSInt{\expose{\hatA}{T'}}{T'}
$$

We may now discharge the assumption $\hatT \leq T'$ with the other premise, and conclude 
$$
\bfw(T') \in \VSInt{\expose{\hatA}{T'}}{T'}
$$

We now proceed with the case analysis. The derivation is syntax driven and contains exactly 
one rule for each connective. We will refer to the rules by listing the connectives involved.

\begin{itemize}
  \item \itemheader{Case $A = \tpunitstycst{t}{p}$ and $B = \tpunitstycst{t}{q}$.} 

  We have 
  $$\bfw(T') \in \VSInt{\expose{\tpunitstycst{t}{\hatp}}{T'}} = \VSInt{\unitstycst}{T'}$$

  We need to show $$\bfw(T') \in \VInt{\expose{\tpunitstycst{t}{\hatq}}{T'}} = \VInt{\unitstycst}{T'}.$$

  See that the definitions of $\VSInt{\unitstycst}{T'}$ and $\VInt{\unitstycst}{T'}$ coincide, therefore we have what 
  we wanted to show.

  \item \itemheader{Case $A = \tptensorstycst{t}{p}{A_1}{A_2}$ and $B = \tptensorstycst{t}{q}{B_1}{B_2}$.}

  We have 
  $$\bfw(T') \in \VSInt{\expose{\tptensorstycst{t}{\hatp}{\hatA_1}{\hatA_2}}{T'}} 
               = \VSInt{\tensorstycst{\Subst{T'}{t}{\hatA_1}}{\Subst{T'}{t}{\hatA_2}}}{T'}$$

  From the assumption we have for all $a$ and fresh $c$, exists $\bfOmega_1$ and $\bfOmega_2$:
  $$\bfw(T')[a] \LLStepsTo{\actsndchancst{a}{c}}{T'} \bfOmega_1[c] \opconc \bfOmega_2[a]$$

  where 
  $$
  \exists \bfw_1. \gl{\bfw_1}{\bfOmega_1} \in \LSInt{\Subst{T'}{t}{\hatA_1}}{T'} \andthat
  \exists \bfw_2. \gl{\bfw_2}{\bfOmega_2} \in \LSInt{\Subst{T'}{t}{\hatA_2}}{T'}
  $$

  We need to show that
  $$\bfw(T') \in \VInt{\expose{\tptensorstycst{t}{\hatq}{\hatB_1}{\hatB_2}}{T'}} 
              = \VInt{\tensorstycst{\Subst{T'}{t}{\hatB_1}}{\Subst{T'}{t}{\hatB_2}}}{T'}.$$

  Fix arbitrary $a$ and $c$. The proof goal concerning reduction sequence is discharged by what we have.

  It remains to be shown that 
  $$
  \gl{\bfw_1}{\bfOmega_1} \in \LInt{\Subst{T'}{t}{\hatB_1}}{T'} \andthat
  \gl{\bfw_2}{\bfOmega_2} \in \LInt{\Subst{T'}{t}{\hatB_2}}{T'}
  $$

  Appeal to our IH with the substitution $\varphi[t \mapsto T']$ to finish the proof.

  \item \itemheader{Case $\wp{A}{t}{p} = \tplarrstycst{t}{p}{A_1}{A_2}$.}

  We have 
  $$\bfw(T') \in \VSInt{\expose{\tplarrstycst{t}{\hatp}{\hatA_1}{\hatA_2}}{T'}} 
              = \VSInt{\larrstycst{\Subst{T'}{t}{\hatA_1}}{\Subst{T'}{t}{\hatA_2}}}{T'}$$

  From the assumption we have for all $a$ and fresh $c$, exists $\bfOmega_2$ such that:
  $$\bfw(T')[a] \LLStepsTo{\actrcvchancst{a}{c}}{T'} \bfOmega_2[a]$$

  where 
  $$
  \forall \bfw_1 \forall \bfOmega_1. \gl{\bfw_1}{\bfOmega_1} \in \LInt{\Subst{T'}{t}{\hatA_1}}{T'} \implies
  \exists \bfw_2. \gl{\bfw_2}{\bfOmega[c] \opconc \bfOmega_2[-]} \in \LSInt{\Subst{T'}{t}{\hatA_2}}{T'}
  $$

  We need to show that
  $$\bfw(T') \in \VInt{\expose{\tplarrstycst{t}{\hatq}{\hatB_1}{\hatB_2}}{T'}} 
              = \VInt{\larrstycst{\Subst{T'}{t}{\hatB_1}}{\Subst{T'}{t}{\hatB_2}}}{T'}.$$

  Fix arbitrary $a$ and $c$. The proof goal regarding stepping is already established. It remains to be shown that
  $$
  \forall \bfw_1 \forall \bfOmega_1. \gl{\bfw_1}{\bfOmega_1} \in \LSInt{\Subst{T'}{t}{\hatB_1}}{T'} \implies
  \exists \bfw_2. \gl{\bfw_2}{\bfOmega[c] \opconc \bfOmega_2[-]} \in \LInt{\Subst{T'}{t}{\hatB_2}}{T'}
  $$

  Let $\bfw_1$ and $\bfOmega_1$ be fixed s.t. 
  $$\gl{\bfw_1}{\bfOmega_1} \in \LSInt{\Subst{T'}{t}{\hatB_1}}{T'}.$$

  With the substitution $\varphi[t \mapsto T']$, appeal to IH of the premise $\FwdComp{B_1}{A_1}{t}$ to show
  $$\gl{\bfw_1}{\bfOmega_1} \in \LInt{\Subst{T'}{t}{\hatA_1}}{T'}$$

  Use this to discharge the assumption and obtain
  $$\exists \bfw_2. \gl{\bfw_2}{\bfOmega[c] \opconc \bfOmega_2[-]} \in \LSInt{\Subst{T'}{t}{\hatA_2}}{T'}$$

  Appeal to IH of the premise $\FwdComp{A_2}{B_2}{t}$ finish the proof.

  \item \itemheader{Case $A = \tpwithstycst{t}{p}{A_1}{A_2}$ and $A = \tpwithstycst{t}{p}{B_1}{B_2}$}

  We have 
  $$\bfw(T') \in \VSInt{\expose{\tpwithstycst{t}{\hatp}{\hatA_1}{\hatA_2}}{T'}} 
               = \VSInt{\bwithstycst{\Subst{T'}{t}{\hatA_1}}{\Subst{T'}{t}{\hatA_2}}}{T'}$$

  From the assumption we have for all $a$, $\exists \bfOmega_1, \bfOmega_2$
  $$
  \bfw(T')[a] \LLStepsTo{\actrcvlblcst{a}{\lblleft}}{T'} \bfOmega_1[a] \andthat
  \bfw(T')[a] \LLStepsTo{\actrcvlblcst{a}{\lblright}}{T'} \bfOmega_2[a] 
  $$
  where 
  $$
  \exists \bfw_1 \gl{\bfw_1}{\bfOmega_1} \in \LSInt{\Subst{T'}{t}{\hatA_1}}{T'} \andthat
  \exists \bfw_2 \gl{\bfw_2}{\bfOmega_2} \in \LSInt{\Subst{T'}{t}{\hatA_2}}{T'}
  $$

  We need to show that
  $$\bfw(T') \in \VInt{\expose{\tpwithstycst{t}{\hatq}{\hatB_1}{\hatB_2}}{T'}} 
              = \VInt{\bwithstycst{\Subst{T'}{t}{\hatB_1}}{\Subst{T'}{t}{\hatB_2}}}{T'}.$$

  Fix arbitrary $a$. The proof goal concerning stepping has been established previously. 
  
  It remains to be shown that 
  $$
  \gl{\bfw_1}{\bfOmega_1} \in \LInt{\Subst{T'}{t}{\hatB_1}}{T'} \andthat
  \gl{\bfw_2}{\bfOmega_2} \in \LInt{\Subst{T'}{t}{\hatB_2}}{T'}
  $$

  Appeal to our IH with the substitution $\varphi[t \mapsto T']$ to finish the proof.

  \item \itemheader{Case $\wp{A}{t}{p} = \tpsumstycst{t}{p}{A_1}{A_2}$.}

  We have 
  $$\bfw(T') \in \VSInt{\expose{\tpsumstycst{t}{\hatp}{\hatA_1}{\hatA_2}}{T'}} 
               = \VSInt{\bsumstycst{\Subst{T'}{t}{\hatA_1}}{\Subst{T'}{t}{\hatA_2}}}{T'}$$

  From the assumption we have for all $a$, either 
  \begin{align}
    \exists \bfOmega_1. \bfw(T')[a] & \LLStepsTo{\actsndlblcst{a}{\lblleft}}{T'} \bfOmega_1[a] 
    \andthat \exists \bfw_1 \gl{\bfw_1}{\bfOmega_1} \in \LSInt{\Subst{T'}{t}{\hatA_1}}{T'} \label{apx:sumcls:a1}
    \\
    \exists \bfOmega_2. \bfw(T')[a] & \LLStepsTo{\actsndlblcst{a}{\lblright}}{T'} \bfOmega_2[a] 
    \andthat \exists \bfw_2 \gl{\bfw_2}{\bfOmega_2} \in \LSInt{\Subst{T'}{t}{\hatA_2}}{T'} \label{apx:sumcls:a2}
  \end{align}

  We need to show that
  $$\bfw(T') \in \VInt{\expose{\tpsumstycst{t}{\hatq}{\hatB_1}{\hatB_2}}{T'}} 
              = \VInt{\bsumstycst{\Subst{T'}{t}{\hatB_1}}{\Subst{T'}{t}{\hatB_2}}}{T'}.$$

  For the case ($i$), fix arbitrary $a$. The proof goal concerning stepping has been established previously by reduction shown 
  in ($i$). It remains to be shown that 
  $$
  \gl{\bfw_i}{\bfOmega_i} \in \LInt{\Subst{T'}{t}{\hatB_i}}{T'} 
  $$

  Appeal to the respective IH with the substitution $\varphi[t \mapsto T']$ to finish the proof.
\end{itemize}
We have considered all cases.
\end{proof}

\clearpage

\begin{theorem}[Cut Retyping]\label{apx:thm:cutcomp}
If $\calG; \calF \entails{\CutComp{\wp{A}{t}{p}}{\wp{B}{t}{q}}{T}}$ then for all $\calF$ satisfying temporal variable assignment $\varphi$,
if $\gl{\bfw}{\bfOmega} \in \LInt{\wp{\hatA}{t}{\hatp}}{\hatT}$ then $\gl{\bfw}{\bfOmega} \in \LSInt{\wp{\hatB}{t}{\hatq}}{\hatT}$.
\end{theorem}
\begin{proof}
Proof proceeds by induction over the derivation $\calG; \calF \entails{\CutComp{\wp{A}{t}{p}}{\wp{B}{t}{q}}{T}}$.

In every case, let $\varphi$ be fixed. Our goal is to show that, 
$$
\forall T'. \Subst{T'}{t}{\hatq} \land (\hatT \leq T') \implies \bfw(T') \in \VSInt{\expose{\hatB}{T'}}{T'}
$$

Let $T'$ be fixed so that $\Subst{T'}{t}{\hatq}$ holds. Assume further $\hatT \leq T'$. It remains to be shown that 
$$
\bfw(T') \in \VSInt{\expose{\hatB}{T'}}{T'}
$$

On the other hand, our assumption is
$$
\forall T'. \Subst{T'}{t}{\hatp} \implies (\hatT \leq T') \land \bfw(T') \in \VInt{\expose{\hatA}{T'}}{T'}
$$

In every one of the cases we have as premise
$$ \calG, t; \calF, T \leq t, q \entails{p} $$

With an extended assignment $\varphi[t \mapsto T']$, we see that
$$(\hatT \leq T') \land \Subst{T'}{t}{\hatq} \implies \Subst{T'}{t}{\hatp}.$$

Both assumptions can be discharged with what we have. Therefore, we conclude $\Subst{T'}{t}{\hatp}$. 

This allows to discharge the premise in our assumption and conclude
$$
\bfw(T') \in \VInt{\expose{\hatA}{T'}}{T'}
$$

We now proceed with the case analysis. The derivation is syntax driven and contains exactly 
one rule for each connective. We will refer to the rules by listing the connectives involved. 

The remainder of the proof greatly resembles that of \cref*{apx:thm:fwdcomp}. The main difference
being there is one more assumption to discharge in applying IHs. 

\begin{itemize}
  \item \itemheader{Case $A = \tpunitstycst{t}{p}$ and $B = \tpunitstycst{t}{q}$.} 

  We have 
  $$\bfw(T') \in \VInt{\expose{\tpunitstycst{t}{\hatp}}{T'}} = \VInt{\unitstycst}{T'}$$

  We need to show $$\bfw(T') \in \VSInt{\expose{\tpunitstycst{t}{\hatq}}{T'}} = \VSInt{\unitstycst}{T'}.$$

  See that the definitions of $\VInt{\unitstycst}{T'}$ and $\VSInt{\unitstycst}{T'}$ coincide, therefore we have what 
  we wanted to show.

  \item \itemheader{Case $A = \tptensorstycst{t}{p}{A_1}{A_2}$ and $B = \tptensorstycst{t}{q}{B_1}{B_2}$.}

  We have 
  $$\bfw(T') \in \VInt{\expose{\tptensorstycst{t}{\hatp}{\hatA_1}{\hatA_2}}{T'}} 
               = \VInt{\tensorstycst{\Subst{T'}{t}{\hatA_1}}{\Subst{T'}{t}{\hatA_2}}}{T'}$$

  From the assumption we have for all $a$ and fresh $c$, exists $\bfOmega_1$ and $\bfOmega_2$:
  $$\bfw(T')[a] \LLStepsTo{\actsndchancst{a}{c}}{T'} \bfOmega_1[c] \opconc \bfOmega_2[a]$$

  where 
  $$
  \exists \bfw_1. \gl{\bfw_1}{\bfOmega_1} \in \LInt{\Subst{T'}{t}{\hatA_1}}{T'} \andthat
  \exists \bfw_2. \gl{\bfw_2}{\bfOmega_2} \in \LInt{\Subst{T'}{t}{\hatA_2}}{T'}
  $$

  We need to show that
  $$\bfw(T') \in \VSInt{\expose{\tptensorstycst{t}{\hatq}{\hatB_1}{\hatB_2}}{T'}} 
               = \VSInt{\tensorstycst{\Subst{T'}{t}{\hatB_1}}{\Subst{T'}{t}{\hatB_2}}}{T'}.$$

  Fix arbitrary $a$ and $c$. The proof goal concerning reduction sequence is discharged by what we have.

  It remains to be shown that 
  $$
  \gl{\bfw_1}{\bfOmega_1} \in \LSInt{\Subst{T'}{t}{\hatB_1}}{T'} \andthat
  \gl{\bfw_2}{\bfOmega_2} \in \LSInt{\Subst{T'}{t}{\hatB_2}}{T'}
  $$

  Appeal to our IH with the substitution $\varphi[t \mapsto T']$ to finish the proof.

  \item \itemheader{Case $\wp{A}{t}{p} = \tplarrstycst{t}{p}{A_1}{A_2}$.}

  We have 
  $$\bfw(T') \in \VInt{\expose{\tplarrstycst{t}{\hatp}{\hatA_1}{\hatA_2}}{T'}} 
               = \VInt{\larrstycst{\Subst{T'}{t}{\hatA_1}}{\Subst{T'}{t}{\hatA_2}}}{T'}$$

  From the assumption we have for all $a$ and fresh $c$, exists $\bfOmega_2$ such that:
  $$\bfw(T')[a] \LLStepsTo{\actrcvchancst{a}{c}}{T'} \bfOmega_2[a]$$

  where 
  $$
  \forall \bfw_1 \forall \bfOmega_1. \gl{\bfw_1}{\bfOmega_1} \in \LSInt{\Subst{T'}{t}{\hatA_1}}{T'} \implies
  \exists \bfw_2. \gl{\bfw_2}{\bfOmega[c] \opconc \bfOmega_2[-]} \in \LInt{\Subst{T'}{t}{\hatA_2}}{T'}
  $$

  We need to show that
  $$\bfw(T') \in \VSInt{\expose{\tplarrstycst{t}{\hatq}{\hatB_1}{\hatB_2}}{T'}} 
               = \VSInt{\larrstycst{\Subst{T'}{t}{\hatB_1}}{\Subst{T'}{t}{\hatB_2}}}{T'}.$$

  Fix arbitrary $a$ and $c$. The proof goal regarding stepping is already established. It remains to be shown that
  $$
  \forall \bfw_1 \forall \bfOmega_1. \gl{\bfw_1}{\bfOmega_1} \in \LInt{\Subst{T'}{t}{\hatB_1}}{T'} \implies
  \exists \bfw_2. \gl{\bfw_2}{\bfOmega[c] \opconc \bfOmega_2[-]} \in \LSInt{\Subst{T'}{t}{\hatB_2}}{T'}
  $$

  Let $\bfw_1$ and $\bfOmega_1$ be fixed s.t. 
  $$\gl{\bfw_1}{\bfOmega_1} \in \LInt{\Subst{T'}{t}{\hatB_1}}{T'}.$$

  With the substitution $\varphi[t \mapsto T']$, appeal to IH of the premise $\CutComp{B_1}{A_1}{t}$ to show
  $$\gl{\bfw_1}{\bfOmega_1} \in \LSInt{\Subst{T'}{t}{\hatA_1}}{T'}$$

  Use this to discharge the assumption and obtain
  $$\exists \bfw_2. \gl{\bfw_2}{\bfOmega[c] \opconc \bfOmega_2[-]} \in \LInt{\Subst{T'}{t}{\hatA_2}}{T'}$$

  Appeal to IH of the premise $\CutComp{A_2}{B_2}{t}$ finish the proof.

  \item \itemheader{Case $A = \tpwithstycst{t}{p}{A_1}{A_2}$ and $A = \tpwithstycst{t}{p}{B_1}{B_2}$}

  We have 
  $$\bfw(T') \in \VInt{\expose{\tpwithstycst{t}{\hatp}{\hatA_1}{\hatA_2}}{T'}} 
               = \VInt{\bwithstycst{\Subst{T'}{t}{\hatA_1}}{\Subst{T'}{t}{\hatA_2}}}{T'}$$

  From the assumption we have for all $a$, $\exists \bfOmega_1, \bfOmega_2$
  $$
  \bfw(T')[a] \LLStepsTo{\actrcvlblcst{a}{\lblleft}}{T'} \bfOmega_1[a] \andthat
  \bfw(T')[a] \LLStepsTo{\actrcvlblcst{a}{\lblright}}{T'} \bfOmega_2[a] 
  $$
  where 
  $$
  \exists \bfw_1 \gl{\bfw_1}{\bfOmega_1} \in \LInt{\Subst{T'}{t}{\hatA_1}}{T'} \andthat
  \exists \bfw_2 \gl{\bfw_2}{\bfOmega_2} \in \LInt{\Subst{T'}{t}{\hatA_2}}{T'}
  $$

  We need to show that
  $$\bfw(T') \in \VSInt{\expose{\tpwithstycst{t}{\hatq}{\hatB_1}{\hatB_2}}{T'}} 
              = \VSInt{\bwithstycst{\Subst{T'}{t}{\hatB_1}}{\Subst{T'}{t}{\hatB_2}}}{T'}.$$

  Fix arbitrary $a$. The proof goal concerning stepping has been established previously. 
  
  It remains to be shown that 
  $$
  \gl{\bfw_1}{\bfOmega_1} \in \LSInt{\Subst{T'}{t}{\hatB_1}}{T'} \andthat
  \gl{\bfw_2}{\bfOmega_2} \in \LSInt{\Subst{T'}{t}{\hatB_2}}{T'}
  $$

  Appeal to our IH with the substitution $\varphi[t \mapsto T']$ to finish the proof.

  \item \itemheader{Case $\wp{A}{t}{p} = \tpsumstycst{t}{p}{A_1}{A_2}$.}

  We have 
  $$\bfw(T') \in \VInt{\expose{\tpsumstycst{t}{\hatp}{\hatA_1}{\hatA_2}}{T'}} 
               = \VInt{\bsumstycst{\Subst{T'}{t}{\hatA_1}}{\Subst{T'}{t}{\hatA_2}}}{T'}$$

  From the assumption we have for all $a$, either 
  \begin{align}
    \exists \bfOmega_1. \bfw(T')[a] & \LLStepsTo{\actsndlblcst{a}{\lblleft}}{T'} \bfOmega_1[a] 
    \andthat \exists \bfw_1 \gl{\bfw_1}{\bfOmega_1} \in \LInt{\Subst{T'}{t}{\hatA_1}}{T'} \label{apx:fwdcls:a1}
    \\
    \exists \bfOmega_2. \bfw(T')[a] & \LLStepsTo{\actsndlblcst{a}{\lblright}}{T'} \bfOmega_2[a] 
    \andthat \exists \bfw_2 \gl{\bfw_2}{\bfOmega_2} \in \LInt{\Subst{T'}{t}{\hatA_2}}{T'} \label{apx:fwdcls:a2}
  \end{align}

  We need to show that
  $$\bfw(T') \in \VSInt{\expose{\tpsumstycst{t}{\hatq}{\hatB_1}{\hatB_2}}{T'}} 
              = \VSInt{\bsumstycst{\Subst{T'}{t}{\hatB_1}}{\Subst{T'}{t}{\hatB_2}}}{T'}.$$

  For the case ($i$), fix arbitrary $a$. The proof goal concerning stepping has been established previously by reduction shown 
  in ($i$). It remains to be shown that 
  $$
  \gl{\bfw_i}{\bfOmega_i} \in \LSInt{\Subst{T'}{t}{\hatB_i}}{T'} 
  $$

  Appeal to the respective IH with the substitution $\varphi[t \mapsto T']$ to finish the proof.
\end{itemize}
We have considered all cases.
\end{proof}

\clearpage
\subsection{Forward and Backward Closure}
\label{apx:semantics:fbc}
\emph{Remark.} Operations on trajectories naturally point-wise extends to families.

\begin{theorem}[Backward closure]
If $\bfw_1 \in \cwls{T_1}{T_2}{\bfOmega_1}{\bfOmega_2}$ and $\gl{\bfw_2}{\bfOmega_2} \in_a \LInt{A}{T_2}$,
then $\gl{(\concat{\bfw_1}{\bfw_2})}{\bfOmega'} \in \LInt{A}{T_1}$
\end{theorem}
\begin{proof}
Let $t.p$ be the predicate associated with $A$.

Our goal is to show that
$$\forall T'. (\Subst{T'}{t}{p}) \implies (T_1 \leq T') \land \concat{\bfw_1}{\bfw_2}(t) \in \VInt{\expose{A}{T'}}{T'}.$$

Let $T'$ be fixed such that $\Subst{T'}{t}{p}$ holds. We then need to show 
$$
T_1 \leq T' \andthat 
(\concat{\bfw_1}{\bfw_2})(T') \in \VInt{\expose{A}{T'}}{T'}
$$

On the other hand, our assumption is 
$$\forall T'. (\Subst{T'}{t}{p}) \implies (T_2 \leq T') \land \bfw_2(t) \in \VInt{\expose{A}{T'}}{T'}.$$

Discharge the assumption with what we have and see
$$T_2 \leq T' \andthat \bfw_2(t) \in \VInt{\expose{A}{T'}}{T'}.$$

Because $T_1 \leq T_2$, with $T_2 \leq T'$ we conclude $T_1 \leq T'$. This finishes our first proof goal. 

Recall that 
$$\cwlequivt{\TIntInf{T_2}}{\concat{\bfw_1}{\bfw_2}}{\bfw_2}.$$

With $T_2 \leq T'$, we compute $$(\concat{\bfw_1}{\bfw_2})(T') = \bfw_2(T').$$

This concludes what we want to show.
\end{proof}

\begin{theorem}[Forward closure]
If $\gl{\bfw}{\bfOmega} \in \LSInt{A}{T}$ then 
$\forall T'$, $T \leq T' \implies \gl{(\rpar{\bfw}{T'})}{\bfw(T')} \in \LSInt{A}{T'}$
\end{theorem}
\begin{proof}
Let $t.p$ be the predicate associated with $A$.

Let $T'$ be fixed such that $T \leq T'$. Our goal is to show $\gl{(\rpar{\bfw}{T'})}{\bfw(T')} \in \LSInt{A}{T'}$. That is:
$$
\forall T''. \Subst{T''}{t}{p} \land (T' \leq T'') \implies (\rpar{\bfw}{T'})(T'') \in \VSInt{\expose{A}{T''}}{T''}
$$

Let $T''$ be fixed such that 
$$
\Subst{T''}{t}{p} \andthat T' \leq T''.
$$

Our goal is then to show
$$
(\rpar{\bfw}{T'})(T'') \in \VSInt{\expose{A}{T''}}{T''}
$$

On the other hand, our assumption is 
$$
\forall T''. \Subst{T''}{t}{p} \land (T \leq T'') \implies \bfw(T'') \in \VSInt{\expose{A}{T''}}{T''}
$$

Because we have $T \leq T'$ therefore conclude $T \leq T' \leq T''$. This allows us to discharge the second premise in our 
assumption. Therefore 
$$\bfw(T'') \in \VSInt{\expose{A}{T''}}{T''}$$

Recall that for $T'' \geq T'$,
$$(\rpar{\bfw}{T'})(T'') = \bfw(T'').$$  

This concludes our proof.
\end{proof}

\clearpage
\subsection{Fundamental Theorem of the Logical Relation}
\label{apx:semantics:ftlr}
\begin{theorem}[FTLR]
If $\calG; \calF \mid \Delta \entails{\IsOfProc{P}{T}{A}}$ then $\calG; \calF \mid \Delta \sementails{\IsOfProc{P}{T}{A}}$.
\end{theorem}
\begin{small}
\begin{proof}
Proof by induction on the typing derivation. Before we dive into the cases, a few remarks: 

\begin{enumerate}
\item Each case starts with almost identical steps of expanding various definitions.
  
In all following cases, start by fixing an $\calF$ satisfying assignment $\varphi$ of variables in $\calG$. 

To show $ \hat{\Delta} \sementails{\IsOfProc{\hat{P}}{\hat{A}}{\hat{T}}} $

Further, fix $\delta = \gl{\bfw_x}{\bfOmega_x} \in \LSInt{\hat{\Delta}}{\hat{T}}$ and the channel variable names $\gamma$.

Since in the majority of the situation $\gamma$ is fixed, 
we will refer to the process $\hat{\gamma}(\hat{P})$ as just $\hat{P}$.


Therefore, the proof goal for each case is: 
$$
\exists \bfw. \gl{\bfw}{\conc{(\concfam{x \in \hatDelta}{\bfOmega_x[\gamma(x)]})}{\RunProc{-}{\hat{P}}}} \in \LInt{\wp{\hat{A}}{t}{\hatp}}{\hat{T}}
$$

\item In various places, including when appealing to backwards closure lemmas, we often need to come up with 
computable trajectories (or families of). It suffices to present a (nameless family of) reduction sequence then 
rely on the relation $\calR$ (defined \cref{main:defn:ctrj})
and context to derive the function component of the pair.

\item When we try to apply to inductive hypothesis (IHs), we need to supply the configuration and trajectories. For the 
sake of brevity we will omit the exact arguments, relying on reader to infer them from the result of the appeal. However, 
these arguments are available (i.e. already constructed) at the point of appeal. 
\end{enumerate}

Now we continue with the case analysis: 

\begin{itemize}

  \item \itemheader{Case $P = \tpfwdcst{T'}{x}$}

    In this case $\hatDelta = \IsOf{x}{A'}$.  
    In particular, $\gamma$ contains exactly one assignment from $x$ to some $a$. 
    
    We have $\calG; \calF \entails{\FwdComp{A'}{A}{T'}}$.  That is $\FwdComp{\hatA'}{\hatA}{\hatT'}$.

    By forward closure then forward retyping lemma we conclude 
    $\gl{\bfw_x}{\bfOmega_x} \in \LInt{\hatA}{\hatT'}$. 

    The goal is to find $\bfw$ s.t. 
    $$\gl{\bfw}{\conc{\RunProc{-}{\tpfwdcst{\hatT}{a}}}{\bfOmega_x[a]}} \in \LInt{\hatA}{\hatT}.$$

    By definition of a nameless family, $\bfOmega_x[a] = \RunProc{a}{P_0} \opconc \Omega_0$ for some $P_0$ and $\Omega_0$. 

    Therefore, 
    $$\conc{\RunProc{-}{\tpfwdcst{\hatT'}{a}}}{\bfOmega_x[a]} 
      \LLStepsTo{}{\hatT'}
      \conc{\FwdProc{-}{a}}{\bfOmega_x[a]}
      \LLStepsTo{}{\hatT'}
     \conc{\RunProc{-}{P_0}}{\Omega_0} = \bfOmega_x[-].$$

    By forward closure and $T' \geq T$, we know $\gl{\rpar{\bfw_x}{\hatT'}}{\bfw_x(\hatT')} \in \LInt{\hatA}{\hatT'}$. 
    
    Use backward closure to finish the proof.

  \item \itemheader{Case $P = \tpspawnproccst{T'}{P}{x}{Q}$}

    In this case $\hatDelta = \hatDelta_1, \hatDelta_2$ and we have $\calG; \calF \entails{\CutComp{A}{B}{T'}}$.

    Induction hypothesis: 

    \begin{enumerate}
      \item $\calG; \calF \mid \Delta_1 \sementails{\IsOfProc{P}{T'}{A}}$
      \item $\calG; \calF \mid \Delta_2, \IsOf{x}{B} \sementails{\IsOfProc{Q}{T'}{C}}$
    \end{enumerate}

    Instantiate IH1 by extracting parts of $\gamma$ and $\delta$ that pertains to $\Delta_1$. 

    Then by cut retyping lemma, conclude 
    $$\exists \bfw_1. \gl{\bfw_1}{\conc{(\concfam{x \in \Delta_1}{\bfOmega_x[\gamma(x)]})}{\RunProc{-}{\hatP}}} \in \LSInt{\hatB}{\hatT'}$$

    We now apply IH2. From $\gamma$ and $\delta$ extract parts relevant to $\Delta_2$. 
    Take above temporal computability pair for $x$ and assign it some fresh channel $b$, conclude:
    $$\exists \bfw. \gl{\bfw}{
      (\concfam{x \in \Delta_2}{\Omega_x[\gamma(x)]}) \opconc
      ((\concfam{x \in \Delta_1}{\Omega_x[\gamma(x)]}) \opconc \RunProc{b}{\hatP}) \opconc
      \RunProc{-}{\Subst{b}{x}{\hatQ}}
      } \in \LInt{\hatC}{\hatT'}$$

    Notice that 
    $$(\concfam{x \in \Delta_2}{\Omega_x[\gamma(x)]}) \opconc ((\concfam{x \in \Delta_1}{\Omega_x[\gamma(x)]})) = (\concfam{x \in \Delta}{\Omega_x[\gamma(x)]})$$.

    Therefore 
    $$\gl{\bfw}{ (\concfam{x \in \Delta}{\Omega_x[\gamma(x)]}) \opconc \RunProc{b}{\hatP} \opconc \RunProc{-}{\Subst{b}{x}{\hatQ}} } \in \LInt{\hatC}{\hatT'}$$

    By dynamics:
    $$\conc{(\concfam{x \in \Delta}{\Omega_x[\gamma(x)]})}{\RunProc{-}{\tpspawnproccst{\hatT'}{P}{x}{Q}}} 
      \LLStepsTo{}{\hatT'}  
    (\concfam{x \in \Delta}{\Omega_x[\gamma(x)]}) \opconc \RunProc{b}{\hatP} \opconc \RunProc{-}{\Subst{b}{x}{\hatQ}}$$

    Conclude what we want to show by backwards closure.

  \item \itemheader{Case $P = \tpclosecst{t}{p}$}

    In this case $\hatDelta$ is empty. Set 
    $$\bfw \triangleq \constcwl{\TIntInf{\hatT}}{\RunProc{-}{\tpclosecst{t}{\hat{p}}}}.$$

    To show: for all $T'$ satisfying $\Subst{T'}{t}{\hatp}$, (1) $T' \geq \hatT$ and (2) $\bfw(T') \in \VInt{\expose{\tpunitstycst{t}{\hatp}}{T'}}{T'}$.

    Fix such $T'$. (1) is given by the temporal premise in the typing rule. 
    
    By definition $\bfw(T') = \RunProc{-}{\tpclosecst{t}{\hatp}}$. To show 
    $$
    \RunProc{-}{\tpclosecst{t}{\hatp}}\in \VInt{\expose{\tpunitstycst{t}{\hatp}}{T'}}{T'}
    $$

    Fix arbitrary $a$. From the premise we know $\Subst{\hatT'}{t}{\hatp}$. By dynamics: 
    $$
    \RunProc{a}{\tpclosecst{t}{\hatp}} \LLStepsTo{\actrcvclosecst{a}}{\hatT'} \StopConf
    $$

    This completes our proof.

  \item \itemheader{Case $P = \tpwaitcst{T'}{x}{Q}$}

    In this case the context is $\Delta, \IsOf{x}{\tpunitstycst{t}{p}}$. 

    We have $\gl{w_x}{\bfOmega_x} \in \LSInt{\tpunitstycst{t}{p}}{T}$ and $\delta_\Delta \in \LSInt{\hatDelta}{T}$.


    From the premise we have $\Subst{\hatT'}{t}{\hatp}$. Therefore, by definition we conclude 
    $$\bfw_x(\hatT') \in \VSInt{\expose{\tpunitstycst{t}{\hatp}}{\hatT'}}{\hatT'} = \VSInt{\unitstycst}{\hatT'}.$$

    Therefore, 
    $$\bfw_x(\hatT')[\gamma(x)] \LLStepsTo{\actrcvclosecst{\gamma(x)}}{\hatT'} \StopConf.$$

    By dynamics
    $$
    \RunProc{-}{\tpwaitcst{\hatT'}{\gamma(x)}{\hatQ}} \LLStepsTo{\actsndclosecst{\gamma(x)}}{\hatT'} \RunProc{-}{\hatQ}
    $$

    Therefore,
    $$
    \bfw_x(\hatT') \opconc \RunProc{-}{\tpwaitcst{\hatT'}{\gamma(x)}{\hatQ}}
    \LLStepsTo{}{\hatT'}
    \RunProc{-}{\hatQ}
    $$

    Therefore,
    $$
    \left(\concfam{y \in (\hatDelta, \IsOf{x}{\tpunitstycst{t}{\hatp}} )}{\bfw_y(\hatT')[\gamma(y)]}\right)
    \opconc \RunProc{-}{\tpwaitcst{\hatT'}{\gamma(x)}{\hatQ}}
    \LLStepsTo{}{\hatT'}
    \left(\concfam{y \in \hatDelta}{\bfw_y(\hatT')[\gamma(y)]} \right)
    \opconc \RunProc{-}{\hatQ}
    $$

    Inductively we have $\hatDelta \sementails{\IsOfProc{\hatQ}{\hatT'}{\hatC}}$.

    Furthermore, by assumption, the premise that $\hatT' \geq \hatT$ and forward closure we have 
    $$
    \mathop{\forall}_{z \in \hatDelta}
    \gl{\rpar{\bfw_z}{\hatT'}}{\bfw_z(\hatT')} \in \LSInt{\hatDelta}{\hatT'}
    $$

    Apply the IH and conclude
    $$
    \exists \bfw'. 
    \gl{\bfw'}{\concfam{z \in \hatDelta}{\bfw_z(\hatT')[\gamma(z)]} \opconc \RunProc{-}{\hatQ}} 
    \in \LInt{C}{\hatT'}
    $$

    Our final step appeals to backward closure by the constructing the trajectory
    $$
    (\ilfam{z \in (\hatDelta, \IsOf{x}{\tpunitstycst{t}{\hatp}})}{\lpar{\bfw_z}{\hatT'}})
    \opconc 
    \constcwl{\TInt{\hatT}{\hatT'}}{\RunProc{-}{\tpwaitcst{\hatT'}{\gamma(x)}{\hatQ}}}
    $$
    then extend it by the reduction just presented.

  \item \itemheader{Case $P = \tptensorsndcst{t}{p}{P_1}{P_2}$}.

    In this case $\hatDelta = \hatDelta_1, \hatDelta_2$ for some $\hatDelta_1$ and $\hatDelta_2$. Take 
    $$
    \bfw \triangleq 
    \left(\concfam{z \in \hatDelta}{\bfw_z[\gamma(z)]}\right) 
    \opconc
    \constcwl{\TIntInf{\hatT}}{\RunProc{-}{\tptensorsndcst{t}{\hatp}{\hatP_1}{\hatP_2}}}
    $$

    Our goal in this case is to show 
    $$ \gl{\bfw}{
      (\concfam{z \in \hatDelta}{\bfOmega_z}) \opconc \RunProc{-}{\tptensorsndcst{t}{\hatp}{\hatP_1}{\hatP_2}}
    } \in \LInt{\tptensorstycst{t}{\hatp}{\hatA_1}{\hatA_2}}{\hatT}
    $$

    By definition, fix arbitrary $T'$ such that $\Subst{T'}{t}{\hatp}$ is true. From the premise we have $T' \geq \hatT$.

    To show: 
    $$
    \bfw(T') = \VInt{\expose{\tptensorstycst{t}{\hatp}{\hatA_1}{\hatA_2}}{T'}}{T'}
    $$

    That is
    $$
    (\concfam{z \in \hatDelta}{\bfw(T')[\gamma(z)]}) \opconc \RunProc{-}{\tptensorsndcst{t}{\hatp}{\hatP_1}{\hatP_2}}
    \in 
    \VInt{\tensorstycst{\Subst{T'}{t}{\hatA_1}}{\Subst{T'}{t}{\hatA_2}}}{T'}
    $$

    Take
    \begin{align*}
    \bfOmega_1 & \triangleq 
    (\concfam{z \in \hatDelta_1}{\bfw(T')[\gamma(z)]}) \opconc \RunProc{-}{\Subst{T'}{t}{\hatP_1}} \\
    \bfOmega_2 & \triangleq 
    (\concfam{z \in \hatDelta_2}{\bfw(T')[\gamma(z)]}) \opconc \RunProc{-}{\Subst{T'}{t}{\hatP_2}} \\
    \end{align*}

    Remark that 
    $$
    \concfam{z \in \hatDelta}{\bfw(T')[\gamma(z)]} = 
    \concfam{z \in \hatDelta_1, \hatDelta_2}{\bfw(T')[\gamma(z)]} = 
    \left(\concfam{z \in \hatDelta_1}{\bfw(T')[\gamma(z)]}\right) \opconc
    \left(\concfam{z \in \hatDelta_2}{\bfw(T')[\gamma(z)]}\right)
    $$

    By dynamics for all $a$ and fresh $c$,
    $$
      (\concfam{z \in \hatDelta}{\bfw(T')[\gamma(z)]}) \opconc \RunProc{a}{\tptensorsndcst{t}{\hatp}{\hatP_1}{\hatP_2}}
      \LLStepsTo{\actsndchancst{a}{c}}{T'}
      \conc{\bfOmega_1[c]}{\bfOmega_2[a]}
    $$

    It remains to be shown that 

    \begin{align}
    \exists \bfw_1. 
    \gl{\bfw_1}{\bfOmega_1} 
    = \gl{\bfw_1}{(\concfam{z \in \hatDelta_1}{\bfw(T')[\gamma(z)]}) \opconc \RunProc{-}{\Subst{T'}{t}{\hatP_1}}}  
    \in \LInt{\Subst{T'}{t}{\hatA_1}}{T'} \label{apx:ftlr:goal:tensor1} \\
    \exists \bfw_2. 
    \gl{\bfw_2}{\bfOmega_2} 
    = \gl{\bfw_2}{(\concfam{z \in \hatDelta_2}{\bfw(T')[\gamma(z)]}) \opconc \RunProc{-}{\Subst{T'}{t}{\hatP_2}}} 
    \in \LInt{\Subst{T'}{t}{\hatA_2}}{T'} \label{apx:ftlr:goal:tensor2}
    \end{align}

    Inductively, making use of the temporal variable substitution $\varphi[t \mapsto T']$, we have 
    \begin{enumerate} 
      \item[(IH1)] $\hatDelta_1 \sementails{\IsOfProc{\Subst{T'}{t}{\hatP_1}}{T'}{\Subst{T'}{t}{\hatA_1}}}$
      \item[(IH2)] $\hatDelta_2 \sementails{\IsOfProc{\Subst{T'}{t}{\hatP_2}}{T'}{\Subst{T'}{t}{\hatA_2}}}$
    \end{enumerate}

    Appeal to IH1 to goal \ref{apx:ftlr:goal:tensor1} and IH2 to show goal \ref{apx:ftlr:goal:tensor2}, 
    making use of forward closure whenever necessary to move from $\hatT$ to $T'$.

  \item \itemheader{Case $P = \tptensorrcvcst{x}{T'}{y}{Q}$}.

    In this case the context is $\Delta, \IsOf{x}{\tptensorstycst{t}{p}{A_1}{A_2}}$. 

    We have $\gl{w_x}{\bfOmega_x} \in \LSInt{\tptensorstycst{t}{\hatp}{\hatA_1}{\hatA_2}}{\hatT}$ and $\delta_\Delta \in \LSInt{\hatDelta}{T}$.


    From the premise we have $\Subst{\hatT'}{t}{\hatp}$. Therefore, by definition we conclude 
    $$\bfw_x(\hatT') 
    \in \VSInt{\expose{\tptensorstycst{t}{\hatp}{\hatA_1}{\hatA_2}}{\hatT'}}{\hatT'} 
      = \VSInt{\tensorstycst{\Subst{\hatT'}{t}{\hatA_1}}{\Subst{\hatT'}{t}{\hatA_2}}}{\hatT'}.$$

    Therefore, for fresh $c$ and some $\bfOmega_1$ and $\bfOmega_2$, 
    $$\bfw_x(\hatT')[\gamma(x)] \LLStepsTo{\actsndchancst{\gamma(x)}{c}}{\hatT'} \conc{\bfOmega_1[c]}{\bfOmega_2[\gamma(x)]}.$$

    Where 
    \begin{align*}
    \exists \bfw_1. \gl{\bfw_1}{\bfOmega_1} \in \LSInt{\Subst{\hatT'}{t}{\hatA_1}}{\hatT'} & &
    \exists \bfw_2. \gl{\bfw_2}{\bfOmega_2} \in \LSInt{\Subst{\hatT'}{t}{\hatA_2}}{\hatT'}
    \end{align*}

    By dynamics
    $$
    \RunProc{-}{\tptensorrcvcst{\gamma(x)}{\hatT'}{y}{\hatQ}} 
      \LLStepsTo{\actrcvchancst{\gamma(x)}{c}}{\hatT'} 
    \RunProc{-}{\Subst{c}{y}{\hatQ}} 
    $$

    Therefore,
    $$
    \bfw_x(\hatT') \opconc \RunProc{-}{\tptensorrcvcst{\gamma(x)}{\hatT'}{y}{\hatQ}} 
    \LLStepsTo{}{\hatT'}
    \bfOmega_1[c]
    \opconc \bfOmega_2[\gamma(x)]
    \opconc \RunProc{-}{\Subst{c}{y}{\hatQ}} 
    $$

    Therefore,
    \begin{gather*}
    \left(\concfam{z \in (\hatDelta, \IsOf{x}{\tpunitstycst{t}{\hatp}} )}{\bfw_z(\hatT')[\gamma(z)]}\right)
    \opconc \RunProc{-}{\tptensorrcvcst{\gamma(x)}{\hatT'}{y}{\hatQ}} 
    \LLStepsTo{}{\hatT'} \\
    \left(\concfam{z \in \hatDelta}{\bfw_z(\hatT')[\gamma(z)]}\right)
    \opconc \bfOmega_1[c]
    \opconc \bfOmega_2[\gamma(x)]
    \opconc \RunProc{-}{\Subst{c}{y}{\hatQ}} 
    \end{gather*}

    Inductively we have $\hatDelta, \IsOf{y}{\Subst{\hatT'}{t}{\hatA_1}}, \IsOf{x}{\Subst{\hatT'}{t}{\hatA_2}} \sementails{\IsOfProc{\hatQ}{\hatT'}{\hatC}}$.

    Furthermore, by assumption, the premise that $\hatT' \geq \hatT$ and forward closure we have 
    $$
    \mathop{\forall}_{z \in \hatDelta}
    \gl{\rpar{\bfw_z}{\hatT'}}{\bfw_z(\hatT')} \in \LSInt{\hatDelta}{\hatT'}
    $$

    Apply the IH and conclude
    $$
    \exists \bfw'. 
    \gl{\bfw'}{\concfam{z \in \hatDelta}{\bfw_z(\hatT')[\gamma(z)]} \opconc 
    \bfOmega_1[c]
    \opconc \bfOmega_2[\gamma(x)]
    \opconc \RunProc{-}{\Subst{c}{y}{\hatQ}} 
    } 
    \in \LInt{C}{\hatT'}
    $$

    Our final step appeals to backward closure by the constructing the trajectory
    $$
    (\ilfam{z \in (\hatDelta, \IsOf{x}{\tptensorstycst{t}{\hatp}{\hatA_1}{\hatA_2}})}{\lpar{\bfw_z}{\hatT'}})
    \opconc 
    \constcwl{\TInt{\hatT}{\hatT'}}{\RunProc{-}{\tptensorrcvcst{\gamma(x)}{\hatT'}{y}{\hatQ}}}
    $$
    then extend it by the reduction just presented.

  \item \itemheader{Case $P = \tpoffercst{t}{p}{P_1}{P_2}$}

    Take 
    $$
    \bfw \triangleq 
    \left(\concfam{z \in \hatDelta}{\bfw_z[\gamma(z)]}\right) 
    \opconc
    \constcwl{\TIntInf{\hatT}}{\RunProc{-}{\tpoffercst{t}{\hatp}{\hatP_1}{\hatP_2}}}
    $$

    Our goal in this case is to show 
    $$ \gl{\bfw}{
      (\concfam{z \in \hatDelta}{\bfOmega_z}) \opconc \RunProc{-}{\tpoffercst{t}{\hatp}{\hatP_1}{\hatP_2}}
    } \in \LInt{\tpwithstycst{t}{\hatp}{\hatA_1}{\hatA_2}}{\hatT}
    $$

    By definition, fix arbitrary $T'$ such that $\Subst{T'}{t}{\hatp}$ is true. From the premise we have $T' \geq \hatT$.

    To show: 
    $$
    \bfw(T') = \VInt{\expose{\tpwithstycst{t}{\hatp}{\hatA_1}{\hatA_2}}{T'}}{T'}
    $$

    That is
    $$
    (\concfam{z \in \hatDelta}{\bfw(T')[\gamma(z)]}) \opconc \RunProc{-}{\tpoffercst{t}{\hatp}{\hatP_1}{\hatP_2}}
    \in 
    \VInt{\bwithstycst{\Subst{T'}{t}{\hatA_1}}{\Subst{T'}{t}{\hatA_2}}}{T'}
    $$

    Take
    \begin{align*}
    \bfOmega_1 & \triangleq 
    (\concfam{z \in \hatDelta}{\bfw(T')[\gamma(z)]}) \opconc \RunProc{-}{\Subst{T'}{t}{\hatP_1}} \\
    \bfOmega_2 & \triangleq 
    (\concfam{z \in \hatDelta}{\bfw(T')[\gamma(z)]}) \opconc \RunProc{-}{\Subst{T'}{t}{\hatP_2}} \\
    \end{align*}

    By dynamics for all $a$,
    \begin{align*}
    (\concfam{z \in \hatDelta}{\bfw(T')[\gamma(z)]}) \opconc \RunProc{a}{\tpoffercst{t}{\hatp}{\hatP_1}{\hatP_2}}
    & \LLStepsTo{\actrcvlblcst{a}{\lblleft}}{T'} 
    \bfOmega_1[a] \\
    (\concfam{z \in \hatDelta}{\bfw(T')[\gamma(z)]}) \opconc \RunProc{a}{\tpoffercst{t}{\hatp}{\hatP_1}{\hatP_2}}
    & \LLStepsTo{\actrcvlblcst{a}{\lblright}}{T'}
    \bfOmega_2[a]
    \end{align*}

    It remains to be shown that 

    \begin{align}
    \exists \bfw_1. 
    \gl{\bfw_1}{\bfOmega_1} 
    = \gl{\bfw_1}{(\concfam{z \in \hatDelta}{\bfw(T')[\gamma(z)]}) \opconc \RunProc{-}{\Subst{T'}{t}{\hatP_1}}}  
    \in \LInt{\Subst{T'}{t}{\hatA_1}}{T'} \label{apx:ftlr:goal:with1} \\
    \exists \bfw_2. 
    \gl{\bfw_2}{\bfOmega_2} 
    = \gl{\bfw_2}{(\concfam{z \in \hatDelta}{\bfw(T')[\gamma(z)]}) \opconc \RunProc{-}{\Subst{T'}{t}{\hatP_2}}} 
    \in \LInt{\Subst{T'}{t}{\hatA_2}}{T'} \label{apx:ftlr:goal:with2}
    \end{align}

    Inductively, making use of the temporal variable substitution $\varphi[t \mapsto T']$, we have 
    \begin{enumerate} 
      \item[(IH1)] $\hatDelta \sementails{\IsOfProc{\Subst{T'}{t}{\hatP_1}}{T'}{\Subst{T'}{t}{\hatA_1}}}$
      \item[(IH2)] $\hatDelta \sementails{\IsOfProc{\Subst{T'}{t}{\hatP_2}}{T'}{\Subst{T'}{t}{\hatA_2}}}$
    \end{enumerate}

    Appeal to IH1 to goal \ref{apx:ftlr:goal:with1} and IH2 to show goal \ref{apx:ftlr:goal:with2}, 
    making use of forward closure whenever necessary to move from $\hatT$ to $T'$.

  \item \itemheader{Cases $P = \tpselectlcst{T'}{x}{Q}$ and $P = \tpselectrcst{T'}{x}{Q}$}

    In these cases the context is $\Delta, \IsOf{x}{ \tpwithstycst{t}{c}{A_1}{A_2}}$. Without loss of generality, we will prove
    for the case $P = \tpselectlcst{T'}{x}{Q}$. The other case proceeds almost identically.

    We have $\gl{w_x}{\bfOmega_x} \in \LSInt{\tpwithstycst{t}{\hatp}{\hatA_1}{\hatA_2}}{\hatT}$ and $\delta_\Delta \in \LSInt{\hatDelta}{T}$.

    From the premise we have $\Subst{\hatT'}{t}{\hatp}$. Therefore, by definition we conclude 
    $$\bfw_x(\hatT') 
    \in \VSInt{\expose{\tpwithstycst{t}{\hatp}{\hatA_1}{\hatA_2}}{\hatT'}}{\hatT'} 
      = \VSInt{\bwithstycst{\Subst{\hatT'}{t}{\hatA_1}}{\Subst{\hatT'}{t}{\hatA_2}}}{\hatT'}.$$

    Therefore, for some $\bfOmega_1$ (for the other case the message is $\lblright$), 
    $$\bfw_x(\hatT')[\gamma(x)] \LLStepsTo{\actrcvlblcst{\gamma(x)}{\lblleft}}{\hatT'} \bfOmega_1[\gamma(x)].$$

    Where $\exists \bfw_1. \bfOmega_1 \in \LSInt{\Subst{\hatT'}{t}{\hatA_1}}{\hatT'}$.

    By dynamics
    $$
    \RunProc{-}{\tpselectlcst{\hatT'}{\gamma(x)}{\hatQ}} 
      \LLStepsTo{\actsndchancst{\gamma(x)}{\lblleft}}{\hatT'} 
    \RunProc{-}{\hatQ} 
    $$

    Therefore,
    $$
    \bfw_x(\hatT') \opconc \RunProc{-}{\tpselectlcst{\hatT'}{\gamma(x)}{\hatQ}} 
    \LLStepsTo{}{\hatT'}
    \bfOmega_1[\gamma(x)]
    \opconc \RunProc{-}{\hatQ} 
    $$

    Therefore,
    $$
    \left(\concfam{z \in (\hatDelta, \IsOf{x}{\tpunitstycst{t}{\hatp}} )}{\bfw_z(\hatT')[\gamma(z)]}\right)
    \opconc \RunProc{-}{\tpselectlcst{\hatT'}{\gamma(x)}{\hatQ}}
    \LLStepsTo{}{\hatT'} 
    \left(\concfam{z \in \hatDelta}{\bfw_z(\hatT')[\gamma(z)]}\right)
    \opconc \bfOmega_1[\gamma(x)] \opconc \RunProc{-}{\hatQ} 
    $$

    Inductively we have $\hatDelta, \IsOf{x}{\Subst{\hatT'}{t}{\hatA_1}} \sementails{\IsOfProc{\hatQ}{\hatT'}{\hatC}}$.

    Furthermore, by assumption, the premise that $\hatT' \geq \hatT$ and forward closure we have 
    $$
    \mathop{\forall}_{y \in \hatDelta}
    \gl{\rpar{\bfw_y}{\hatT'}}{\bfw_y(\hatT')} \in \LSInt{\hatDelta}{\hatT'}
    $$

    Apply the IH and conclude
    $$
    \exists \bfw'. 
    \gl{\bfw'}{
      \left(\concfam{z \in \hatDelta}{\bfw_z(\hatT')[\gamma(z)]}\right)
      \opconc \bfOmega_1[\gamma(x)]
      \opconc \RunProc{-}{\hatQ} 
    } 
    \in \LInt{C}{\hatT'}
    $$

    Our final step appeals to backward closure by the constructing the trajectory
    $$
    (\ilfam{z \in (\hatDelta, \IsOf{x}{\tpwithstycst{t}{\hatp}{\hatA_1}{\hatA_2}})}{\lpar{\bfw_z}{\hatT'}})
    \opconc 
    \constcwl{\TInt{\hatT}{\hatT'}}{\RunProc{-}{\tpselectlcst{\hatT'}{\gamma(x)}{\hatQ}}}
    $$
    then extend it by the reduction just presented.

\item \itemheader{Cases $P = \tpinlcst{t}{p}{A_1}{A_2}{P}$ and $P = \tpinrcst{t}{p}{A_1}{A_2}{P}$}

    We will do the proof primarily for $P = \tpinlcst{t}{p}{A_1}{A_2}{P}$. The other case is highly analogously, and 
    we will point out the difference whenever noteworthy.

    Take 
    $$
    \bfw \triangleq 
    \left(\concfam{z \in \hatDelta}{\bfw_z[\gamma(z)]}\right) 
    \opconc
    \constcwl{\TIntInf{\hatT}}{\RunProc{-}{\tpinlcst{t}{\hatp}{\hatA_1}{\hatA_2}{\hatP}}}
    $$

    Our goal in this case is to show 
    $$ \gl{\bfw}{
      (\concfam{z \in \hatDelta}{\bfOmega_z}) \opconc \RunProc{-}{\tpinlcst{t}{\hatp}{\hatA_1}{\hatA_2}{\hatP}}
    } \in \LInt{\tpsumstycst{t}{\hatp}{\hatA_1}{\hatA_2}}{\hatT}
    $$

    By definition, fix arbitrary $T'$ such that $\Subst{T'}{t}{\hatp}$ is true. From the premise we have $T' \geq \hatT$.

    To show: 
    $$
    \bfw(T') = \VInt{\expose{\tpsumstycst{t}{\hatp}{\hatA_1}{\hatA_2}}{T'}}{T'}
    $$

    That is
    $$
    (\concfam{z \in \hatDelta}{\bfw(T')[\gamma(z)]}) \opconc \RunProc{-}{\tpinlcst{t}{\hatp}{\hatA_1}{\hatA_2}{\hatP}}
    \in 
    \VInt{\bwithstycst{\Subst{T'}{t}{\hatA_1}}{\Subst{T'}{t}{\hatA_2}}}{T'}
    $$

    Take (the choice is identical for the other case)
    $$
    \bfOmega \triangleq 
    (\concfam{z \in \hatDelta}{\bfw(T')[\gamma(z)]}) \opconc \RunProc{-}{\Subst{T'}{t}{\hatP}}
    $$

    By dynamics for all $a$ (for the other case, the label sent would be $\lblright$),
    $$
    (\concfam{z \in \hatDelta}{\bfw(T')[\gamma(z)]}) \opconc \RunProc{a}{\tpinlcst{t}{\hatp}{\hatA_1}{\hatA_2}{\hatP}}
    \LLStepsTo{\actsndlblcst{a}{\lblleft}}{T'} 
    \bfOmega[a] 
    $$

    It remains to be shown that 
    $$
    \exists \bfw. 
    \gl{\bfw}{\bfOmega} 
    = \gl{\bfw}{(\concfam{z \in \hatDelta}{\bfw(T')[\gamma(z)]}) \opconc \RunProc{-}{\Subst{T'}{t}{\hatP}}}  
    \in \LInt{\Subst{T'}{t}{\hatA_1}}{T'}
    $$

    Inductively, making use of the temporal variable substitution $\varphi[t \mapsto T']$, we have 
    $$
    \hatDelta \sementails{\IsOfProc{\Subst{T'}{t}{\hatP_1}}{T'}{\Subst{T'}{t}{\hatA_1}}}
    $$

    Appeal to IH  
    making use of forward closure whenever necessary to move from $\hatT$ to $T'$.

  \item \itemheader{Case $P = \tpcasecst{T'}{x}{Q_1}{Q_2}$}

    In this case the context is $\Delta, \IsOf{x}{\tpsumstycst{t}{c}{A_1}{A_2}}$. 

    We have $\gl{w_x}{\bfOmega_x} \in \LSInt{\tpsumstycst{t}{\hatp}{\hatA_1}{\hatA_2}}{\hatT}$ and $\delta_\Delta \in \LSInt{\hatDelta}{T}$.

    From the premise we have $\Subst{\hatT'}{t}{\hatp}$. Therefore, by definition we conclude 
    $$\bfw_x(\hatT') 
    \in \VSInt{\expose{\tpsumstycst{t}{\hatp}{\hatA_1}{\hatA_2}}{\hatT'}}{\hatT'} 
      = \VSInt{\bsumstycst{\Subst{\hatT'}{t}{\hatA_1}}{\Subst{\hatT'}{t}{\hatA_2}}}{\hatT'}.$$

    Therefore, either (1) for some $\bfOmega_1$, 
    $$\bfw_x(\hatT')[\gamma(x)] \LLStepsTo{\actsndlblcst{\gamma(x)}{\lblleft}}{\hatT'} \bfOmega_1[\gamma(x)], 
    \where 
    \exists \bfw_1. \gl{\bfw_1}{\bfOmega_1} \in \LSInt{\Subst{\hatT'}{t}{\hatA_1}}{\hatT'};
    $$
    or (2) for some $\bfOmega_2$, 
    $$\bfw_x(\hatT')[\gamma(x)] \LLStepsTo{\actsndlblcst{\gamma(x)}{\lblright}}{\hatT'} \bfOmega_2[\gamma(x)],
    \where 
    \exists \bfw_2. \gl{\bfw_1}{\bfOmega_2} \in \LSInt{\Subst{\hatT'}{t}{\hatA_2}}{\hatT'}.
    $$

    We briefly look at the client side. By dynamics
   \begin{align*}
    \RunProc{-}{\tpcasecst{\hatT'}{\gamma(x)}{\hatQ_1}{\hatQ_2}} 
    & \LLStepsTo{\actrcvlblcst{\gamma(x)}{\lblleft}}{\hatT'} 
    \RunProc{-}{\Subst{c}{y}{\hatQ_1}}  \\
    \RunProc{-}{\tpcasecst{\hatT'}{\gamma(x)}{\hatQ_1}{\hatQ_2}} 
    & \LLStepsTo{\actrcvlblcst{\gamma(x)}{\lblright}}{\hatT'} 
    \RunProc{-}{\Subst{c}{y}{\hatQ_2}} 
   \end{align*} 

    Now we perform the case split. For case ($i$):
    $$
    \bfw_x(\hatT') \opconc \RunProc{-}{\tpcasecst{\hatT'}{\gamma(x)}{\hatQ_1}{\hatQ_2}} 
    \LLStepsTo{}{\hatT'}
    \bfOmega_1[\gamma(x)]
    \opconc \RunProc{-}{\hatQ_i} 
    $$

    and 
    \begin{gather*}
    \left(\concfam{z \in (\hatDelta, \IsOf{x}{\tpunitstycst{t}{\hatp}} )}{\bfw_z(\hatT')[\gamma(z)]}\right)
    \opconc \RunProc{-}{\tpcasecst{\hatT'}{\gamma(x)}{\hatQ_1}{\hatQ_2}}
    \LLStepsTo{}{\hatT'} \\
    \left(\concfam{z \in \hatDelta}{\bfw_z(\hatT')[\gamma(z)]}\right)
    \opconc \bfOmega_1[\gamma(x)]
    \opconc \RunProc{-}{\hatQ_i}.
    \end{gather*}

    Inductively for both cases $i = 1, 2$, we have 
    $\hatDelta, \IsOf{x}{\Subst{\hatT'}{t}{\hatA_i}} \sementails{\IsOfProc{\hatQ}{\hatT'}{\hatC}}$.

    Furthermore, by assumption, the premise that $\hatT' \geq \hatT$ and forward closure we have 
    $$
    \mathop{\forall}_{z \in \hatDelta}
    \gl{\rpar{\bfw_z}{\hatT'}}{\bfw_z(\hatT')} \in \LSInt{\hatDelta}{\hatT'}
    $$

    Apply the IH and conclude for both cases:
    $$
    \exists \bfw'. 
    \gl{\bfw'}{\left(\concfam{z \in \hatDelta}{\bfw_z(\hatT')[\gamma(z)]}\right) 
    \opconc \bfOmega_1[\gamma(x)]
    \opconc \RunProc{-}{\hatQ_i} 
    } 
    \in \LInt{C}{\hatT'}
    $$

    Our final step in both cases appeals to backward closure by the constructing the trajectory
    $$
    (\ilfam{z \in (\hatDelta, \IsOf{x}{ \tpsumstycst{t}{\hatp}{\hatA_1}{\hatA_2} })}{\lpar{\bfw_z}{\hatT'}})
    \opconc 
    \constcwl{\TInt{\hatT}{\hatT'}}{\RunProc{-}{\tpcasecst{\hatT}{\gamma(x)}{\hatQ_1}{\hatQ_2}}}
    $$
    then extend it by the reduction just presented.

\item \itemheader{Case $P = \tplarrrcvcst{t}{p}{A_1}{x}{P}$}

    Take 
    $$
    \bfw \triangleq 
    \left(\concfam{z \in \hatDelta}{\bfw_z[\gamma(z)]}\right) 
    \opconc
    \constcwl{\TIntInf{\hatT}}{\RunProc{-}{\tplarrrcvcst{t}{\hatp}{\hatA_1}{x}{\hatP}}}
    $$

    Our goal in this case is to show 
    $$ \gl{\bfw}{
      (\concfam{z \in \hatDelta}{\bfOmega_z}) \opconc \RunProc{-}{\tplarrrcvcst{t}{\hatp}{\hatA_1}{x}{\hatP}}
    } \in \LInt{\tplarrstycst{t}{\hatp}{\hatA_1}{\hatA_2}}{\hatT}
    $$

    By definition, fix arbitrary $T'$ such that $\Subst{T'}{t}{\hatp}$ is true. From the premise we have $T' \geq \hatT$.

    To show: 
    $$
    \bfw(T') = \VInt{\expose{\tplarrstycst{t}{\hatp}{\hatA_1}{\hatA_2}}{T'}}{T'}
    $$

    That is
    $$
    (\concfam{z \in \hatDelta}{\bfw(T')[\gamma(z)]}) \opconc \RunProc{-}{\tplarrrcvcst{t}{\hatp}{\hatA_1}{x}{\hatP}}
    \in 
    \VInt{\larrstycst{\Subst{T'}{t}{\hatA_1}}{\Subst{T'}{t}{\hatA_2}}}{T'}
    $$

    Fix arbitrary $c$, take 
    $$
    \bfOmega \triangleq 
    (\concfam{z \in \hatDelta}{\bfw(T')[\gamma(z)]}) \opconc \RunProc{-}{\Subst{T', c}{t, x}{\hatP}}
    $$

    By dynamics for all $a$,
    $$
    (\concfam{z \in \hatDelta}{\bfw(T')[\gamma(z)]}) \opconc \RunProc{a}{\tplarrrcvcst{t}{\hatp}{\hatA_1}{x}{\hatP}}
    \LLStepsTo{\actrcvchancst{a}{c}}{T'} 
    \bfOmega[a] 
    $$

    Now fix arbitrary $\bfw_1$ and $\bfOmega_1$ s.t. 
    $$
    \gl{\bfw_1}{\bfOmega_1} \in \LSInt{\Subst{T'}{t}{\hatA_1}}{T'}
    $$

    It remains to be shown that 
    $$
    \exists \bfw. 
    \gl{\bfw}{\bfOmega} 
    = \gl{\bfw}{(\concfam{z \in \hatDelta}{\bfw(T')[\gamma(z)]}) \opconc \RunProc{-}{\Subst{T'}{t}{\hatP}} \opconc \bfOmega_1[c]}  
    \in \LInt{\Subst{T'}{t}{\hatA_2}}{T'}
    $$

    Inductively, making use of the temporal variable substitution $\varphi[t \mapsto T']$, we have 
    $$
    \hatDelta, \IsOf{x}{\Subst{T'}{t}{\hatA_1}}\sementails{\IsOfProc{\Subst{T'}{t}{\hatP_1}}{T'}{\Subst{T'}{t}{\hatA_2}}}
    $$

    Appeal to IH making use of forward closure whenever necessary to move from $\hatT$ to $T'$: 
    \begin{itemize}
      \item The runtime channel substitution assigns $x$ the channel name $c$. 
      \item The configuration component of sub-forest consists of configurations in the $\delta$ and $x \mapsto \bfOmega_1$. 
      \item The trajectory component consists of $\delta$ interleaved with $\bfw_1$.
    \end{itemize}

  \item \itemheader{Case $P = \tplarrsndcst{x}{T'}{P}{Q}$}
  
    In this case the context is $\Delta_1, \Delta_2, \IsOf{x}{\tplarrstycst{t}{c}{A_1}{A_2}}$. Let $\Delta \triangleq \Delta_1, \Delta_2$.

    We have $\gl{w_x}{\bfOmega_x} \in \LSInt{\tplarrstycst{t}{\hatp}{\hatA_1}{\hatA_2}}{\hatT}$, $\delta_{\Delta_1} \in \LSInt{\hatDelta_1}{T}$ and $\delta_{\Delta_2} \in \LSInt{\hatDelta_2}{T}$.

    From the premise we have $\Subst{\hatT'}{t}{\hatp}$. Therefore, by definition we conclude 
    $$\bfw_x(\hatT') 
    \in \VSInt{\expose{\tplarrstycst{t}{\hatp}{\hatA_1}{\hatA_2}}{\hatT'}}{\hatT'} 
      = \VSInt{\larrstycst{\Subst{\hatT'}{t}{\hatA_1}}{\Subst{\hatT'}{t}{\hatA_2}}}{\hatT'}.$$

    Therefore, for all $c$ and some $\bfOmega_1$:
    $$
    \bfw_x(\hatT')[\gamma(x)] \LLStepsTo{\actrcvchancst{\gamma(x)}{c}}{\hatT'} \bfOmega_1[\gamma(x)]
    $$
    and 
    $$\forall \bfw_2 \bfOmega_2. 
    \gl{\bfw_2}{\bfOmega_2} \in \LInt{\Subst{\hatT'}{t}{\hatA_1}}{\hatT'}
    \implies 
    (\exists \bfw'. \gl{\bfw'}{\conc{\bfOmega_1[\gamma(x)]}{\bfOmega_2}} \in \LSInt{\Subst{\hatT'}{t}{\hatA_2}}{\hatT'})
    $$

    By dynamics, for some fresh $c$, 
    $$
    \RunProc{-}{\tplarrsndcst{\gamma(x)}{\hatT'}{\hatP}{\hatQ}} 
    \LLStepsTo{\actsndchancst{\gamma(x)}{c}}{\hatT'} 
    \RunProc{c}{\hatP}
    \opconc \RunProc{-}{\hatQ} \\
    $$

    Therefore
    $$
    \bfw_x(\hatT') \opconc \RunProc{-}{\tplarrsndcst{\gamma(x)}{\hatT'}{\hatP}{\hatQ}} 
    \LLStepsTo{}{\hatT'}
    \bfOmega_1[\gamma(x)]
    \opconc \RunProc{c}{\hatP}
    \opconc \RunProc{-}{\hatQ} \\
    $$

    and 
    \begin{gather*}
    \left(\concfam{z \in (\hatDelta, \IsOf{x}{\tpunitstycst{t}{\hatp}} )}{\bfw_z(\hatT')[\gamma(z)]}\right)
    \opconc \RunProc{-}{\tplarrsndcst{\gamma(x)}{\hatT'}{\hatP}{\hatQ}}
    \LLStepsTo{}{\hatT'} \\
    \left(\concfam{z \in \hatDelta}{\bfw_z(\hatT')[\gamma(z)]}\right)
    \opconc \bfOmega_1[\gamma(x)]
    \opconc \RunProc{c}{\hatP}
    \opconc \RunProc{-}{\hatQ}
    \end{gather*}

    Inductively we have 
    \begin{enumerate}
      \item[IH1:] $\hatDelta_1 \sementails{\IsOfProc{\hatP}{\hatT'}{\Subst{\hatT'}{t}{\hatA_1}}}$
      \item[IH2:] $\hatDelta_2, \IsOf{x}{\Subst{\hatT'}{t}{\hatA_2}} \sementails{\IsOfProc{\hatQ}{\hatT'}{\hatC}}$.
    \end{enumerate}

    Furthermore, by assumption, the premise that $\hatT' \geq \hatT$ and forward closure we have 
    $$
    \mathop{\forall}_{z \in \hatDelta}
    \gl{\rpar{\bfw_z}{\hatT'}}{\bfw_z(\hatT')} \in \LSInt{\hatDelta}{\hatT'}
    $$

    Apply IH and conclude 
    $$
    \exists \bfw_1. 
    \gl{\bfw_2}{(\concfam{z \in \hatDelta_1}{\bfw_z(\hatT')[\gamma(z)]}) \opconc \RunProc{c}{\hatP}}
    \in \LInt{\Subst{\hatT'}{t}{A_1}}{\hatT'}
    $$

    Apply this to we obtained from the definition of the logical relation for $\multimap$:
    $$
    \exists \bfw'. 
    \gl{\bfw'}{\bfOmega_1[\gamma(x)] \opconc (\concfam{z \in \hatDelta_1}{\bfw_z(\hatT')[\gamma(z)]}) \opconc \RunProc{c}{\hatP}}
    \in \LSInt{\Subst{\hatT'}{t}{A_2}}{\hatT'}
    $$

    With this appeal to IH2 and conclude:
    $$
    \exists \bfw''. 
    \gl{\bfw''}{
  \left(\concfam{z \in \hatDelta}{\bfw_z(\hatT')[\gamma(z)]}\right)
      \opconc \bfOmega_1[\gamma(x)]
      \opconc \RunProc{c}{\hatP}
      \opconc \RunProc{-}{\hatQ}
    } 
    \in \LInt{\hatC}{\hatT'}
    $$

    Remark: In this step, we regrouped the consecutive tensor operator using structural congruences: 
    \begin{gather*}
    \left((\concfam{z \in \hatDelta_1}{\bfw_z(\hatT')[\gamma(z)]}) \opconc \RunProc{c}{\hatP} \right)
    \opconc \bfOmega_1[\gamma(x)]
    \opconc \left(\concfam{z \in \hatDelta_2}{\bfw_z(\hatT')[\gamma(z)]}\right)
    \opconc \RunProc{-}{\hatQ} \\ 
    = \\
    \left(\concfam{z \in \hatDelta}{\bfw_z(\hatT')[\gamma(z)]}\right)
    \opconc \bfOmega_1[\gamma(x)]
    \opconc \RunProc{c}{\hatP}
    \opconc \RunProc{-}{\hatQ}
    \end{gather*}

    Our final step in both cases appeals to backward closure by the constructing the trajectory
    $$
    (\ilfam{y \in (\hatDelta, \IsOf{x}{\tplarrstycst{t}{\hatp}{\hatA_1}{\hatA_2}})}{\lpar{\bfw_y}{\hatT'}})
    \opconc 
    \constcwl{\TInt{\hatT}{\hatT'}}{\RunProc{-}{\tplarrsndcst{\gamma(x)}{\hatT'}{\hatP}{\hatQ}}}
    $$
    then extend it by the reduction just presented.

\end{itemize}
\end{proof}
\end{small}

\clearpage
\section{Exemplary Proofs for Semantically Typing the Sensor}
\label{apx:pf-sensor}

The idea is that we build up our proof by consecutively analyzing each state, by proving the following subgoals in order: 
\begin{enumerate}
    \item $\exists \bfw$ s.t. $\gl{\bfw}{\SensorProc{-}{S_5}{T}} \in \LInt{\tpunitstycst{t_4}{T + 20 \leq t_4}}{T}$
    \item $\exists \bfw$ s.t. $\gl{\bfw}{\SensorProc{-}{S_4}{T}} \in \LInt{\tpbangstycst{t_3}{T + 30 \leq t_3}{\tau_\kw{Gas}}{\tpunitstycst{t_4}{t_3 + 20 \leq t_4}}}{T}$
    \item $\exists \bfw$ s.t. $\gl{\bfw}{\SensorProc{-}{S_2}{T}} \in \LInt{\tpbangstycst{t_2}{T \leq t_2}{\tau_\kw{Temp}}{ \tpbangstycst{t_3}{t_2 + 30 \leq t_3 }{\tau_\kw{Gas}}{\tpunitstycst{t_4}{t_3 + 20\kw{ms} \leq t_4}}}}{T}$
    \item $\exists \bfw$ s.t. $\gl{\bfw}{\SensorProc{-}{S_3}{T}} \in \LInt{\tpunitstycst{t_3}{T \leq _3}}{T}$
    \item $\exists \bfw$ s.t. $\gl{\bfw}{\SensorProc{-}{S_1}{T}} \in \LInt{\tpbangstycst{t_2}{T \leq t_2}{\tau_\kw{Temp}}{\tpunitstycst{t_3}{t_2 \leq t_3}}}{T}$
\end{enumerate}

These goals analyze the automaton in reverse topological order, each building up from before. 
In each case, the proof mostly constitutes unfolding definitions and making observations. 

\begin{proof} As an example, consider the proof of goal (1) and (2). 
\begin{enumerate}
    \item Take a constant $\bfw$ s.t. $\bfw(t) \triangleq \SensorProc{-}{S_5}{T}$. 
    Our goal is to show that for all $T'$ s.t. $T + 20 \leq T'$, 
    $\bfw(T') \in \VInt{\expose{\tpunitstycst{t_4}{T + 20 \leq t_4}}{T'}}{T'}$.
    That is to fix $T''$ and show \newline $\SensorProc{-}{S_5}{T} \in \VInt{\unitstycst}{T''}$.
    By definition, suffices to fix $a$ and show that \newline
    $\SensorProc{a}{S_5}{T} \LLStepsTo{\actrcvclosecst{a}}{T''} \StopConf{}$. Immediate by \ruleref{rule:sensor:s5}.

    \item Again take $\bfw(t) \triangleq \SensorProc{-}{S_4}{T}$.  Fix any $T'$ s.t. $T + 30 \leq T'$. To show \newline
    $\bfw(t) \in \VInt{\expose{\bangstycst{\tau_\kw{Gas}}{\tpunitstycst{t_4}{t + 20 \leq t_4}}}{t}}{t}$.
    That is to fix $T''$ and show \newline
    $\SensorProc{-}{S_5}{T} \in \VInt{\bangstycst{\tau_\kw{Gas}}{\tpunitstycst{t_4}{T'' + 20 \leq t_4}}}{T''}$.
    By definition, suffices fix $a$ and show that 
    (a) for some $\bfOmega$,  $\SensorProc{a}{S_5}{T} \LLStepsTo{\actsnd{a}{\kw{val}(v_\kw{gas})}}{T''} \bfOmega[a]$, \newline and 
    (b) $\exists \bfw$ s.t. $\gl{\bfw}{\bfOmega} \in \LInt{t_4}{T'' + 20 \leq t_4}$. Let $\bfOmega$ be $\SensorProc{-}{S_5}{t}$. 
    Obligation (a) is immediate from \ruleref{rule:sensor:s4} and (b) 
    can be proved by appealing to above sub-proof. 
\end{enumerate}

The remaining cases proceed in a very similar fashion.
\end{proof}

\clearpage
\section{Implementation}
\subsection{DSL Encoding}
\label{apx:dsl}
A \tilst process is written by enclosing it inside a \rust{rtsm!~\{~\dots~\}} block.
The \rust{rtsm!} macro invokes our type checking pass on the DSL inside.
Within the DSL block, we can define \tilst types and processes via a syntax analogous to that used in \cref{main:sec:tillst}.
\cref{fig:rustsensor} shows an example encoding of the sensor process from \cref{main:sec:motivation}.
The process is defined as \dsl{fn sensor() { }-> BME680 \{ ... \}} where \dsl{BME680} is the type of the channel it provides.
Where functional layer values are required, we allow arbitrary Rust expressions delimited by \dsl{\$e\$}.
Temporal conditions are specified as \dsl{t where C} and 
are binders for time variable $t$ which must satisfy the proposition $C$.
When the Rust compiler encounters the \rust{rtsm!~\{~\dots~\}} block,
it invokes a procedural macro which runs our parser followed by our type checker,
both written in Rust.


\begin{figure}[h]
  \centering
  \footnotesize
  \input{fig/bme-code}
  \caption{The sensor hub process as implemented in the \tilst Rust DSL}
  \label{fig:rustsensor}
\end{figure}

\subsection{Rust DSL Examples}
\label{apx:rust:example}
In this section, we give details for the examples from the implementation section.
Table \cref{tbl:examples} lists the features exercised for each example. 

\begin{table}
  \caption{Examples expressed using \tilst in our Rust DSL}
  
  \begin{tabular}{| c | c | c | c | c | c | c | c | c |}
  \hline
  {\footnotesize Example}               &
  {\footnotesize Cut}                   &
  {\footnotesize Fwd}                   &

 {\footnotesize Unit ($\unitstycst$)}  &
  {\footnotesize Lolli ($\multimap$)}   &
  {\footnotesize Tensor ($\otimes$)}    &
  {\footnotesize In. choice ($\oplus$)} &
  {\footnotesize Ex. choice ($\&$)}     &
  {\footnotesize Fun. values} 
  \\
  \hline
  Minimum & {\footnotesize $\surd$} &  & {\footnotesize $\surd$} &   &  &  &  &   \\
  Sec. \cref{main:sec:tillst} $P_1, ..., P_4$ & {\footnotesize $\surd$} & {\footnotesize $\surd$} & {\footnotesize $\surd$} 
          & {\footnotesize $\surd$} & {\footnotesize $\surd$} &  &  &  \\
  Keyless entry & {\footnotesize $\surd$} &  & {\footnotesize $\surd$} & {\footnotesize $\surd$} &  
          & {\footnotesize $\surd$} &  & {\footnotesize $\surd$}  \\
  BME680  & {\footnotesize $\surd$} &  & {\footnotesize $\surd$} & {\footnotesize $\surd$} &  
          &  & {\footnotesize $\surd$} & {\footnotesize $\surd$} \\
  Colli. Detector   &  &  & {\footnotesize $\surd$} & {\footnotesize $\surd$} &  
          & {\footnotesize $\surd$} &  & {\footnotesize $\surd$} \\
  \hline
  \end{tabular}
  
  \centering
  \small
  \label{tbl:examples}
\end{table}

\subsubsection{Keyless Car Entry}
In this example, we adopt the keyless entry protocol from Rate-based session types \citep{IraciOOPSLA2023}. 
The main change is we use higher-order channel for the challenge-response part of the protocol. We choose 
this for better fitting the need of temporary inverse of provider-client role between the car and key. 
Note that the car is the client at the beginning, then it becomes a provider for the challenge in the 
authentication phase. 

\begin{dslblock}
type CHALLENGE =
        Produce<u128, t4 where In<t0, t4, Shift<t0, 95>>,
        Request<u128, t5 where In<t4, t5, Shift<t4, 5>>,
        Unit<t6 where Eq<t6, Shift<t0, 100>>>>>

type KEY =
        Request<unit, t1 where In<t0, t1, Shift<t0, 90>>,
        InChoice<t2 where In<t1, t2, Shift<t0, 90>>,
        Lolli<t3 where In<t2, t3, Shift<t0, 90>>,
          CHALLENGE, Unit<t7 where Eq<t7, Shift<t0, 100>>>>,
        Unit<t8 where Eq<t8, Shift<t0, 100>>>>>

type CAR = Unit<t9 where Eq<t9, Shift<t0, 100>>>

fn key() -> KEY {
    Query<t6 where In<t0, t6, Shift<t0, 90>>> {
        wake => SwitchL<t7 where In<t6, t7, Shift<t0, 90>>>;
        Lam<t8 where In<t7, t8, Shift<t0, 90>>> {
            c => Cons <t8> (c) {
                challenge => Supply <Shift<t8, 3>> (c)
                    $ challenge ^ 0x1_02_003_0004_00005 $ ;
                Wait <Shift<t0, 100>>(c);
                Close <t9 where Eq<t9, Shift<t0, 100>>>
            }}}}

fn car() -> CAR {
    Spawn <t0> (key) { k =>
        Supply <Shift<t0, 0>> (k) $ () $;
        Case <Shift<t0, 0>> (k)
        { L =>
            App <Shift<t0, 0>> ( k <= {
                Prod<t4 where In<t0, t4, Shift<t0, 95>>> 
                  $ 0x1337d00d_12345678 $;
                Query<t5 where In<t4, t5, Shift<t4, 5>>> { 
                  resp => 
                    Close <t6 where Eq<t6, Shift<t0, 100>>>
                }});
            Close <t7 where Eq<t7, Shift<t0, 100>>> }
        { R =>
            Wait <Shift<t0, 100>> (k);
            Close <t8 where Eq<t8, Shift<t0, 100>>>
        }}}
\end{dslblock}

\subsubsection{Collision Detector}
\usetikzlibrary{graphs}

In an Air Traffic Control (ATC) system, air planes employs a radar based system to detect 
near presence of other aircrafts in order to avoid collion. The aptly named collision detection system has three components, as shown in \cref{fig:atc-overall}. The \emph{Radar} scans surrounding airspace and periodically sends data to a processing unit \emph{CDx} for analysis. 
The \emph{Client} representing the on-board flight computer takes actions according 
to the result of the analysis. 

\begin{figure}[htbp]
    \begin{tikzpicture}
        \graph [nodes={draw, rectangle} ] 
            { Radar[at=((0:0))] -> CDx[xshift=1cm] -> Client[xshift=2cm] };
    \end{tikzpicture}
    \centering
    \caption{System overview of the ATC collision detection system}
    \label{fig:atc-overall}
\end{figure}

For the system to be effective, it is crucial actions are taken in a timely manner according to the sensory data. 
In particular, results need to be available as soon as possible in case of a potential collision. 
This imposes a hard deadline for the analysis component \emph{CDx}. \cref{fig:cdx} shows the processing pipeline of the CDx. 
Once the data has been received from the radar, it preforms some initial analysis to determine if a collision is immenent. 
If true then CDx has to return a result within 5ms. Therefore  preforms a fast-tracked rough analysis.
Otherwise, then it preforms a detailed analysis with a deadline of 10ms. 

\begin{figure}[htbp]
\begin{verbatim}
process CDx (x : Input) = 
  if InitialAnalysis(x) then     -- Initial analysis for immenent collision
    return FastAnalysis(x)       -- Yes; Provide rough but fast analysis  (5ms)
  else 
    return QuickAnalysis(x)      -- No; Provide detailed analysis (10ms)
\end{verbatim}
    \centering
    \caption{Dataflow chart for CDx}
    \label{fig:cdx}
\end{figure}

In this work we are interested in modelling CDx for a single iteration of this process.
Assuming the current time is $t_0$. The component may be assigned the following type: 

\begin{align*}
A_\kw{Radar} &\triangleq \tpbangstycst{t_2}{t_1 = t_2}{\tau_\kw{input}}{\tpunitstycst{t_3}{t_3 = t_2}} \\
\\
A_\kw{CDx}  &\triangleq \tplarrstycst{t_1}{t_0 \leq t_1}{A_\kw{Radar}}{(\tpsumstycst{t_2}{t_1 \leq t_2 \leq t_1 + 5ms}{A_\kw{Fast}}{A_\kw{Slow}})}  \\
A_\kw{Fast} &\triangleq \tpbangstycst{t_3}{t_1 + 5ms \leq t_3 \leq t_1 + 10ms}{\tau_\kw{result}}{\tpunitstycst{t_4}{t_3 \leq t_4 \leq t_1 + 10ms}} \\
A_\kw{Slow} &\triangleq \tpbangstycst{t_3}{t_3 = t_1 + 10ms}{\tau_\kw{result}}{\tpunitstycst{t_4}{t_4 = t_1 + 10ms}} \\
\end{align*}

The type stipulates that the CDx component is ready to receive data (of functional type $\tau_\kw{input}$) at any moment
$t_1$ in the future of now ($t_0$). 
Because the initial analysis takes negligible time, result is available immediately and is made available to 
the client right after the data is received and before $t_1 + 5ms$, as stipulated by the condition on $t_2$.
In the truth branch then it may take the remaining time to produce a quick 
analysis, as shown in $A_\kw{Fast}$. Otherwise, it has an extended deadline of $t_1 + 10ms$, as shown in $A_\kw{Slow}$. 
In either case, the process may terminate anytime after result is produced ($t_3$) and before $t_1 + 10ms$, 
which is the time that a new cycle would have initiated.

In our system, we may implement the CDx component using the following process term: 



\begin{align*}
\IsOfProc{P_\kw{CDx}}{t_0}{A_\kw{CDx}} & \triangleq
\begin{aligned}
& \tplarrrcvcst{t_1}{t_0 \leq t_1}{A_\kw{Radar}}{r}{} \\
& x \leftarrow \kw{consume}^{t_1}\,r; \kw{close}^{t_1} r; \\
& \kw{if}\;\kw{Initial}(x)\; \\
& \kw{then}\; (\tpinlcst{t_2}{t_1 \leq t_2 \leq t_1 + 5ms}{A_1}{A_2}{P_\kw{Fast}}) \\
& \kw{else}\;(\tpinrcst{t_2}{t_1 \leq t_2 \leq t_1 + 5ms}{A_1}{A_2}{P_\kw{Slow}}) \\
\end{aligned}
\\
\IsOfProc{P_\kw{Fast}}{t_1}{A_\kw{Fast}} & \triangleq 
    \tpproduceproccst{t_3}{t_1 + 5ms \leq t_3 \leq t_1 + 10ms}{\kw{Fast}(x)}{\tpclosecst{t_4}{t_3 \leq t_4 \leq t_1 + 10ms}}
 \\
\IsOfProc{P_\kw{Slow}}{t_1}{A_\kw{Slow}} & \triangleq 
    \tpproduceproccst{t_3}{t_3 = t_1 + 10ms}{\kw{Slow}(x)}{\tpclosecst{t_4}{t_4 = t_1 + 10ms}} \\
\end{align*}

Next, we provide the process terms implemented in Rust DSL, note that we also provide the terms for Radar.

\begin{dslblock} 
type Tradar = Produce <sort_input, t1 where Geq <t1, t0>, 
              Unit<t2 where Eq <t1, t2>>>
type Tcdx = Lolli <t1 where Geq <t1, t0>, Tradar,
            InChoice <t2 where In <t1, t2, Shift<t1, 5>>, 
              Tfast, Tslow>>
type Tfast = Produce <sort_result, t3 where 
                       In <Shift <t1, 5>, t3, Shift<t1, 10>>, 
             Unit<t4 where In<t3, t4, Shift<t1, 10>>>>
type Tslow = Produce <sort_result, t3 where 
                       Eq <t3, Shift<t1, 10>>, 
             Unit<t4 where Eq <t4, Shift<t1, 10>>>>

fn radar () -> Tradar {
    Prod <t2 where Geq<t2, t0>> $ radar_input() $;
    Close <t3 where Eq<t2, t3>>
}

fn cdx () -> Tcdx {
    Lam <t1 where Geq <t1, t0>> { r =>
        Cons <t1> (r) { x =>
            Wait <t1> (r);
            if $ initial(x) $
            then
                SwitchL <t2 where In <t1, t2, Shift <t1, 5>>>;
                Prod 
                  <t3 where In<Shift<t1, 5>, t3, Shift<t1, 10>>> 
                  $ fast() $;
                Close <t4 where In<t3, t4, Shift<t1, 10>>>
            else
                SwitchR <t2 where In <t1, t2, Shift <t1, 5>>>;
                Prod 
                  <t3 where Eq <t3, Shift<t1, 10>>> 
                  $ slow() $;
                Close <t4 where Eq <t4, Shift<t1, 10>>>
        }
    }
}
\end{dslblock}

\fi

\end{document}